\documentclass[twocolumn,superscriptaddress,preprintnumbers,amsmath,amssymb]{revtex4}
\usepackage[pdftex]{graphicx}% Include figure files
\usepackage{dcolumn}% Align table columns on decimal point
\usepackage{bm}% bold math
\usepackage{color}
\usepackage{ulem}
\usepackage{amsmath,amssymb}
\begin{document}

\title{Rotational motion of a camphor disk in a circular region}

\author{Yuki~Koyano\footnote{Corresponding author. E-mail: y.koyano@chiba-u.jp.}}
\affiliation{Department of Physics, Chiba University, 1-33 Yayoi-cho, Inage-ku, Chiba 263-8522, Japan}

\author{Nobuhiko~J.~Suematsu}
\affiliation{Graduate School of Advanced Mathematical Sciences, Meiji University, 4-21-1 Nakano, Tokyo 164-8525, Japan}
\affiliation{Meiji Institute of Advanced Study of Mathematical Sciences (MIMS), Meiji University, 4-21-1 Nakano, Tokyo 164-8525, Japan}

\author{Hiroyuki~Kitahata}
\affiliation{Department of Physics, Chiba University, 1-33 Yayoi-cho, Inage-ku, Chiba 263-8522, Japan}

\begin{abstract}
In a two-dimensional axisymmetric system, the system symmetry allows rotational or oscillatory motion as stable stationary motion for a symmetric self-propelled particle.
In the present paper, we studied the motion of a camphor disk confined in a two-dimensional circular region.
By reducing the mathematical model describing the dynamics of the motion of a camphor disk and the concentration field of camphor molecules on a water surface, we analyzed the reduced equations around a bifurcation point where the rest state at the center of the system becomes unstable.
As a result, we found that rotational motion is stably realized through the double-Hopf bifurcation from the rest state.
The theoretical results were confirmed by numerical calculation and well corresponded to the experimental results.
\end{abstract}

\maketitle

\section{Introduction}

There are many systems which show self-propelled motion~\cite{Ramaswamy2010, Ohta2017, Marchetti2013, Bechinger2016}.
In such self-propelled systems, particles move consuming and dissipating free energy, where the symmetric property is one of the classificatory criteria.
First of all, self-propelled systems are classified into symmetric and asymmetric systems.
A symmetric self-propelled particle moves through spontaneous symmetry breaking, while an asymmetric one moves in the direction predetermined by the asymmetry.
Here we focus on a symmetric self-propelled system.

In a two-dimensional axisymmetric system, a rotational or oscillatory motion is often observed~\cite{Wioland2013, Ariel2015, Sumino2008, Pimienta2011, Pimienta2014, Takabatake2014, Tanaka2015}.
Here we define a rotational motion as the motion where the particle moves at a constant distance from the system center with a nonzero constant speed, while
an oscillatory motion as a motion where the particle reciprocally moves back and forth along a line through the system center.
Bacteria confined in a quasi-two dimensional circular chamber exhibit rotational motion~\cite{Wioland2013}.
A laser-heated oil droplet exhibits oscillatory or rotational motion depending on the laser intensity~\cite{Takabatake2014}.
Phenomenological models which can reproduce rotational or oscillatory motion have been proposed~\cite{Mikhailov2002, Sumino2008, Schweitzer1998, Schweitzer2003}.
To understand which motion, rotation or oscillation, is realized in a two-dimensional axisymmetric system, we have discussed criteria by a theoretical approach based on a weakly nonlinear analysis~\cite{Koyano2015}.
In the analysis, only two assumptions are imposed; one is that a self-propelled particle is confined by a harmonic potential.
The other is that the system is near the bifurcation point where rest state at the minimum of the harmonic potential becomes unstable.
However, our proposed criteria have not been applied to actual systems.
A camphor-water system is one of the candidates to which our criteria can be applied.

In a camphor-water system~\cite{Skey1878, Tomlinson1862, Rayleigh1889, Nakata1997, Nakata2015}, self-propulsion of a camphor particle is induced in the physicochemical processes; when a camphor particle is placed onto the water surface, it releases camphor molecules.
The released camphor molecules are distributed at the water surface and reduce the surface tension.
In other words, a camphor particle releases its repellants (camphor molecules) and moves to the region with less repellent concentration.
The camphor-water system is so simple that various geometries can be realized~\cite{Nakata2016, Koyano2016, Gorecki2017, Kitahata2013, Iida2014, Ei2018, Koyano2017, Mimura2007, Chen2009, Miyaji2017, Soh2008, Nishimori2017, Nakata2018}.
For example, the size of the water chamber can be one of the parameters which can affect the motion of the camphor disk~\cite{Nakata2016, Koyano2016}.
The shape of the particle can also be changed, e.g., an elliptic camphor disk~\cite{Kitahata2013, Iida2014, Ei2018} and a string-shaped camphor~\cite{Nakata2018}.
The mathematical model reflecting the elemental processes is also available~\cite{Nagayama2004, Nakata2015, Nakata2018_book}.
The model consists of two equations; one is the equation of motion describing the dynamics of the position of the camphor particle, and the other is the reaction-diffusion equation describing the concentration field of camphor molecules distributed at the water surface.
From the viewpoint of the theoretical analysis, the model for the camphor-water system is easy to analyze since the reaction-diffusion equation for the concentration field of camphor molecules at water surface has a reaction term described by a piecewise linear function.
The advantage of the camphor-water system is thus that the theoretical and experimental approaches are both available~\cite{Nagayama2004, Nakata2015, Koyano2016, Koyano2017, Ei2018, Soh2008, Nishimori2017}.
Besides, a camphor-water system is considered to be one of the simplest negative-chemotactic system~\cite{Mikhailov2002, Iida2010, Nagai2013, Pimienta2011, Pimienta2014, Yamamoto2015, Bansagi2013, Lagzi2010}, and can be applied to other systems~\cite{Vecchiarelli2014, Banigan2016}.

In the present paper, we consider the dynamics of a camphor disk confined in a two-dimensional circular water phase as a good example of a symmetric self-propelled particle in a two-dimensional axisymmetric system.
We apply our previously proposed criteria \cite{Koyano2015} to it in order to discuss which motion, rotation or oscillation, is selected.
First, we introduce the mathematical model constructed based on the elemental physicochemical processes in Sec.~II.
Then we reduce it into a dynamical system described by a single ordinary differential equation and apply the result of the weakly nonlinear analysis in Sec.~III.
The theoretical result has been compared with the numerical and experimental results in Secs.~IV and V, respectively.

\section{Mathematical Model}

\begin{figure}
	\begin{center}
		\includegraphics[bb=0 0 184 154]{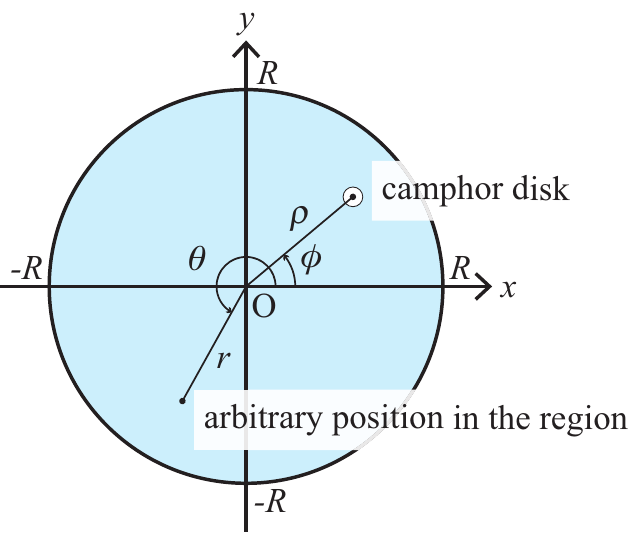}
	\end{center}
	\caption{Schematic illustration of the considered system.
		The position of the camphor disk and an arbitrary position are denoted as $\bm{\rho} = (\rho, \phi)$ and $\bm{r} = (r, \theta)$, respectively, in the two-dimensional polar coordinates.}
	\label{def_region}
\end{figure}

We consider the motion of a camphor disk confined in a two-dimensional circular region with a radius of $R$.
Here we introduce a mathematical model, which is almost the same as that in Ref.~\cite{Koyano2016} except for the boundary condition of the system.
The center position of the camphor disk is represented by $\bm{\rho}(t) = (\rho(t), \phi(t))$ in the two-dimensional polar coordinates.
Here, $\rho$ and $\phi$ are in domains $[0,R]$ and $[0, 2 \pi)$, respectively.
The configuration of the system is illustrated in Fig.~\ref{def_region}.

The equation of motion with regard to the center position of a camphor disk is described as:
\begin{equation}
\label{eq.of.m.}
\sigma S \frac{d^2 \bm{\rho}}{dt^2} = - \xi S \frac{d \bm{\rho}}{dt} + \bm{F}_d (c;\bm{\rho}),
\end{equation}
where $\sigma$ and $\xi$ are the mass and resistant coefficient per unit area, $S (= \pi \epsilon^2)$ is the surface area of the camphor disk, and $\bm{F}_d$ denotes the driving force originating from the surface tension difference.
Here, we set the radius of the camphor disk to be $\epsilon$.
The resistant force proportional to the velocity was confirmed in the previous study~\cite{Suematsu2014}.

The driving force $\bm{F}_d$ is modeled by summing up the driving force working on the periphery of the camphor disk.
To avoid the dependence of $\epsilon$, we divide the both sides of Eq.~\eqref{eq.of.m.} by $S$, and then we take the limit that $\epsilon$ goes to $+0$ as follows:
\begin{equation}
\label{eq.of.m._2}
\sigma \frac{d^2 \bm{\rho}}{dt^2} = - \xi \frac{d \bm{\rho}}{dt} + \bm{F} (c;\bm{\rho}).
\end{equation}
The driving force per unit area of the camphor disk is given by:
\begin{align}
\label{def.drv.f}
\bm{F} (c; \bm{\rho} (t)) =& \lim_{\epsilon \to +0} \frac {1}{S} \bm{F}_d
= \lim_{\epsilon \to +0} \frac {1}{S} \int_{\partial \Omega} \gamma \left ( c (\bm{\rho} + \epsilon \bm{n}) \right ) \bm{n} d\ell,
\end{align}
where $\partial \Omega$ is the periphery of the region $\displaystyle{ \Omega = \left \{ \bm{r} \middle | \left| \bm{r}-\bm{\rho} \right | \leq \epsilon \right \} }$, which is a circular region corresponding to the camphor disk with a radius of $\epsilon$, and $\bm{n} = \bm{n}(\theta)$ is a unit vector represented as $\bm{n}(\theta) = (\cos \theta, \sin \theta)$ in the Cartesian coordinates.
When the concentration field can be expanded around $\bm{r} = \bm{\rho}$, the definition of driving force in Eq.~\eqref{def.drv.f} is expressed as:
\begin{align}
\bm{F} (c; \bm{\rho} (t)) =& \lim_{\epsilon \to +0} \frac {-k}{\pi \epsilon^2} \int_{0}^{2 \pi} \left [ c (\bm{\rho}) + \epsilon \bm{n}(\theta) \cdot \nabla c (\bm{\rho}) \right ] \bm{n}(\theta) \epsilon d\theta \\
=& -k \left . \nabla c \right |_{\bm{r}=\bm{\rho}}.
\end{align}
The dependence of surface tension on camphor concentration field was experimentally measured (20 mN/m decrease in surface tension for the saturated camphor aqueous solution)~\cite{Ikura2012}.
The sufrace tension decrease around a moving camphor disk was also measured as 4 mN/m~\cite{Kayahara2018}.
Here, we assume that the surface tension $\gamma$ is a linear decreasing function with regard to the camphor concentration $c$, i.e., $\gamma = - k c + \gamma_0$, where $k$ and $\gamma_0$ are positive constants.
$\gamma_0$ gives the surface tension of pure water.
Hereafter, we consider Eq.~\eqref{eq.of.m._2} as for the motion of a camphor disk.

The time evolution for concentration field $c = c(\bm{r},t)$ at the water surface is described by the following equation:
\begin{equation}
\label{d.eq.}
\frac{\partial c}{\partial t} = D \left ( \frac{\partial^2}{\partial r^2} + \frac{1}{r} \frac{\partial}{\partial r} + \frac{1}{r^2} \frac{\partial^2}{\partial \theta^2} \right ) c - \alpha c + f,
\end{equation}
where $\bm{r} = (r, \theta)$ is an arbitrary position in the circular region in polar coordinates, $D$ is the effective diffusion constant, $\alpha$ is the sublimation rate, and $f$ denotes the release of camphor molecules from the camphor disk.
Here we consider a two-dimensional system corresponding to the water surface since camphor molecules hardly dissolve into water.
Actually, the saturated concentration of camphor aqueous solution was reported as around 10 mM~\cite{Ikura2012}.
The source term $f$ is given by
\begin{align}
\label{source_finite}
f (\bm{r};\bm{\rho}) = \frac{c_0}{\pi \epsilon^2} \Theta \left ( \epsilon - | \bm{r} - \bm{\rho} | \right ) 
=
\left \{
\begin{array}{ll}
\displaystyle{\frac{c_0}{\pi \epsilon^2}}, & (| \bm{r} - \bm{\rho} | \leq \epsilon), \\
\hspace{-10mm}\\
0, & (| \bm{r} - \bm{\rho} | > \epsilon),
\end{array}
\right .
\end{align}
where $\Theta (\cdot)$ denotes the Heaviside step function and $c_0$ is the total amount of the release of camphor molecules from the camphor disk per unit time.
By taking the limit that $\epsilon$ goes to $+0$, the source term is expressed as
\begin{equation}
\label{source}
f (\bm{r};\bm{\rho}) = c_0 \delta (\bm{r}-\bm{\rho}) =
\left \{
\begin{array}{ll}
\displaystyle{\frac{c_0}{r} \delta (r-\rho) \delta (\theta-\phi)}, & (\rho > 0), \\
\hspace{-10mm}\\
\displaystyle{\frac{c_0}{\pi r} \delta (r-\rho)}, & (\rho = 0),
\end{array}
\right .
\end{equation}
where $\delta(\cdot)$ denotes the Dirac's delta function.
Here we assume that the camphor molecules are released constantly at the position of the camphor disk $\bm{\rho} = (\rho(t), \phi(t))$ in polar coordinates.
The Neumann boundary condition:
\begin{equation}
\left . \frac{\partial c}{\partial r} \right |_{r=R} = 0, \label{Neumann}
\end{equation}
is imposed to Eq.~\eqref{d.eq.}.

In the following analysis, dimensionless forms of Eqs.~\eqref{eq.of.m._2}, \eqref{def.drv.f}, \eqref{d.eq.}, \eqref{source}, and \eqref{Neumann},
\begin{equation}
\label{d_eq.of.m._2}
\tilde{\sigma} \frac{d^2 \tilde{\bm{\rho}}}{d\tilde{t}^2} = - \tilde{\xi} \frac{d \tilde{\bm{\rho}}}{d\tilde{t}} + \tilde{\bm{F}} (\tilde{c}; \tilde{\bm{\rho}}),
\end{equation}
\begin{align}
\label{d_def.drv.f}
\tilde{\bm{F}} (\tilde{c}; \tilde{\bm{\rho}} (\tilde{t})) =& \lim_{\tilde{\epsilon} \to +0} \frac {1}{\tilde{S}} \int_{\partial \tilde{\Omega}} \tilde{\gamma} \left ( \tilde{c} (\tilde{\bm{\rho}} + \tilde{\epsilon} \bm{n}) \right ) \bm{n} d\tilde{\ell} \\
=& -\left . \tilde{\nabla} \tilde{c} \right |_{\tilde{\bm{r}}=\tilde{\bm{\rho}}}.
\end{align}
\begin{equation}
\label{d_d.eq.}
\frac{\partial \tilde{c}}{\partial \tilde{t}} = \left ( \frac{\partial^2}{\partial \tilde{r}^2} + \frac{1}{\tilde{r}} \frac{\partial}{\partial \tilde{r}} + \frac{1}{\tilde{r}^2} \frac{\partial^2}{\partial \theta^2} \right ) \tilde{c} - \tilde{c} + \tilde{f},
\end{equation}
\begin{equation}
\label{d_source}
\tilde{f} (\tilde{\bm{r}};\tilde{\bm{\rho}}) = \delta (\tilde{\bm{r}}-\tilde{\bm{\rho}}),
\end{equation}
\begin{equation}
\label{d_Neumann}
\left . \frac{\partial \tilde{c}}{\partial \tilde{r}} \right | _{\tilde{r} = \tilde{R}} = 0,
\end{equation}
are used.
The details of nondimensionalization are shown in Appendix~\ref{appA}.
Hereafter, we omit tildes ($\tilde{}$).

It should be noted that the hydrodynamic effect is neglected in the present model.
In this system, the inhomogeneities of surface tension are brought about by the camphor concentration gradient, and then the Marangoni flow occurs~\cite{Soh2008, Lauga2012, Kitahata2018, Suematsu2014}.
The profile of the concentration field of camphor molecules becomes broader due to the Marangoni flow, which can be included as the effective diffusion constant in Eq.~\eqref{d.eq.}~\cite{Kitahata2018, Suematsu2014}.

\section{Theoretical Analysis}

The equations are reduced around the rest state where the camphor disk rests at the center position of the circular region.
First, the concentration field $c$ is expanded with Bessel function and Fourier series for radial and angular directions, respectively, as
\begin{align}
c(r,\theta,t) = \frac{1}{2 \pi} \sum_{m = -\infty}^{\infty} \sum_{n = 0}^{\infty} a_{|m|n} c_{mn}(t) \mathcal{J}_{|m|} (k_{|m|n} r) e^{i m \theta},
\end{align}
where $\mathcal{J}_{m}$ is the first-kind Bessel function of $m$-th order.
Here $k_{mn}$ and $a_{mn}$ are the wavenumber and the normalization constant, respectively. 
We adopt wavenumbers $k_{mn}$ such that the bases satisfy the Neumann boundary condition in Eq.~\eqref{d_Neumann}.
The explicit representations of $k_{mn}$ and $a_{mn}$ are provided in Appendix~\ref{appB}.
The source term in Eq.~\eqref{d_source} is also expanded as
\begin{align}
& f(r,\theta;\rho,\phi) = \frac{1}{2 \pi} \sum_{m = -\infty}^{\infty} \sum_{n = 0}^{\infty} a_{|m|n} \mathcal{J}_{|m|} (k_{|m|n} \rho(t)) \nonumber \\
& \qquad \qquad \qquad \times \mathcal{J}_{|m|} (k_{|m|n} r) e^{i m (\theta-\phi(t))}.
\end{align}
Thus we have the equation for concentration in wavenumber space:
\begin{align}
\frac{d c_{mn}}{dt} = - ({k_{|m|n}}^2 + 1) c_{mn} + \mathcal{J}_{|m|} (k_{|m|n} \rho(t)) e^{-i m \phi(t)}.
\end{align}
Then, the Green's function $g_{mn}(t)$ is calculated.
Here, the Green's function is defined as a function that satisfies the following equation:
\begin{align}
\frac{d g_{mn}}{dt} = - ({k_{|m|n}}^2 + 1) g_{mn} + \delta(t).
\end{align}
By solving the above equation, we have
\begin{align}
g_{mn}(t) = e^{ -({k_{|m|n}}^2 + 1) t} \Theta (t)
= \left \{
\begin{array}{ll}
e^{-({k_{|m|n}}^2 + 1) t}& (t \geq 0), \\
0& (t<0),
\end{array}
\right .
\end{align}
where $\Theta (t)$ is the Heaviside step function.

Using the Green's function $g_{mn}$, the concentration field $c_{mn}$ in wavenumber space is described as
\begin{align}
&c_{mn}(\bm{\rho}(t)) \nonumber \\
&= \int_{-\infty}^{t} \mathcal{J}_{|m|} (k_{|m|n} \rho({t}')) e^{-i m \phi({t}')} e^{ -({k_{|m|n}}^2 + 1) (t-{t}')} d{t}'. \label{integration_of_green.f}
\end{align}
By expanding the above integration with regard to the time using partial integration, we have the concentration field in the wavenumber space expanded with regard to the current position, velocity, accerelation, and so on~\cite{Ohta2009_2, Koyano2016, Koyano2017}.
Then the expanded concentration field is converted into the real space.
The detailed calculations are presented in Appendix~\ref{appC}.
Using the definition of the driving force in Eq.~\eqref{d_def.drv.f}, the driving force $\bm{F}$ is calculated as follows:
\begin{align}
\label{DRIVING_FORCE}
\bm{F} (\bm{\rho}, \dot{\bm{\rho}}, \ddot{\bm{\rho}}) =& a(R) \bm{\rho} + b(R) \dot{\bm{\rho}} + c(R) |\bm{\rho}|^2 \bm{\rho} + g(R) \ddot{\bm{\rho}} \nonumber \\
& + h(R) |\dot{\bm{\rho}}|^2 \bm{\rho} + j(R) (\bm{\rho} \cdot \dot{\bm{\rho}}) \bm{\rho} + k(R) |\dot{\bm{\rho}}|^2 \dot{\bm{\rho}} \nonumber \\
& + n(R) |\bm{\rho}|^2 \dot{\bm{\rho}} + p(R) (\bm{\rho} \cdot \dot{\bm{\rho}}) \dot{\bm{\rho}},
\end{align}
where $\bm{F}$ is now the function of $\bm{\rho}$, $\dot{\bm{\rho}}$, and $\ddot{\bm{\rho}}$.
The coefficient of each term in Eq.~\eqref{DRIVING_FORCE} is explicitly obtained as:
\begin{align}
&a (R) = \frac{1}{4\pi} \left ( \frac{\mathcal{K}'_{0}(R)}{\mathcal{I}'_{0}(R)} + \frac{\mathcal{K}'_{1}(R)}{\mathcal{I}'_{1}(R)} \right ), \label{AR}
\end{align}
\begin{align}
&b (R) = \frac{1}{4\pi} \left ( -\gamma_{\mathrm{Euler}} + \log \frac{2}{\epsilon} \right ) \nonumber \\
& \qquad + \frac{1}{8\pi} \left [ 2 \frac{\mathcal{K}'_{1}(R)}{\mathcal{I}'_{1}(R)} + \left ( 1 + \frac{1}{R^2} \right ) \frac{1}{(\mathcal{I}'_{1}(R))^2} \right ], \label{BR}
\end{align}
\begin{align}
&c (R) = \frac{1}{32\pi} \left ( 3 \frac{\mathcal{K}'_{0}(R)}{\mathcal{I}'_{0}(R)} + 4 \frac{\mathcal{K}'_{1}(R)}{\mathcal{I}'_{1}(R)} + \frac{\mathcal{K}'_{2}(R)}{\mathcal{I}'_{2}(R)} \right ), \label{CR}
\end{align}
\begin{align}
g (R) = - \frac{1}{16\pi} \left [ 1 + \left ( R + \frac{1}{R} \right ) \frac{\mathcal{I}''_{1}(R)}{(\mathcal{I}'_{1}(R))^3} - \frac{1}{(\mathcal{I}'_{1}(R))^2} \right ], \label{GR}
\end{align}
\begin{align}
h (R) &= \frac{1}{64\pi} \left [ 8 \frac{\mathcal{K}'_{0}(R)}{\mathcal{I}'_{0}(R)} + 4 \frac{\mathcal{K}'_{1}(R)}{\mathcal{I}'_{1}(R)} - 4 \frac{\mathcal{K}'_{2}(R)}{\mathcal{I}'_{2}(R)} \right. \nonumber \\
& \quad - 2 R \frac{\mathcal{I}''_{0}(R)}{(\mathcal{I}'_{0}(R))^3} - \left ( R + \frac{1}{R} \right ) \frac{\mathcal{I}''_{1}(R)}{(\mathcal{I}'_{1}(R))^3} \nonumber \\
& \quad + \left ( R + \frac{4}{R} \right ) \frac{\mathcal{I}''_{2}(R)}{(\mathcal{I}'_{2}(R))^3} + \frac{6}{(\mathcal{I}'_{0}(R))^2} \nonumber \\
& \quad \left . + \left ( \frac{2}{R^2} + 3 \right ) \frac{1}{(\mathcal{I}'_{1}(R))^2} - \left ( \frac{8}{R^2} + 3 \right ) \frac{1}{(\mathcal{I}'_{2}(R))^2} \right ], \label{HR}
\end{align}
\begin{align}
j (R) &= \frac{1}{16\pi} \left [ 4 \frac{\mathcal{K}'_{0}(R)}{\mathcal{I}'_{0}(R)} + 4 \frac{\mathcal{K}'_{1}(R)}{\mathcal{I}'_{1}(R)} + \frac{1}{(\mathcal{I}'_{0}(R))^2} \right . \nonumber \\
& \quad \left . + \left ( 1 + \frac{1}{R^2} \right ) \frac{1}{(\mathcal{I}'_{1}(R))^2} \right ], \label{JR}
\end{align}
\begin{align}
k (R) &= \frac{1}{128\pi} \left [ - 4 + 3 \left ( 1 + R^2 \right ) \frac{(\mathcal{I}''_{1}(R))^2}{(\mathcal{I}'_{1}(R))^4} + \frac{4}{(\mathcal{I}'_{1}(R))^2} \right . \nonumber \\
& \quad \left . - \left ( \frac{3}{R} + 7 R \right ) \frac{\mathcal{I}''_{1}(R)}{(\mathcal{I}'_{1}(R))^3} - \left ( 1 + R^2 \right ) \frac{\mathcal{I}'''_{1}(R)}{(\mathcal{I}'_{1}(R))^3} \right ], \label{KR}
\end{align}
\begin{align}
n (R) &= \frac{1}{32\pi} \left [ \left ( 1 + \frac{1}{R^2} \right ) \frac{1}{(\mathcal{I}'_{1}(R))^2} \right . \nonumber \\
& \quad \left . + \left ( 1 + \frac{4}{R^2} \right ) \frac{1}{(\mathcal{I}'_{2}(R))^2} + 4 \frac{\mathcal{K}'_{2}(R)}{\mathcal{I}'_{2}(R)} + 4 \frac{\mathcal{K}'_{1}(R)}{\mathcal{I}'_{1}(R)} \right ], \label{NR}
\end{align}
\begin{align}
p (R) &= \frac{1}{32\pi} \left [ 4 \frac{\mathcal{K}'_{1}(R)}{\mathcal{I}'_{1}(R)} + 4 \frac{\mathcal{K}'_{2}(R)}{\mathcal{I}'_{2}(R)} - \left ( \frac{1}{R} + R \right ) \frac{\mathcal{I}''_{1}(R)}{(\mathcal{I}'_{1}(R))^3} \right . \nonumber \\
& \quad - \left ( \frac{4}{R} + R \right ) \frac{\mathcal{I}''_{2}(R)}{(\mathcal{I}'_{2}(R))^3} + \left ( \frac{2}{R^2} + 3 \right ) \frac{1}{(\mathcal{I}'_{1}(R))^2} \nonumber \\
& \quad \left . + \left ( \frac{8}{R^2} + 3 \right ) \frac{1}{(\mathcal{I}'_{2}(R))^2} \right ], \label{PR}
\end{align}
where $\mathcal{I}_m$ and $\mathcal{K}_m$ are the first- and second-kinds modified Bessel function of $m$-th order, and $\gamma_{\mathrm{Euler}}$ is the Euler-Mascheroni constant ($\gamma_{\mathrm{Euler}} \simeq 0.577$).
The coefficients $a(R)$ and $g(R)$ are both negative independent of $R (\geq 0)$.
Thus the terms, $a(R) \bm{\rho}$ and $g(R) \ddot{\bm{\rho}}$, mean harmonic potential force and inertia, respectively.
The plots of Eqs.~\eqref{AR} to \eqref{PR} against $R$ are shown in Supplemental Material.

We confirmed that, when $R$ goes to infinity, $a(R)$, $c(R)$, $h(R)$, $j(R)$, $n(R)$, and $p(R)$ go to zero, and $b(R)$, $g(R)$, and $k(R)$ go to $(-\gamma_{\mathrm{Euler}} + \log (2/\epsilon))/(4\pi)$, $-1/(16\pi)$, and $-1/(32\pi)$, respectively.
The coefficients for $R \to \infty$ are consistent with the results for the case where a camphor disk is in an infinite two-dimensional region~\cite{Koyano2017}.

\begin{figure}
	\begin{center}
		\includegraphics[bb=0 0 143 115]{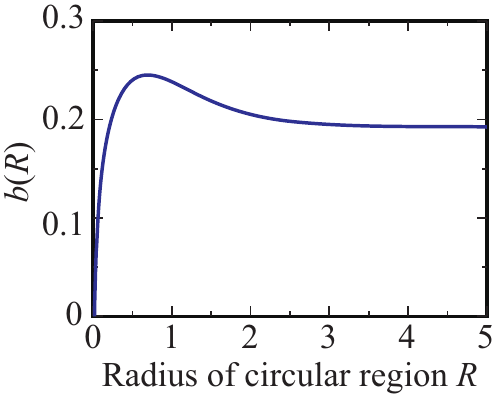}
		\caption{Profile of $b(R)$.
		Here we set $\epsilon = 0.1$.}
		\label{br}
	\end{center}
\end{figure}

As the result of the reduction, we have the following dynamical system as
\begin{align}
& (\sigma - g(R)) \ddot{\bm{\rho}} \nonumber \\
& = a(R) \bm{\rho} + (b(R)-\xi) \dot{\bm{\rho}} + c(R) |\bm{\rho}|^2 \bm{\rho} + h(R) |\dot{\bm{\rho}}|^2 \bm{\rho} \nonumber \\
& \quad + j(R) ( \bm{\rho} \cdot \dot{\bm{\rho}} ) \bm{\rho} + k(R) |\dot{\bm{\rho}}|^2 \dot{\bm{\rho}} + n(R) |\bm{\rho}|^2 \dot{\bm{\rho}} \nonumber \\
& \quad + p(R) ( \bm{\rho} \cdot \dot{\bm{\rho}} ) \dot{\bm{\rho}}. \label{reduced_equation}
\end{align}
In this dynamical system, a double-Hopf bifurcation occurs at $b(R) = \xi$, where the coefficient of $\dot{\bm{\rho}}$ is zero.
When $\xi$ is greater than $b(R)$, the rest state where the camphor disk rests at the center position of the circular region is stable.
On the other hand, the rest state becomes unstable for $\xi < b(R)$.
The profile of $b(R)$ has a single peak as shown in Fig.~\ref{br}, which is similar to the case for the one-dimensional finite system~\cite{Koyano2016}.

To clarify what motion occurs in the case that the rest state becomes unstable through the double-Hopf bifurcation, we apply the results obtained by weakly nonlinear analysis reported in our previous paper~\cite{Koyano2015}.
Here we assume that the terms, $(\sigma - g(R)) \ddot{\bm{\rho}}$ and $a(R) \bm{\rho}$, which cause harmonic oscillation, are the main terms, and the other terms are the perturbative terms.
Such an assumption is valid when the bifurcation parameters are close to the bifurcation point.
We convert time $t$ to $\tau = \omega(R,\sigma) t$, where $\omega(R,\sigma) = \sqrt{-a(R)/(\sigma - g(R))}$, in order to rescale the time with regard to the period of the harmonic oscillation.
Then we have
\begin{align}
\label{dyn_sys}
\ddot{\bm{\rho}} =& -\bm{\rho} + (B(R,\sigma)-\Xi) \dot{\bm{\rho}} + C(R,\sigma) |\bm{\rho}|^2 \bm{\rho} + H(R) |\dot{\bm{\rho}}|^2 \bm{\rho} \nonumber \\
& + J(R,\sigma) ( \bm{\rho} \cdot \dot{\bm{\rho}} ) \bm{\rho} + K(R,\sigma) |\dot{\bm{\rho}}|^2 \dot{\bm{\rho}} + N(R,\sigma) |\bm{\rho}|^2 \dot{\bm{\rho}} \nonumber \\
& + P(R) ( \bm{\rho} \cdot \dot{\bm{\rho}} ) \dot{\bm{\rho}},
\end{align}
where $B(R,\sigma) = b(R) / \omega(R,\sigma)$, $\Xi = \xi / \omega(R,\sigma)$, $C(R,\sigma) = c(R) / \omega(R,\sigma)^2$, $H(R) = h(R)$, $J(R,\sigma) = j(R) / \omega(R,\sigma)$, $K(R,\sigma) = k(R) \omega(R,\sigma)$, $N(R,\sigma) = n(R) / \omega(R,\sigma)$, and $P(R) = p(R)$.

In our previous paper~\cite{Koyano2015}, we have derived the conditions for stable rotation:
\begin{equation}
\label{criterion_rot}
\left \{
\begin{array}{l}
F_{\mathrm{rot}}(R, \sigma) < 0, \\
F_{\mathrm{crt}}(R,\sigma) < 0,
\end{array}
\right .
\end{equation}
and for stable oscillation:
\begin{equation}
\label{criterion_osc}
\left \{
\begin{array}{l}
F_{\mathrm{osc}}(R, \sigma) < 0, \\
F_{\mathrm{crt}}(R,\sigma) > 0,
\end{array}
\right .
\end{equation}
where we define the following functions:
\begin{align}
F_{\mathrm{rot}}(R, \sigma) =& K(R,\sigma) + N(R,\sigma),\\
F_{\mathrm{crt}}(R,\sigma) =& K(R,\sigma) - N(R,\sigma) + J(R,\sigma),\\
F_{\mathrm{osc}}(R, \sigma) =& 3K(R,\sigma) + N(R,\sigma) + J(R,\sigma).
\end{align}
Based on the conditions in Eqs.~\eqref{criterion_rot} and \eqref{criterion_osc}, we can learn which motion appears, rotational or oscillatory motion, when the rest state becomes unstable, i.e., $\xi < b(R)$.
The conditions include the radius of the system $R$ and the mass density $\sigma$ as parameters.
When $F_{\mathrm{rot}}(R,\sigma)$ and $F_{\mathrm{osc}}(R, \sigma)$ are negative, the solution corresponding to a small-amplitude rotational and oscillatory motion exists, respectively.
As for $F_{\mathrm{crt}}(R,\sigma)$, the rotational motion is stable if it is negative, while the oscillatory motion is stable if it is positive.

\begin{figure}
	\begin{center}
		\includegraphics[bb=0 0 211 153]{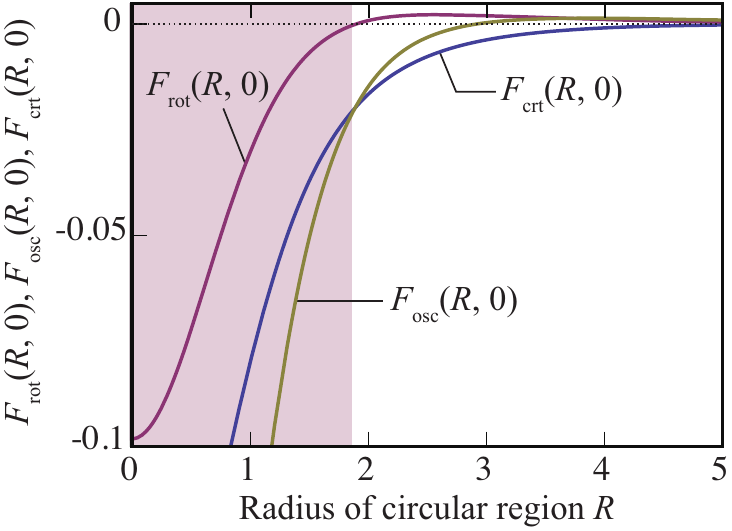}
		\caption{Plots of $F_{\mathrm{rot}}(R,0)$, $F_{\mathrm{osc}}(R,0)$, and $F_{\mathrm{crt}}(R,0)$ against the radius of the circular region $R$.
		When the rest state is unstable, i.e. $\xi < b(R)$, rotational motion is linearly stable in the range of $R$ indicated by coloring with magenta.}
		\label{pd_2D}
	\end{center}
\end{figure}

\begin{figure}
	\begin{center}
		\includegraphics[bb=0 0 142 136]{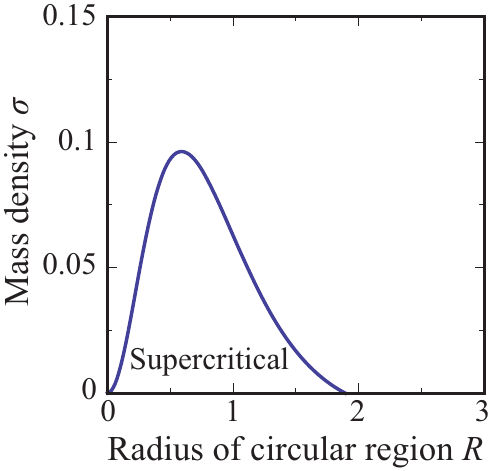}
		\caption{Boundary between the regions on $R$-$\sigma$ plane.
		The region with ``supercritical'' indicates the parameter region where the stable rotational motion through the supercritical double-Hopf bifurcation at $\xi = b(R)$ can be seen.
		The curve is given by $F_{\mathrm{rot}}(R, \sigma) = 0$.
		}
		\label{supsub}
	\end{center}
\end{figure}

In Fig.~\ref{pd_2D}, the functions important to check the conditions are plotted for $\sigma = 0$.
It is suggested that the rotational motion is stably observed when the radius of the circular region is smaller than ca.~2, when $\sigma = 0$.
In case that $\sigma > 0$, the region where the rotational motion is stable, i.e., the region of $R$ which gives $F_{\mathrm{rot}}(R,\sigma) < 0$, is smaller than that for $\sigma = 0$, though the oscillatory motion never becomes stable.
Figure~\ref{supsub} shows the curve, $F_{\mathrm{rot}}(R, \sigma) = 0$, on $R$-$\sigma$ plane separating the parameter regions where the stable rotational motion through the supercritical double-Hopf bifurcation can be seen or not.
By increasing $\sigma$, the region of $R$ which gives $F_{\mathrm{rot}}(R,\sigma) < 0$ becomes smaller and smaller, and finally disappears.

\section{Numerical Calculation}

To confirm the validity of the theoretical results, we performed numerical calculations based on the original model before the reduction.
For the concentration field, we used Eq.~\eqref{d_d.eq.} with the source term:
\begin{align}
\label{d_source_finite}
f (\bm{r};\bm{\rho}) = \frac{1}{2 \pi {\epsilon_R}^2} \left [ 1 + \tanh \left ( \frac{\epsilon_R - | \bm{r} - \bm{\rho} |}{\delta} \right ) \right ],
\end{align}
instead of Eq.~\eqref{d_source}.
Here the size of the camphor disk was considered to be finite ($\epsilon_R = 0.1$) in order to avoid the difficulty originating from the treatment of the Dirac's delta function in numerical calculation.
The smoothing parameter $\delta$ was set to be $0.02$.
The explicit method was used for Eq.~\eqref{d_d.eq.}.
The time and spatial steps were $\Delta t = 10^{-5}$ and $\Delta x = 10^{-2}$, respectively.

The potential force originating from the confinement by the Neumann boundary condition decays exponentially as a function of the distance from the boundary.
Since the potential force toward the center is too small near the center of the circular region for the system with larger $R$, the camphor disk was sometimes stuck to the asymmetric position due to the spatial mesh for concentration field.
The noise term can be considered to be originating from the fluctuation due to the nonuniformity of the actual experimental system such as the camphor disk shape.
Thus we adopted the equation of motion~\eqref{d_eq.of.m._2} with a random noise:
\begin{equation}
\sigma \frac{d^2 \bm{\rho}}{dt^2} = - \xi \frac{d \bm{\rho}}{dt} + \bm{F} (c; \bm{\rho}) + \bm{\eta} (t),
\label{eq.m._w.noise}
\end{equation}
where $\bm{\eta} (t)$ describes the small noise.
The noise satisfies the relations $\left < \bm{\eta} (t) \right > = \bm{0}$ and $\left < \eta_i (t) \eta_j (s) \right > = 2 Y \delta_{ij} \delta (t-s)$, where the bracket $\left < \cdots \right >$ indicates an average over time.
The noise intensity $Y$ was set to be $10^{-5}$.
The mass per unit area $\sigma$ was fixed to be $\sigma = 0$ to avoid the camphor disk from going out of the circular region due to the inertia.
It is noted that the inertia-like effect coming from the concentration field (cf.~$-g(R) \ddot{\bm{\rho}}$ in Eq.~\eqref{reduced_equation}) remains.
The Euler method was used for Eq.~\eqref{eq.m._w.noise}.
For the discretization, the driving force in Eq.~\eqref{d_def.drv.f} was replaced with:
\begin{align}
\label{d_def.drv.f_finite}
\bm{F} (c; \bm{\rho} (t)) = \frac {1}{S} \int_{\partial \Omega} \gamma \left ( c (\bm{\rho} + \epsilon \bm{n}) \right ) \bm{n} d\ell.
\end{align}
For the integration in Eq.~\eqref{d_def.drv.f_finite}, we adopted the summation over $40$ arc elements.

\begin{figure}
	\begin{center}
		\includegraphics[bb=0 0 247 400]{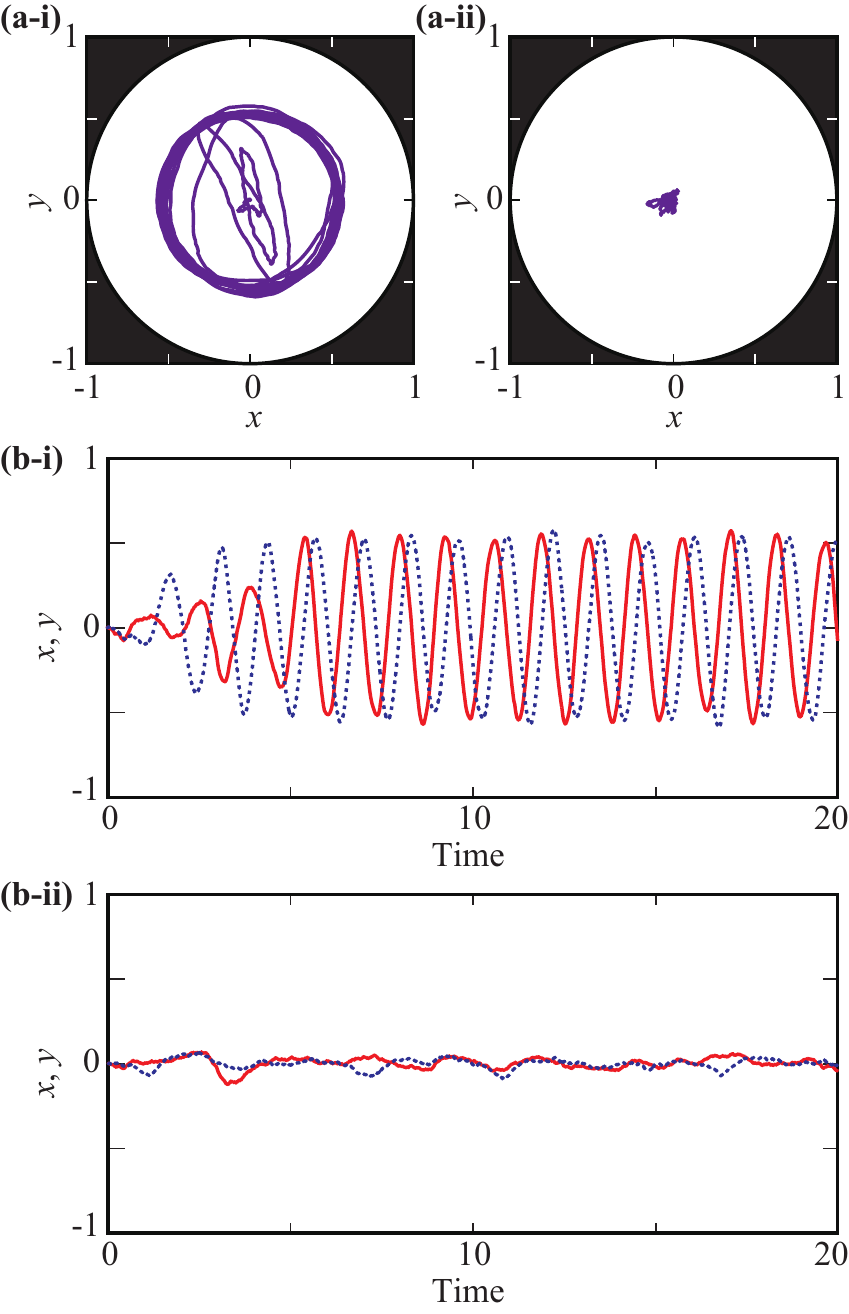}
		\caption{Numerical results on the trajectories of a camphor disk in a circular region for the resistance coefficient per unit area (i) $\xi=0.15$ and (ii) $0.2$.
		The camphor disk exhibited rotational motion or staying near the center of the circular region.
		(a) The trajectory on the $x$-$y$ plane. (b) Time evolutions of $x(t)$ and $y(t)$ shown in red- and blue-colored curves (solid and broken curves), respectively.
		The initial condition for the position of the camphor disk was set to be the origin; $x=0$ and $y=0$.
		The initial concentration field $c$ was zero at every point in the region.
		}
		\label{2d_trajectory}
	\end{center}
\end{figure}

Here we show the numerical results for the radius of the circular region $R = 1$.
The results for the resistance coefficient per unit area $\xi = 0.15$ and $0.2$ are shown in Fig.~\ref{2d_trajectory}.
A camphor disk was initially located at the system center, and the concentration field $c$ was initially zero at every point in the circular region.
We obtained the trajectory toward the circular orbit around the center of the circular region for $\xi = 0.15$ and that staying near the center of the circular region for $\xi = 0.2$.
To clarify the transition between rotational motion and rest at the center, we introduce the averaged distance $\left < |\bm{\rho}(t)| \right >$ over the time from $200$ to $300$.
Here we call it a rotational radius.
In an ideal condition, i.e., averaging sufficiently long time trajectories without noise, rotational motion gives a finite rotation radius, while rest at the origin gives zero.
Here we calculated the rotation radii from the finite time series with noise, and thus larger and smaller rotation radii indicated rotation and rest at the origin, respectively.
We obtained rotation radii as $0.54 \pm 0.02$ for $\xi = 0.15$ and $0.05 \pm 0.02$ for $\xi = 0.2$.
Therefore it is expected that the bifurcation point exists between $\xi = 0.15$ and $\xi = 0.2$ from the numerical results.
On the other hand, the expected bifurcation point $\xi_0$ by the theoretical analysis is \textit{ca.}~$0.218$, which is given by $\xi_0 = b(1)$.
It is noted that $\epsilon$ in $b(R)$ is set to be $\epsilon = \epsilon_R \exp(1/4) = 0.1 \exp(1/4)$ (cf.~Footnote 36 in Ref.~\cite{Koyano2017}).
The order of the bifurcation point is the same as that obtained by the numerical results though there remains some discrepancy between them.
We guess the dominant reason for the discrepancy between theoretical and numerical results in the value of bifurcation point is the difference in the size of the camphor disk; in the theoretical analysis, the size of the camphor disk is assumed to be $\epsilon_R \to +0$, but in the numerical calculation, $\epsilon_R$ has a finite value $\epsilon_R = 0.1$.

\begin{figure}
	\begin{center}
		\includegraphics[bb=0 0 181 134]{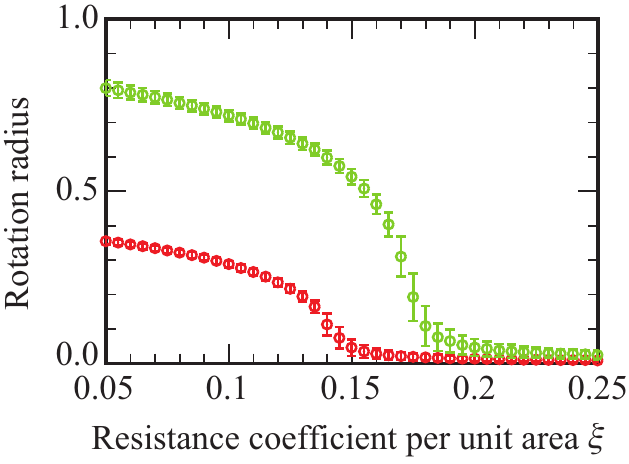}
		\caption{Rotation radius depending on the resistance coefficient per unit area $\xi$.
		The red (dark gray) and green (light gray) plots show the rotation radius for $R = 0.5$ and $1$, respectively.
		The rest state at the origin became unstable around $\xi \simeq 0.15$ and $0.18$ for $R=0.5$ and $1$, respectively, which correspond to the theoretical result on the double-Hopf bifurcation point $\xi = b(R)$.
		The error bars show standard deviation.
		The initial conditions were the same as those in Fig.~\ref{2d_trajectory}.
		}
		\label{fig_xi_vs_r}
	\end{center}
\end{figure}

The bifurcation structure is seen more clearly in Fig.~\ref{fig_xi_vs_r}, which shows the rotation radii depending on the resistance coefficient per unit area $\xi$.
By decreasing resistance coefficient per unit area $\xi$, the rest state became unstable, and the stable rotational motion appeared.
Such bifurcation structure is consistent with the linear stability analysis of the rest state as shown in Fig.~\ref{br}, and the linear stability analysis of rotational motion for $\xi < b(R)$ as shown in Fig.~\ref{pd_2D}.
Because the stability of the rest state is close to neutral and the motion is sensitive to the noise, the standard deviation of rotation radii becomes larger around the bifurcation point $\xi = b(R)$.

\begin{figure}
	\begin{center}
		\includegraphics[bb=0 0 178 134]{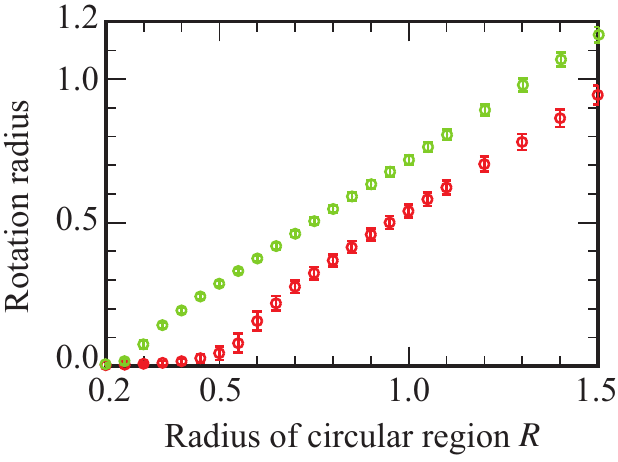}
		\caption{Rotation radius depending on the radius of the circular region $R$.
		The green (light gray) and red (dark gray) plots show the rotation radius for $\xi = 0.1$ and $0.15$, respectively.
		The rest state at the origin became unstable around $R \simeq 0.25$ and $0.5$ for $\xi=0.1$ and $0.15$, respectively, which correspond to the theoretical result $\xi = b(R)$.
		The error bars show standard deviation.
		The initial conditions were the same as those in Fig.~\ref{2d_trajectory}.
		}
		\label{fig_r_vs_r}
	\end{center}
\end{figure}

As shown in the phase diagram in Fig.~\ref{br}, the radius of the circular region $R$ can also be a bifurcation parameter.
Figure~\ref{fig_r_vs_r} shows a bifurcation diagram where the bifurcation parameter is the radius of the circular region $R$.
By increasing the system size, the rest state became unstable, and the stable rotational motion appeared.
Again, the bifurcation structure is consistent with the linear stability analysis of the rest state, and the linear stability analysis of rotational motion for $\xi < b(R)$.
The rotation radii linearly increased when the radius of the circular region was sufficiently larger than the bifurcation point.

\section{Experiments}

We also made experiments to confirm the theoretical results.
Here we observed the motion of a camphor disk in a circular water phase whose radius was continuously controlled.

A camphor gel disk, whose diameter was 4.0 mm and thickness was 0.5 mm, was made of agar gel in which water was replaced with camphor methanol solution.
After the methanol dried up, the camphor disk was floated on a water phase (15 mm in depth)~\cite{Soh2008}.
To achieve a variable-sized water phase, an optical focus (diaphragm) was put on the large water surface.
The camphor disk was placed in the open area of the diaphragm to gradually increase or decrease the radius $R$ of the circular open area.
The camphor molecules are blocked at the edge of the diaphragm consistently with the Neumann boundary condition employed in the theoretical analysis.
The details of the experimental setup are shown in Appendix~\ref{appE}.

\begin{figure}
	\begin{center}
		\includegraphics[bb=0 0 247 400]{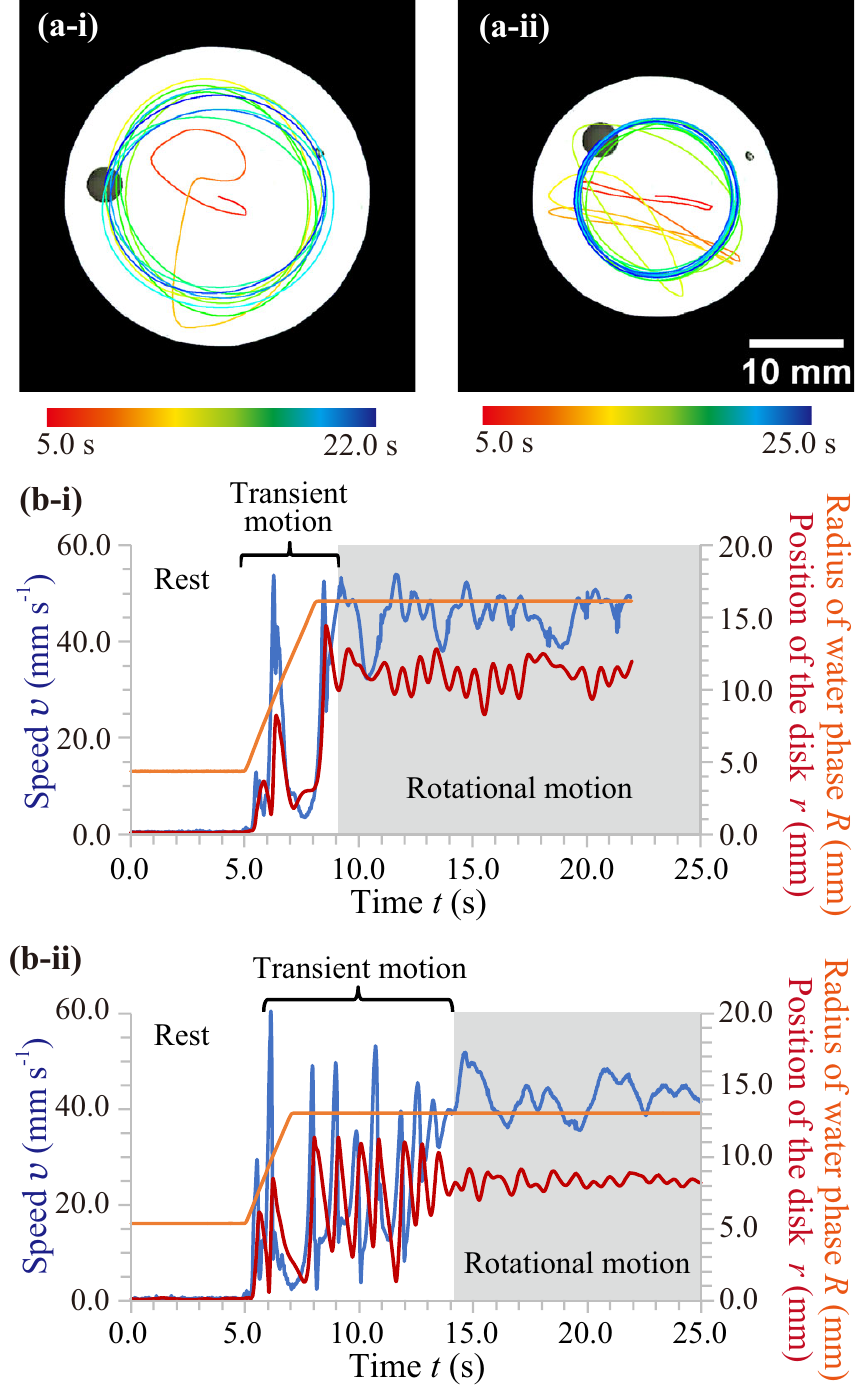}
		\caption{Experimental results on the motion of a camphor disk in a circular water phase.
		(a) Trajectory of the moving camphor disk.
		(b) Time series of speed $v$, the distance between the camphor disk and the circular water phase center $r$, and the radius of the circular water phase $R$.
		The radius of the circular water phase was gradually changed from 5.0~mm to (i) 16.1~mm and (ii) 13.0~mm.
		The camphor disks rotated for (i) 10 times and (ii) 15 times.
		}
		\label{Trajectory_180104}
	\end{center}
\end{figure}

At the initial stage with the small radius of a water phase ($R \sim$ 5.0~mm), the disk was in the rest state at the center of the water phase.
With an increase in the radius of the circular water phase $R$, the disk started to move and finally showed rotational motion as shown in Fig.~\ref{Trajectory_180104}(a).
For rotational motion, both the moving speed $v$ and the position of the disk $r$ were almost constant in time as shown in Fig.~\ref{Trajectory_180104}(b).
The theoretical and numerical results qualitatively match the transition from the rest state at the center position of the circular water phase to the rotational motion with an increase in the radius of the circular water phase $R$.

\section{Discussion}

In the present study, the mathematical model for the motion of a camphor disk in a two-dimensional circular region was reduced into the ordinary differential equation of the second order.
The reduction is valid when the distance between the camphor disk and the center of the system is infinitesimally small.
This assumption is justified when the system is close to the bifurcation point.
We also assume that the higher-order time derivative can be neglected.
This assumption cannot be justified theoretically.
For a one-dimensional case, we have checked the validity of the truncation numerically in Ref.~\cite{Koyano2016}.
Therefore, we assumed the truncation of the higher-order time derivative is valid also in the present two-dimensional case.
The reduced equation has the same form with the dynamical system proposed in Ref.~\cite{Koyano2015}, so the bifurcation structure of it was analyzed using the results in Ref.~\cite{Koyano2015}.
By changing the radius of the circular region or the resistance coefficient, the rest state at the center of the circular region becomes unstable through the double-Hopf bifurcation~\cite{Kuznetsov2004}.

When the rest state becomes unstable, the camphor disk moves toward the boundary in a certain direction to avoid the region with a higher concentration of camphor molecules.
Then the camphor disk is reflected by interaction with the boundary through the concentration field, and moves toward the opposite side about the center of the circular region.
Thus the camphor disk first moves back and forth, which is transient due to the instability of the oscillatory motion.
Finally, a rotational motion is realized.
Such transient behaviors were seen in both numerical calculations and experiments in Figs.~\ref{2d_trajectory}(a-i) and \ref{Trajectory_180104}(a-ii).
The mass of the camphor disk was measured to be 4.9 mg $\pm$ 0.2 mg in the experiments, though it is neglected in the numerical calculations.
The characteristic time scale where the inertia is dominant is estimated by the ratio $m/\eta$, where $m$ and $\eta$ are mass and resistance coefficient, respectively.
Here the characteristic time scale is estimated as $m/\eta = 5 \times 10^{-2}$ s.
Thus, we can neglect the inertia term $m \ddot{\bm{\rho}}$ as far as we focus on the motion whose characteristic time scale is longer than $5 \times 10^{-2}$ s.

In Fig.~\ref{supsub}, the parameter region where the rotational motion with a small radius is stable is shown.
Outside of the region, i.e., the circular region with a larger radius or the heavier mass of the camphor disk, the camphor disk might rotate with a large radius through a discontinuous transition.
To clarify this, we should discuss the higher order terms appearing in Eq.~\eqref{reduced_equation}, which is left for future work.
It should be noted that for a camphor disk motion in such a larger chamber the discussion based on the interaction with the wall may work~\cite{Mimura2007,Miyaji2017}.

\section{Conclusion}

In the present paper, we discussed the motion of a camphor disk confined in a two-dimensional circular water phase.
We reduced the mathematical model for the camphor disk motion and applied our previous results based on the weakly nonlinear analysis to the reduced equation.
The theoretical results suggest that the rotational motion occurs when the rest state becomes unstable.
The stability of the rest state is determined by the resistance coefficient, which affects the mobility of the camphor disk, and the radius of the circular region, which determines the intensity of the confinement.
The theoretical results were confirmed by numerical calculation and experiments.

\begin{acknowledgments}

The authors acknowledge Professor J.~Gorecki and Professor M.~Nagayama for their helpful discussion, and Mr.~S.~Kubo for his preliminary experimental results.
This work was supported by JSPS KAKENHI Grants Numbers JP17J05270, JP25103008, JP15K05199, JP16H03949, and JP16K05486, and the Cooperative Research Program of ``Network Joint Research Center for Materials and Devices'' (Nos.~20181023 and 20183003).
This work was also supported by JSPS and PAN under the Japan-Poland Research Cooperative Program ``Spatio-temporal patterns of elements driven by self-generated, geometrically constrained flows''.

\end{acknowledgments}

\begin{appendix}

\section{Dimensionless form of the mathematical model\label{appA}}

We consider the nondimensionalization of Eq.~\eqref{d.eq.}.
The dimensions of $\alpha$, $D$, and $c_0$ are [1/T], [L${}^2$/T], and [C L${}^2$/T], respectively.
Here, T, L, and C represent the dimensions of time, length, and concentration, respectively.
Thus, we introduce the dimensionless time, position, and concentration as $\tilde{t} = \alpha t$, $\tilde{r} = \sqrt{\alpha/D} \; r$, and $\tilde{c} = c D /c_0$, respectively.
By substituting the three dimensionless variables into Eq.~\eqref{d.eq.} and dividing the both sides of it with $c_0 \alpha / D$, we obtain
\begin{align}
& \frac{\partial \tilde{c} \left ( \tilde{r}, \theta, \tilde{\rho}, \phi \right )}{\partial \tilde{t}} = \left ( \frac{\partial^2}{\partial \tilde{r}^2} + \frac{1}{\tilde{r}} \frac{\partial}{\partial \tilde{r}} + \frac{1}{\tilde{r}^2} \frac{\partial^2}{\partial \theta^2} \right ) \tilde{c} \left ( \tilde{r}, \theta, \tilde{\rho}, \phi \right ) \nonumber \\
& - \tilde{c} \left ( \tilde{r}, \theta, \tilde{\rho}, \phi \right ) + \frac{D}{c_0 \alpha} f \left ( \sqrt{\frac{D}{\alpha}} \tilde{r}, \theta; \sqrt{\frac{D}{\alpha}} \tilde{\rho} \left ( \tilde{t} \right ), \phi \left ( \tilde{t} \right ) \right ).
\end{align}
The source term is considered as follows:
\begin{align}
& \frac{D}{c_0 \alpha} f \left ( \sqrt{\frac{D}{\alpha}} \tilde{r}, \theta; \sqrt{\frac{D}{\alpha}} \rho \left ( \tilde{t} \right ), \phi \left ( \tilde{t} \right ) \right ) \nonumber \\
&\quad = \frac{1}{\tilde{r}} \delta \left ( \tilde{r} - \tilde{\rho} \left ( \tilde{t} \right ) \right )
\delta \left ( \theta - \phi \left ( \tilde{t} \right ) \right ) \equiv \tilde{f} \left ( \tilde{r},\theta; \tilde{\rho},\phi \right ).
\end{align}
Here we use $\delta(ax) = \delta(x)/|a|$.
Then, we have
\begin{align}
\frac{\partial \tilde{c} \left ( \tilde{r}, \theta, \tilde{\rho}, \phi \right )}{\partial \tilde{t}} =&
\left ( \frac{\partial^2}{\partial \tilde{r}^2} + \frac{1}{\tilde{r}} \frac{\partial}{\partial \tilde{r}} + \frac{1}{\tilde{r}^2} \frac{\partial^2}{\partial \theta^2} \right ) \tilde{c} \left ( \tilde{r}, \theta, \tilde{\rho}, \phi \right ) \nonumber \\
& - \tilde{c} \left ( \tilde{r}, \theta, \tilde{\rho}, \phi \right )
+ \tilde{f} \left ( \tilde{r}, \theta; \tilde{\rho}, \phi \right ),
\end{align}
where $\tilde{\rho} = \sqrt{\alpha/D} \rho$.

Next, Eq.~\eqref{eq.of.m._2} is nondimensionalized.
The variables $t$, $\bm{r}$, $\bm{\rho}$, and $c$ are replaced with $\tilde{t}$, $\tilde{\bm{r}} (= \sqrt{\alpha/D} \bm{r})$, $\tilde{\bm{\rho}} (= \sqrt{\alpha/D} \bm{\rho})$, and $\tilde{c}$, and then we have
\begin{align}
\label{eq.m.}
\sigma \sqrt{D \alpha^3} \frac{d^2 \tilde{\bm{\rho}}(\tilde{t})}{d \tilde{t}^2} = - \xi \sqrt{D \alpha}\frac{d \tilde{\bm{\rho}}(\tilde{t})}{d \tilde{t}} + \bm{F} \left ( \sqrt{\frac{D}{\alpha}} \tilde{\bm{\rho}}; \frac{c_0}{D} \tilde{c} \right ).
\end{align}
In Eq.~\eqref{eq.m.}, we can eliminate all coefficients but one.
Here, the driving force is nondimensionalized as,
\begin{align}
&\bm{F}(c; \bm{\rho}) = c_0 \sqrt{\frac{\alpha}{D^3}} \times \nonumber \\
& \lim_{\tilde{\epsilon} \to +0} \frac {-k}{\pi \tilde{\epsilon}^2} \int_{0}^{2 \pi} \tilde{c} \left ( \sqrt{\frac{D}{\alpha}} \left ( \tilde{\bm{\rho}} + \tilde{\epsilon} \bm{n}(\theta) \right ) ; \sqrt{\frac{D}{\alpha}} \tilde{\bm{\rho}} \right ) \tilde{\epsilon} d\theta \nonumber \\
&\qquad \quad \equiv k c_0 \sqrt{\frac{\alpha}{D^3}} \tilde{\bm{F}} \left ( \tilde{\bm{\rho}}; \tilde{c} \right ).
\end{align}
Here, $\tilde{\bm{F}}$ is a dimensionless driving force, and $\tilde{\epsilon} = \sqrt{\alpha/D} \; \epsilon$.
Then we obtain
\begin{equation}
\frac{\sigma D^2 \alpha}{k c_0} \frac{d^2 \tilde{\bm{\rho}}}{d \tilde{t}^2} = - \frac{\xi D^2}{k c_0} \frac{d \tilde{\bm{\rho}}}{d \tilde{t}} + \tilde{\bm{F}} \left ( \tilde{c}; \tilde{\bm{\rho}} \right ).
\end{equation}
Thus we define the dimensionless mass and resistance coefficient per unit area as:
\begin{align}
\tilde{\sigma} \equiv \frac{\sigma D^2 \alpha}{k c_0}, \quad \tilde{\xi} \equiv\frac{\xi D^2}{k c_0}.
\end{align}
The dimensionless forms in Eqs.~\eqref{d_eq.of.m._2} to \eqref{d_Neumann} are obtained.

\section{Discrete Hankel transform\label{appB}}

Here, we provide some notes on ``discrete Hankel transform'' for a function which satisfies the Neumann boundary condition~\cite{Bowman1958, Duffy2001, Alhargan1991}.
It is noted that discrete Hankel transform with Neumann boundary condition corresponds to the special case of Dini's expansion, which is the expansion for a function satisfying the boundary condition: $z^{-m} [z \mathcal{J}'_m (z) + H \mathcal{J}_m (z)] = 0$ (cf.~Chap.~XVIII in Ref.~\cite{Watson}).
When the constant $H$ is zero, then Dini's expansion is discrete Hankel transform for a function which satisfies the Neumann boundary condition.
In the book by Bowman~\cite{Bowman1958}, the discrete Hankel transform for a function which satisfies the Dirichlet boundary condition is denoted, and as for the case with Neumann boundary condition, the calculation can be proceeded in almost a parallel manner.

The Bessel differential equation is given as
\begin{equation}
\left ( \frac{d^2}{d r^2} + \frac{1}{r} \frac{d}{dr} + \left( 1 - \frac{m^2}{r^2} \right ) \right ) \mathcal{J}_m (r)= 0.
\end{equation}
By replacing $r$ with $k r$, we have
\begin{equation}
\left ( \frac{d^2}{d r^2} + \frac{1}{r} \frac{d}{d r} + \left( k^2 - \frac{m^2}{r^2} \right ) \right ) \mathcal{J}_m (k r)= 0. \label{eq:bde2}
\end{equation}
Equation \eqref{eq:bde2} is transformed into the following form:
\begin{equation}
\frac{d}{d r} \left ( r \frac{d \mathcal{J}_m (k r)}{d r} \right ) - \frac{m^2}{r} \mathcal{J}_m (k r) + k^2 r \mathcal{J}_m (k r) = 0. \label{eq:bde3}
\end{equation}
Here we consider the subtraction between equations in Eq.~\eqref{eq:bde3} with different $k$.
We have
\begin{align}
& \mathcal{J}_m(h r) \frac{d}{d r} \left ( r \frac{d \mathcal{J}_m (k r)}{d r} \right ) - \mathcal{J}_m(k r) \frac{d}{d r} \left ( r \frac{d \mathcal{J}_m (h r)}{d r} \right ) \nonumber \\
& \qquad + ( k^2 - h^2) r \mathcal{J}_m (k r) \mathcal{J}_m (h r) = 0. \label{eq:bde5}
\end{align}
Using Eq.~\eqref{eq:bde5}, we have
\begin{align}
& - ( k^2 - h^2) \int_{0}^{R} r \mathcal{J}_m (k r) \mathcal{J}_m (h r) dr \nonumber \\
& = \left [ \mathcal{J}_m (h r) \left ( r \frac{d \mathcal{J}_m (k r)}{d r} \right ) - \mathcal{J}_m (k r) \left ( r \frac{d \mathcal{J}_m (h r)}{d r} \right ) \right ]_{0}^{R} \label{eq:bde6}.
\end{align}
Hereafter, we consider the nonnegative integer $m$.

For $m \geq 1$, we set $\displaystyle{ k_{m n} \equiv \zeta_{m n} / R > 0}$ instead of arbitrary $k$ and $h$, where $\zeta_{m n} > 0$ is the element of the zero set of $\mathcal{J}'_m$.
Here $\mathcal{J}'_m (r)$ means $\partial \mathcal{J}_m (r) / \partial r$.
The second index of $\zeta_{mn}$ is set to be $n = 0, 1, 2, \cdots$ so that $\zeta_{m i} < \zeta_{m j}$ for $i<j$.

Since $\mathcal{J}'_m (k_{mn} R) = 0$ holds from the definition of $k_{mn}$, we have
\begin{equation}
( {k_{m l}}^2 - {k_{m n}}^2) \int_{0}^{R} r \mathcal{J}_{m} (k_{m l} r) \mathcal{J}_{m} (k_{m n} r) dr = 0
\end{equation}
from Eq.~\eqref{eq:bde6}.
For $l \neq n$, we have
\begin{equation}
\int_{0}^{R} r \mathcal{J}_{m} (k_{m l} r) \mathcal{J}_{m} (k_{m n} r) dr = 0,
\end{equation}
and thus $\mathcal{J}_{m} (k_{m l} r)$ and $\mathcal{J}_{m} (k_{m n} r)$ whose domains are $[0,R]$ are orthogonal to each other for $l \neq n$.

To obtain the norm of $\mathcal{J}_{m} (k_{m n} r)$, we calculate the following integration using Eq.~\eqref{eq:bde6}:
\begin{align}
& \int_{0}^{R} r \mathcal{J}_{m} (k_{m n} r) \mathcal{J}_{m} (k_{m n} r) dr \nonumber \\
& = \lim_{k \to k_{m n}} \int_{0}^{R} r \mathcal{J}_{m} (k r) \mathcal{J}_{m} (k_{m n} r) dr \nonumber \\
& = \lim_{k \to k_{m n}} \nonumber \\
& \quad \frac{R \left ( k_{m n} \mathcal{J}_{m} (k R) \mathcal{J}'_{m} (k_{m n} R) - k \mathcal{J}_{m} (k_{m n} R) \mathcal{J}'_{m} (k R) \right )}{k^2 - {k_{m n}}^2}.
\end{align}
By applying L'H\^opital's rule, we have
\begin{align}
& \int_{0}^{R} r \mathcal{J}_m (k_{m n} r) \mathcal{J}_m (k_{m n} r) dr \nonumber \\
& = \frac{R^2}{2} \left ( \mathcal{J}'_m (k_{m n} R) \mathcal{J}'_m (k_{m n} R) \frac{}{} \right . \nonumber \\
& \quad \left . - \left ( \frac{1}{k_{m n} R} \mathcal{J}'_m (k_{m n} R) + \mathcal{J}''_m (k_{m n} R) \right ) \mathcal{J}_m(k_{m n} R) \right ). \label{eq:bde13}
\end{align}
Since $\mathcal{J}'_m(k_{m n} R) = 0$ holds, we have
\begin{align}
&\int_{0}^{R} r \mathcal{J}_m (k_{m n} r) \mathcal{J}_m (k_{m n} r) dr \nonumber \\
&= - \frac{R^2}{2} \mathcal{J}''_m (k_{m n} R) \mathcal{J}_m(k_{m n} R) \nonumber \\
&= - \frac{R^2}{2} \mathcal{J}''_m (\zeta_{m n}) \mathcal{J}_m(\zeta_{m n}) \equiv \frac{1}{a_{m n}}.
\end{align}

For $m = 0$, we set $\displaystyle{ k_{0n} \equiv \zeta_{0 n} / R \geq 0}$, where $\zeta_{0 n} \geq 0$ is the element of the zero set of $\mathcal{J}'_0$.
The second index of $\zeta_{0n}$ is set to be $n = 0, 1, 2, \cdots$ so that $\zeta_{0 i} < \zeta_{0 j}$ for $i<j$.
The only difference from the case for $m \geq 1$ is that $k_{00} = \zeta_{00} / R$ is exceptionally set to be $0$.
This is because $\mathcal{J}_{0} (0) \neq 0$ while $\mathcal{J}_{m} (0) = 0$ for $m \geq 1$, and $\sqrt{a_{00}} \mathcal{J}_{0} (0)$ is a linearly independent component in the basis.
In the case of $k_{00}=0$, Eq.~\eqref{eq:bde13} is not valid.
By setting $k_{00}=0$ and then applying L'H\^opital's rule, we have
\begin{align}
\int_{0}^{R} r \mathcal{J}_0 (k_{00} r) \mathcal{J}_0 (k_{00} r) dr = \frac{R^2}{2} \equiv \frac{1}{a_{00}}. \label{eq:bde14}
\end{align}

Thus, for every integer $m \geq 0$, the functions $\displaystyle{ \left \{ \sqrt{a_{m n}} \mathcal{J}_m (k_{m n} r) \right \} }$ are the basis of the function space for $[0,R]$.
An arbitrary function $f(r)$ which satisfies the Neumann condition at $r = R$ is expressed in the discrete Hankel transform as
\begin{equation}
f(r) = \sum_{n = 0}^{\infty} a_{m n} f_{m n} \mathcal{J}_m (k_{m n} r),
\end{equation}
where
\begin{equation}
f_{m n} \equiv \int_{0}^{R} f(r) \mathcal{J}_m (k_{m n} r) r dr.
\end{equation}

Before closing this Appendix, we note the comments on the application of discrete Hankel transform with Neumann condition to some actual problem.
Previously, the discrete Hankel transform with Neumann condition was applied to the Helmholtz equation as in Eq.~(5.3.34) in Ref.~\cite{Duffy2001}.
In the present study, the discrete Hankel transform with Neumann condition was applied to the reaction-diffusion equation in Eq.~\eqref{d_d.eq.}, the modified Helmholtz equation, and we succeeded in the expansion of the concentration field with respect to the source position and its time derivatives.

\section{Reduction of the mathematical model\label{appC}}

Equation~\eqref{integration_of_green.f} is expanded using the partial integration as follows~\cite{Ohta2009_2}:
\begin{widetext}
\begin{align}
& c(\bm{r}, \bm{\rho}(t)) \nonumber \\
&= \frac{1}{2 \pi} \sum_{m = -\infty}^{\infty} \sum_{n = 0}^{\infty} \frac{a_{|m|n}}{k_{|m|n}^2+1} \mathcal{J}_{|m|} (k_{|m|n} \rho(t)) \mathcal{J}_{|m|} (k_{|m|n} r) e^{i m (\theta-\phi(t))} \nonumber \\
& \quad + \frac{1}{2 \pi} \sum_{m = -\infty}^{\infty} \sum_{n = 0}^{\infty} \frac{a_{|m|n}}{(k_{|m|n}^2+1)^2} \left \{ - k_{|m|n} \dot{\rho}(t) \mathcal{J}'_{|m|} (k_{|m|n} \rho(t)) + i m \dot{\phi}(t) \mathcal{J}_{|m|} (k_{|m|n} \rho(t)) \right \} \mathcal{J}_{|m|} (k_{|m|n} r) e^{i m (\theta-\phi(t))} \nonumber \\
& \quad + \frac{1}{2 \pi} \sum_{m = -\infty}^{\infty} \sum_{n = 0}^{\infty} \frac{a_{|m|n}}{(k_{|m|n}^2+1)^3} \left \{ k_{|m|n} \ddot{\rho}(t) \mathcal{J}'_{|m|} (k_{|m|n} \rho(t)) + {k_{|m|n}}^2 (\dot{\rho}(t))^2 \mathcal{J}''_{|m|} (k \rho(t)) \right . \nonumber \\
& \quad \qquad \left . - 2 i k_{|m|n} m \dot{\rho}(t)\dot{\phi}(t) \mathcal{J}'_{|m|} (k_{|m|n} \rho(t)) - i m \ddot{\phi}(t) \mathcal{J}_{|m|} (k_{|m|n} \rho(t)) - m^2 (\dot{\phi}(t))^2 \mathcal{J}_{|m|} (k_{|m|n} \rho(t)) \right \} \mathcal{J}_{|m|} (k_{|m|n} r) e^{i m (\theta-\phi(t))} \nonumber \\
& \quad + \frac{1}{2 \pi} \sum_{m = -\infty}^{\infty} \sum_{n = 0}^{\infty} \frac{a_{|m|n}}{(k_{|m|n}^2+1)^4} \left \{ - k_{|m|n} \dddot{\rho}(t) \mathcal{J}'_{|m|} (k_{|m|n} \rho(t)) - 3 k_{|m|n}^2 \dot{\rho}(t) \ddot{\rho}(t) \mathcal{J}''_{|m|} (k_{|m|n} \rho(t)) \right . \nonumber \\
& \quad \qquad + 3 i k_{|m|n} m \ddot{\rho}(t) \dot{\phi}(t) \mathcal{J}'_{|m|} (k_{|m|n} \rho(t)) - k_{|m|n}^3 (\dot{\rho}(t))^3 \mathcal{J}'''_{|m|} (k_{|m|n} \rho(t)) + 3 i k_{|m|n}^2 m (\dot{\rho}(t))^2 \dot{\phi}(t) \mathcal{J}''_{|m|} (k_{|m|n} \rho(t)) \nonumber \\
& \quad \qquad + 3 i k_{|m|n} m \dot{\rho}(t) \ddot{\phi}(t) \mathcal{J}'_{|m|} (k_{|m|n} \rho(t)) + 3 k_{|m|n} m^2 \dot{\rho}(t) (\dot{\phi}(t))^2 \mathcal{J}'_{|m|} (k_{|m|n} \rho(t)) + i m \dddot{\phi}(t) \mathcal{J}_{|m|} (k_{|m|n} \rho(t)) \nonumber \\
& \quad \qquad \left . + 3 m^2 \dot{\phi}(t) \ddot{\phi}(t) \mathcal{J}_{|m|} (k_{|m|n} \rho(t)) - i m^3 (\dot{\phi}(t))^3 \mathcal{J}_{|m|} (k_{|m|n} \rho(t)) \right \} \mathcal{J}_{|m|} (k_{|m|n} r) e^{i m (\theta-\phi(t))} \nonumber \\
& \quad + \cdots. \label{expansion}
\end{align}
\end{widetext}
The first term,
\begin{align}
c_0 (\bm{r}, \bm{\rho}(t)) =& \frac{1}{2 \pi} \sum_{m = -\infty}^{\infty} \sum_{n = 0}^{\infty} \frac{a_{|m|n}}{k_{|m|n}^2+1} \nonumber \\
& \times \mathcal{J}_{|m|} (k_{|m|n} \rho(t)) \mathcal{J}_{|m|} (k_{|m|n} r) e^{i m (\theta-\phi(t))}, \label{steady.state.w_boundary_J}
\end{align}
only depends on the current position of the camphor disk, and corresponds to the steady concentration field when the camphor disk is fixed at the current position for long time.
The explicit form of the steady state is obtained as
\begin{align}
&c_0(r,\theta, \rho(t), \phi(t)) \nonumber \\
&= \frac{1}{2 \pi} \mathcal{K}_0 \left ( \sqrt{r^2 + \rho^2 - 2 r \rho \cos(\theta-\phi)} \right ) \nonumber \\
& \quad - \frac{1}{2 \pi}\frac{{\mathcal{K}_0}' (R)}{{\mathcal{I}_0}' (R)} \mathcal{I}_0 (\rho) \mathcal{I}_0 (r) \nonumber \\
& \quad - \frac{1}{\pi} \sum_{m=1}^{\infty} \frac{{\mathcal{K}_m}' (R)}{{\mathcal{I}_m}' (R)} \mathcal{I}_m (\rho) \mathcal{I}_m (r) \cos m (\theta-\phi). \label{steady.state.w_boundary}
\end{align}
The detailed calculation is presented in Appendix~\ref{appD}.
First, length scale appeared in Eqs.~\eqref{steady.state.w_boundary_J} and \eqref{steady.state.w_boundary} is rescaled with $\lambda$ as $r \to \lambda r$, $\rho \to \lambda \rho$, $R \to \lambda R$, $k_{mn} \to k_{mn}/\lambda$.
Then the both sides of the rescaled Eqs.~\eqref{steady.state.w_boundary_J} and \eqref{steady.state.w_boundary} are differentiated with regard to $\lambda$, $\rho$, and $\phi$.
By setting $\lambda = 1$, the expressions with modified Bessel functions for the other terms in Eq.~\eqref{expansion} are obtained.
\begin{align}
& c(\bm{r}; \bm{\rho}) = c_{\mathrm{infinite}} (\bm{r}; \bm{\rho}) + c_{\mathrm{correction}} (\bm{r}; \bm{\rho}). \label{eqc4}
\end{align}
\begin{align}
& c_{\mathrm{infinite}} (\bm{r}; \bm{\rho}) \nonumber \\
& = c_{00} \left ( |\bm{r} - \bm{\rho}| \right ) + c_{10} \left ( |\bm{r} - \bm{\rho}| \right ) (\bm{r} - \bm{\rho}) \cdot \dot{\bm{\rho}} \nonumber \\
& \quad + c_{20} \left ( |\bm{r} - \bm{\rho}| \right ) (\bm{r} - \bm{\rho}) \cdot \ddot{\bm{\rho}} + c_{21} \left ( |\bm{r} - \bm{\rho}| \right ) \left| \dot{\bm{\rho}} \right|^2 \nonumber \\
& \quad + c_{22} \left ( |\bm{r} - \bm{\rho}| \right ) \left[(\bm{r} - \bm{\rho}) \cdot \dot{\bm{\rho}} \right]^2+ c_{30} \left ( |\bm{r} - \bm{\rho}| \right ) (\bm{r} - \bm{\rho}) \cdot \dddot{\bm{\rho}} \nonumber \\
& \quad + c_{31} \left ( |\bm{r} - \bm{\rho}| \right ) \left| \dot{\bm{\rho}} \right|^2 (\bm{r} - \bm{\rho}) \cdot \dot{\bm{\rho}} + c_{33} \left ( |\bm{r} - \bm{\rho}| \right ) \dot{\bm{\rho}} \cdot \ddot{\bm{\rho}} \nonumber \\
& \quad + c_{32} \left ( |\bm{r} - \bm{\rho}| \right ) \left[(\bm{r} - \bm{\rho}) \cdot \dot{\bm{\rho}} \right]^3 \nonumber \\
& \quad + c_{34} \left ( |\bm{r} - \bm{\rho}| \right ) \left[(\bm{r} - \bm{\rho}) \cdot \dot{\bm{\rho}} \right] \left[(\bm{r} - \bm{\rho}) \cdot \ddot{\bm{\rho}} \right], \label{EQC5}
\end{align}
\begin{widetext}
\begin{align}
&c_{\mathrm{correction}} (\bm{r}; \bm{\rho}) \nonumber \\
&= c_{0}^{00} (R,r) + c_{0}^{10} (R,r) (\bm{r} \cdot \bm{\rho}) + c_{0}^{20} (R,r) (\bm{r} \cdot \bm{\rho})^2 + c_{1}^{20} (R,r) |\bm{\rho}|^2 + c_{0}^{11} (R,r) \left (\bm{r} \cdot \dot{\bm{\rho}} \right ) + c_{0}^{30} (R,r) (\bm{r} \cdot \bm{\rho})^3 \nonumber \\
& \quad + c_{1}^{30} (R,r) |\bm{\rho}|^2 (\bm{r} \cdot \bm{\rho}) + c_{0}^{21} (R,r) \left ( \bm{\rho} \cdot \dot{\bm{\rho}} \right ) + c_{1}^{21} (R,r) \left ( \bm{r} \cdot \bm{\rho} \right ) \left ( \bm{r} \cdot \dot{\bm{\rho}} \right ) + c_{0}^{12} (R,r) \left ( \bm{r} \cdot \ddot{\bm{\rho}} \right ) + c_{0}^{31} (R,r) |\bm{\rho}|^2 \left (\bm{r} \cdot \dot{\bm{\rho}} \right ) \nonumber \\
& \quad + c_{1}^{31} (R,r) \left (\bm{r} \cdot \bm{\rho} \right ) \left (\bm{\rho} \cdot \dot{\bm{\rho}} \right ) + c_{2}^{31} (R,r) \left (\bm{r} \cdot \bm{\rho} \right )^2 \left (\bm{r} \cdot \dot{\bm{\rho}} \right ) + c_{0}^{22} (R,r) \left (\bm{\rho} \cdot \ddot{\bm{\rho}} \right ) + c_{1}^{22} (R,r) \left | \dot{\bm{\rho}} \right |^2 + c_{2}^{22} (R,r) \left (\bm{r} \cdot \bm{\rho} \right ) \left (\bm{r} \cdot \ddot{\bm{\rho}} \right ) \nonumber \\
& \quad + c_{3}^{22} (R,r) \left (\bm{r} \cdot \dot{\bm{\rho}} \right )^2 + c_{0}^{13} (R,r) \left (\bm{r} \cdot \dddot{\bm{\rho}} \right ) + c_{0}^{32} (R,r) \left | \dot{\bm{\rho}} \right |^2 \left (\bm{r} \cdot \bm{\rho} \right ) + c_{1}^{32} (R,r) \left (\bm{r} \cdot \dot{\bm{\rho}} \right ) \left (\bm{\rho} \cdot \dot{\bm{\rho}} \right ) + c_{2}^{32} (R,r) \left (\bm{r} \cdot \bm{\rho} \right ) \left (\bm{r} \cdot \dot{\bm{\rho}} \right )^2 \nonumber \\
& \quad + c_{3}^{32} (R,r) \left (\bm{r} \cdot \bm{\rho} \right ) \left (\bm{\rho} \cdot \ddot{\bm{\rho}} \right ) + c_{4}^{32} (R,r) \left | \bm{\rho} \right |^2 \left (\bm{r} \cdot \ddot{\bm{\rho}} \right ) + c_{5}^{32} (R,r) \left (\bm{r} \cdot \bm{\rho} \right )^2 \left (\bm{r} \cdot \ddot{\bm{\rho}} \right ) + c_{0}^{23} (R,r) \left (\bm{\rho} \cdot \dddot{\bm{\rho}} \right ) + c_{1}^{23} (R,r) \left (\dot{\bm{\rho}} \cdot \ddot{\bm{\rho}} \right ) \nonumber \\
& \quad + c_{2}^{23} (R,r) \left (\bm{r} \cdot \dot{\bm{\rho}} \right ) \left (\bm{r} \cdot \ddot{\bm{\rho}} \right ) + c_{3}^{23} (R,r) \left (\bm{r} \cdot \bm{\rho} \right ) \left (\bm{r} \cdot \dddot{\bm{\rho}} \right ) + c_{0}^{33} (R,r) \left | \bm{\rho} \right |^2 \left (\bm{r} \cdot \dddot{\bm{\rho}} \right ) + c_{1}^{33} (R,r) \left | \dot{\bm{\rho}} \right |^2 \left (\bm{r} \cdot \dot{\bm{\rho}} \right ) \nonumber \\
& \quad + c_{2}^{33} (R,r) \left (\bm{r} \cdot \bm{\rho} \right ) \left (\bm{\rho} \cdot \dddot{\bm{\rho}} \right ) + c_{3}^{33} (R,r) \left (\bm{r} \cdot \dot{\bm{\rho}} \right )^3 + c_{4}^{33} (R,r) \left (\bm{r} \cdot \bm{\rho} \right ) \left (\dot{\bm{\rho}} \cdot \ddot{\bm{\rho}} \right ) + c_{5}^{33} (R,r) \left (\bm{r} \cdot \bm{\rho} \right )^2 \left (\bm{r} \cdot \dddot{\bm{\rho}} \right ) \nonumber \\
& \quad + c_{6}^{33} (R,r) \left (\bm{r} \cdot \dot{\bm{\rho}} \right ) \left (\bm{\rho} \cdot \ddot{\bm{\rho}} \right ) + c_{7}^{33} (R,r) \left (\bm{r} \cdot \ddot{\bm{\rho}} \right ) \left (\bm{\rho} \cdot \dot{\bm{\rho}} \right ) + c_{8}^{33} (R,r) \left (\bm{r} \cdot \bm{\rho} \right ) \left (\bm{r} \cdot \dot{\bm{\rho}} \right ) \left (\bm{r} \cdot \ddot{\bm{\rho}} \right ) + \mathcal{O} (\rho^4). \label{RESULT_CONC_APP}
\end{align}
\end{widetext}
The coefficients $c_{**} (R,r)$ and $c_{*}^{**} (R,r)$ are analytically obtained, whose explicit forms are shown in Sec.~\ref{SecS1} in Supplemental Material.
From Eq.~\eqref{RESULT_CONC_APP} and the definition of the driving force in Eq.~\eqref{d_def.drv.f}, the driving force is calculated as
\begin{align}
\bm{F} = \bm{F}_{\mathrm{infinite}} + \bm{F}_{\mathrm{correction}},
\end{align}
\begin{align}
\bm{F}_{\mathrm{infinite}} =& \frac{1}{4\pi} \left ( -\gamma_{\mathrm{Euler}} +\log \frac{2}{\epsilon} \right ) \dot{\bm{\rho}} - \frac{1}{16\pi} \ddot{\bm{\rho}} \nonumber \\
& - \frac{1}{32\pi} \left | \dot{\bm{\rho}} \right |^2 \dot{\bm{\rho}} + \frac{1}{48\pi} \dddot{\bm{\rho}},
\end{align}
\begin{widetext}
\begin{align}
& \bm{F}_{\mathrm{correction}} = - \left . \nabla_{\bm{r}} c(\bm{r}; \bm{\rho}) \right |_{\bm{r}=\bm{\rho}} \nonumber \\
&= - \left [ \frac{1}{\rho} {c'}_{0}^{00} (R,\rho) + {c'}_{0}^{10} (R,\rho) \rho + c_{0}^{10} (R,\rho) + {c'}_{0}^{20} (R,\rho) \rho^3 + 2 c_{0}^{20} (R,\rho) \rho^2 + {c'}_{1}^{20} (R,\rho) \rho + {c'}_{0}^{30} (R,\rho) \rho^5 + 3 c_{0}^{30} (R,\rho) \rho^4 \right . \nonumber \\
& \quad \left . + {c'}_{1}^{30} (R,\rho) \rho^3 + c_{1}^{30} (R,\rho) \rho^2 \right ] \bm{\rho} - \left [ \frac{1}{\rho} {c'}_{0}^{11} (R,\rho) + \frac{1}{\rho} {c'}_{0}^{21} (R,\rho) + {c'}_{1}^{21} (R,\rho) \rho + c_{1}^{21} (R,\rho) + {c'}_{0}^{31} (R,\rho) \rho + {c'}_{1}^{31} (R,\rho) \rho \right. \nonumber \\
& \quad \left . + c_{1}^{31} (R,\rho) + {c'}_{2}^{31} (R,\rho) \rho^3 + 2 c_{2}^{31} (R,\rho) \rho^2 \right ] \left (\bm{\rho} \cdot \dot{\bm{\rho}} \right ) \bm{\rho} - \left [ c_{0}^{11} (R,\rho) + c_{1}^{21} (R,\rho) \rho^2 + c_{0}^{31} (R,\rho) \rho^2 + c_{2}^{31} (R,\rho) \rho^4 \right ] \dot{\bm{\rho}} \nonumber \\
& \quad - \left [ \frac{1}{\rho} {c'}_{0}^{12} (R,\rho) + \frac{1}{\rho} {c'}_{0}^{22} (R,\rho) + {c'}_{2}^{22} (R,\rho) \rho + c_{2}^{22} (R,\rho) + {c'}_{3}^{32} (R,\rho) \rho + c_{3}^{32} (R,\rho) + {c'}_{4}^{32} (R,\rho) \rho + {c'}_{5}^{32} (R,\rho) \rho^3 \right . \nonumber \\
& \quad \left . + 2 c_{5}^{32} (R,\rho) \rho^2 \right ] \left ( \bm{\rho} \cdot \ddot{\bm{\rho}} \right ) \bm{\rho} - \left [ c_{0}^{12} (R,\rho) + c_{2}^{22} (R,\rho) \rho^2 + c_{4}^{32} (R,\rho) \rho^2 + c_{5}^{32} (R,\rho) \rho^4 \right ] \ddot{\bm{\rho}} - \left [ \frac{1}{\rho} {c'}_{1}^{22} (R,\rho) + {c'}_{0}^{32} (R,\rho) \rho \right. \nonumber \\
& \quad \left . + c_{0}^{32} (R,\rho) \right ] \left | \dot{\bm{\rho}} \right |^2 \bm{\rho} - \left [ \frac{1}{\rho} {c'}_{3}^{22} (R,\rho) + \frac{1}{\rho} {c'}_{1}^{32} (R,\rho) + {c'}_{2}^{32} (R,\rho) \rho + c_{2}^{32} (R,\rho) \right ] \left (\bm{\rho} \cdot \dot{\bm{\rho}} \right )^2 \bm{\rho} - \left[ 2 c_{3}^{22} (R,\rho) + c_{1}^{32} (R,\rho) \right. \nonumber \\
& \quad \left . + 2 c_{2}^{32} (R,\rho) \rho^2 \right ] \left (\bm{\rho} \cdot \dot{\bm{\rho}} \right ) \dot{\bm{\rho}} - \left [ \frac{1}{\rho} {c'}_{0}^{13} (R,\rho) + \frac{1}{\rho} {c'}_{0}^{23} (R,\rho) + {c'}_{3}^{23} (R,\rho) \rho + c_{3}^{23} (R,\rho) + {c'}_{0}^{33} (R,\rho) \rho + {c'}_{2}^{33} (R,\rho) \rho \right . \nonumber \\
& \quad \left . + c_{2}^{33} (R,\rho) + {c'}_{5}^{33} (R,\rho) \rho^3 + 2 c_{5}^{33} (R,\rho) \rho^2 \right ] \left (\bm{\rho} \cdot \dddot{\bm{\rho}} \right ) \bm{\rho} - \left [ c_{0}^{13} (R,\rho) + c_{3}^{23} (R,\rho) \rho^2 + c_{0}^{33} (R,\rho) \rho^2 + c_{5}^{33} (R,\rho) \rho^4 \right ] \dddot{\bm{\rho}} \nonumber \\
& \quad + \left [ \frac{1}{\rho} {c'}_{1}^{23} (R,\rho) + {c'}_{4}^{33} (R,\rho) \rho + c_{4}^{33} (R,\rho) \right ] \left (\dot{\bm{\rho}} \cdot \ddot{\bm{\rho}} \right ) \bm{\rho} - \left [ \frac{1}{\rho} {c'}_{2}^{23} (R,\rho) + \frac{1}{\rho} {c'}_{6}^{33} (R,\rho) + \frac{1}{\rho} {c'}_{7}^{33} (R,\rho) + {c'}_{8}^{33} (R,\rho) \rho \right. \nonumber \\
& \quad \left . + c_{8}^{33} (R,\rho) \right ] \left (\bm{\rho} \cdot \dot{\bm{\rho}} \right ) \left (\bm{\rho} \cdot \ddot{\bm{\rho}} \right ) \bm{\rho} - \left [ c_{2}^{23} (R,\rho) + c_{6}^{33} (R,\rho) + c_{8}^{33} (R,\rho) \rho^2 \right ] \left (\bm{\rho} \cdot \ddot{\bm{\rho}} \right ) \dot{\bm{\rho}} - \left [ c_{2}^{23} (R,\rho) + c_{7}^{33} (R,\rho) \right. \nonumber \\
& \quad \left . + c_{8}^{33} (R,\rho) \rho^2 \right ] \left (\bm{\rho} \cdot \dot{\bm{\rho}} \right ) \ddot{\bm{\rho}} - \frac{1}{\rho} {c'}_{1}^{33} (R,\rho) \left | \dot{\bm{\rho}} \right |^2 \left (\bm{\rho} \cdot \dot{\bm{\rho}} \right ) \bm{\rho} - c_{1}^{33} (R,\rho) \left | \dot{\bm{\rho}} \right |^2 \dot{\bm{\rho}} - \frac{1}{\rho} {c'}_{3}^{33} (R,\rho) \left (\bm{\rho} \cdot \dot{\bm{\rho}} \right )^3 \bm{\rho} - 3 c_{3}^{33} (R,\rho) \left (\bm{\rho} \cdot \dot{\bm{\rho}} \right )^2 \dot{\bm{\rho}} \nonumber \\
& \quad + \mathcal{O} (\rho^4).
\end{align}
\end{widetext}
By taking the terms up to the third-order of $\bm{\rho}$ and $\dot{\bm{\rho}}$, and first-order of $\ddot{\bm{\rho}}$, the driving force as a function of the trajectory of the camphor disk and the radius of the circular region is obtained as in Eq.~\eqref{DRIVING_FORCE}.

\section{Derivation of the steady concentration field\label{appD}}

In this section, the derivation of the steady concentration field for a fixed camphor disk at an arbitrary position in the circular region is obtained.
The steady state in an infinite region is obtained as follows:
\begin{align}
c(r,\theta) = \frac{1}{2 \pi} \mathcal{K}_0 \left ( \sqrt{r^2 + \rho^2 - 2 r \rho \cos(\theta-\phi)} \right ).
\end{align}
The detailed derivation is referred in Ref.~\cite{Koyano2017}.
To satisfy the Neumann boundary condition, we adequately add the general solution for Eq.~\eqref{d_d.eq.} without the source term, i.e., the homogeneous form of Eq.~\eqref{d_d.eq.}:
\begin{equation}
\left ( \frac{\partial^2}{\partial r^2} + \frac{1}{r} \frac{\partial}{\partial r} + \frac{1}{r^2} \frac{\partial^2}{\partial \theta^2} \right ) c(r,\theta) - c(r,\theta) = 0. \label{2.216}
\end{equation}
From the definition of the modified Bessel functions, the general solution of Eq.~\eqref{2.216} is expressed as
\begin{align}
c(r,\theta) =& A_0 \mathcal{K}_0 (r) + B_0 \mathcal{I}_0 (r) \nonumber \\
& + \sum_{m=1}^{\infty} \left ( A_m \mathcal{K}_m (r) + B_m \mathcal{I}_m (r) \right ) \cos m (\theta - \phi) \nonumber \\
& + \sum_{m=1}^{\infty} \left ( C_m \mathcal{K}_m (r) + D_m \mathcal{I}_m (r) \right ) \sin m (\theta - \phi).
\end{align}
By considering the symmetric property of the system, the $m$-th mode term should be expressed only by $\cos m (\theta-\phi)$, i.e., $C_m$ and $D_m$ should be zero.
Furthermore, $\mathcal{K}_n (r)$ $(n \geq 1)$ is not suitable for representing the concentration field of camphor since the integration, $\displaystyle{\int_{0}^{2\pi} \int_{0}^{R} \mathcal{K}_n (r) r dr d\theta}$, diverges for $n \geq 1$.
$\mathcal{K}_0 (r)$ diverges at $r=0$ and is not suitable when a camphor disk is off the origin.
When a camphor disk is located at the origin, $\mathcal{K}_0 (r)$ is already included as the steady state without the Neumann boundary.
Thus, the concentration field with the correction terms should be given by the following form:
\begin{align}
c(r,\theta) =& \frac{1}{2\pi} \mathcal{K}_0 \left ( \sqrt{r^2 + \rho^2 - 2 r \rho \cos(\theta-\phi)} \right ) \nonumber \\
& + \sum_{m=0}^{\infty} B_m \mathcal{I}_m (r) \cos m (\theta-\phi).
\end{align}
Then, the coefficients $B_m$ are determined by the boundary condition in Eq.~\eqref{d_Neumann}, that is
\begin{align}
& \frac{1}{2 \pi} \left . \frac{\partial }{\partial r} \mathcal{K}_0 \left ( \sqrt{r^2 + \rho^2 - 2 r \rho \cos(\theta-\phi)} \right ) \right |_{r=R} \nonumber \\
&= - \sum_{m=0}^{\infty} B_m \left . \frac{\partial \mathcal{I}_m (r)}{\partial r} \cos m (\theta-\phi) \right |_{r=R}.
\end{align}
If $\partial \mathcal{K}_0 \left ( \sqrt{r^2 + \rho^2 - 2 r \rho \cos(\theta-\phi)} \right ) / \partial r$ at $r = R$ is expanded with regard to $\cos m (\theta - \phi)$, $B_m$ is explicitly obtained.
By using the formula for $R>r$ which is represented in Eq.~(8) in p.361 of Ref.~\cite{Watson}:
\begin{equation}
\mathcal{K}_0 \left ( \sqrt{R^2 + r^2 - 2 R r \cos \theta} \right ) = \sum_{n=-\infty}^{\infty} \mathcal{K}_n (R) \mathcal{I}_n (r) \cos n \theta,
\end{equation}
we have
\begin{align}
& \frac{\partial}{\partial R} \int_{0}^{2 \pi} \mathcal{K}_0 \left ( \sqrt{R^2 + \rho^2 - 2 R \rho \cos(\theta-\phi)} \right ) \cos n (\theta - \phi) d \theta \nonumber \\
& \quad = 2 \pi \frac{\partial \mathcal{K}_n (R)}{\partial R} \mathcal{I}_n (\rho), \quad (n = 0, 1, 2, \cdots).
\end{align}
Here we use $\mathcal{K}_{-m} (r) = \mathcal{K}_{m} (r)$ and $\mathcal{I}_{-m} (r) = \mathcal{I}_{m} (r)$.
As a consequence, we have
\begin{align}
B_0 =& \frac{1}{2\pi} \frac{\mathcal{K}'_0 (R)}{\mathcal{I}'_0 (R)} \mathcal{I}_0 (\rho), \\
B_n =& \frac{1}{\pi} \frac{\mathcal{K}'_n (R)}{\mathcal{I}'_n (R)} \mathcal{I}_n (\rho), \quad (n = 1, 2, \cdots)
\end{align}
Thus we have Eq.~\eqref{steady.state.w_boundary}.
It should be noted that the conservation of integration of concentration over the circular region,
\begin{equation}
\int_{0}^{R} \int_{0}^{2\pi} c(r,\theta) r dr d\theta = 1,
\end{equation}
is satisfied.

\section{Detailed Experimental Conditions\label{appE}}

Milli-Q water (360~mL) was poured into a plastic container (150~mm in width, 150~mm in length, and 30~mm in depth) and the optical focus (Sigma-koki, IDC-025) was placed on the water surface with silicone sheet as a spacer (15~mm in depth).
The optical focus has a circular hole, whose radius $R$ was able to be changed from 0.5 to 20.0~mm.
For the preparation of the camphor disk~\cite{Soh2008}, agar gel sheet was soaked into methanol (MeOH) at first, and then, into a camphor MeOH solution (1.0~g mL${}^{-1}$) for more than 12~hours.
After the MeOH in an agar gel was completely replaced with camphor MeOH solution, the gel sheet was rinsed with water and was cut into a circular shape.
A camphor gel disk was floated on the water phase, and its dynamic behavior was observed with a video camera (Handycam, Sony, video rate: 30~fps).
The radius of the water phase $R$ was about 5.0~mm at first, and then, increased to 13.0 mm or 16.1 mm with a constant speed (4.0~mm s${}^{-1}$).

\end{appendix}

\clearpage

\begin{widetext}
\section*{Supplemental Material: Rotational motion of a camphor disk in a circular region}

\setcounter{figure}{0}
\setcounter{section}{0}

\renewcommand{\thesection}{S\arabic{section}}

\renewcommand{\thefigure}{S\arabic{figure}}

\renewcommand{\theequation}{S\arabic{equation}}

\section{Derivation of Eqs.~\eqref{EQC5} and \eqref{RESULT_CONC_APP} \label{SecS1}}

As expressed in Eq.~\eqref{eqc4}, the expanded concentration field $c$ is composed of the concentration field for infinite system size $c_{\mathrm{infinite}}$ and boundary effect $c_{\mathrm{correction}}$:
\begin{align}
c(\bm{r}; \bm{\rho}) = c_{\mathrm{infinite}} (\bm{r}; \bm{\rho}) + c_{\mathrm{correction}} (\bm{r}; \bm{\rho}).
\end{align}

The expanded concentration field for infinite system size $c_{\mathrm{infinite}}$ is obtained as follows (cf.~Ref.~\cite{Koyano2017}):
\begin{align}
c_{\mathrm{infinite}} (\bm{r}; \bm{\rho})
& = c_{00} \left ( |\bm{r} - \bm{\rho}| \right ) + c_{10} \left ( |\bm{r} - \bm{\rho}| \right ) (\bm{r} - \bm{\rho}) \cdot \dot{\bm{\rho}} + c_{20} \left ( |\bm{r} - \bm{\rho}| \right ) (\bm{r} - \bm{\rho}) \cdot \ddot{\bm{\rho}} + c_{21} \left ( |\bm{r} - \bm{\rho}| \right ) \left| \dot{\bm{\rho}} \right|^2 \nonumber \\
& \quad + c_{22} \left ( |\bm{r} - \bm{\rho}| \right ) \left[(\bm{r} - \bm{\rho}) \cdot \dot{\bm{\rho}} \right]^2+ c_{30} \left ( |\bm{r} - \bm{\rho}| \right ) (\bm{r} - \bm{\rho}) \cdot \dddot{\bm{\rho}} + c_{31} \left ( |\bm{r} - \bm{\rho}| \right ) \left| \dot{\bm{\rho}} \right|^2 (\bm{r} - \bm{\rho}) \cdot \dot{\bm{\rho}} \nonumber\\
& \quad + c_{32} \left ( |\bm{r} - \bm{\rho}| \right ) \left[(\bm{r} - \bm{\rho}) \cdot \dot{\bm{\rho}} \right]^3 + c_{33} \left ( |\bm{r} - \bm{\rho}| \right ) \dot{\bm{\rho}} \cdot \ddot{\bm{\rho}}+ c_{34} \left ( |\bm{r} - \bm{\rho}| \right ) \left[(\bm{r} - \bm{\rho}) \cdot \dot{\bm{\rho}} \right] \left[(\bm{r} - \bm{\rho}) \cdot \ddot{\bm{\rho}} \right], \label{concentration_field_2D_infinite_2}
\end{align}
where
\begin{align}
& c_{00} \left ( |\bm{r} - \bm{\rho}| \right ) = \frac{1}{2\pi} \mathcal{K}_0 \left ( |\bm{r} - \bm{\rho}| \right ), && c_{10} \left ( |\bm{r} - \bm{\rho}| \right ) = - \frac{1}{4\pi} \mathcal{K}_0 \left ( |\bm{r} - \bm{\rho}| \right ), \nonumber \\
& c_{20} \left ( |\bm{r} - \bm{\rho}| \right ) = \frac{1}{16\pi} |\bm{r} - \bm{\rho}| \mathcal{K}_1 \left ( |\bm{r} - \bm{\rho}| \right ), && c_{21} \left ( |\bm{r} - \bm{\rho}| \right ) = - \frac{1}{16\pi} |\bm{r} - \bm{\rho}| \mathcal{K}_1 \left ( |\bm{r} - \bm{\rho}| \right ), \nonumber \\
& c_{22} \left ( |\bm{r} - \bm{\rho}| \right ) = \frac{1}{16\pi} \mathcal{K}_0 \left ( |\bm{r} - \bm{\rho}| \right ), && c_{30} \left ( |\bm{r} - \bm{\rho}| \right ) = -\frac{1}{96\pi} |\bm{r} - \bm{\rho}|^2 \mathcal{K}_2 \left ( |\bm{r} - \bm{\rho}| \right ), \nonumber \\
& c_{31} \left ( |\bm{r} - \bm{\rho}| \right ) = \frac{1}{32\pi} |\bm{r} - \bm{\rho}| \mathcal{K}_1 \left ( |\bm{r} - \bm{\rho}| \right ), && c_{32} \left ( |\bm{r} - \bm{\rho}| \right ) = -\frac{1}{96\pi} \mathcal{K}_0 \left ( |\bm{r} - \bm{\rho}| \right ), \nonumber \\
&c_{33} \left ( |\bm{r} - \bm{\rho}| \right ) = \frac{1}{32\pi} |\bm{r} - \bm{\rho}|^2 \mathcal{K}_2 \left ( |\bm{r} - \bm{\rho}| \right ), && c_{34} \left ( |\bm{r} - \bm{\rho}| \right ) = -\frac{1}{32\pi} |\bm{r} - \bm{\rho}| \mathcal{K}_1 \left ( |\bm{r} - \bm{\rho}| \right ).\label{concentration_field_2D_infinite_each_term}
\end{align}

Here we derive Eq.~\eqref{RESULT_CONC_APP}.
The first term in the righthand side in Eq.~\eqref{expansion} should correspond to the steady state with a camphor particle fixed at $(\rho, \phi)$ in the two-dimensional polar coordinates.
The steady state is independently obtained as shown in Eq.~\eqref{steady.state.w_boundary}.
Thus we have
\begin{align}
&\frac{1}{2 \pi} \sum_{m=-\infty}^{\infty} \sum_{n=0}^{\infty} a_{|m|n} \frac{1}{{k_{|m|n}}^2+1} \mathcal{J}_{|m|} (k_{|m|n} \rho) \mathcal{J}_{|m|} (k_{|m|n} r) e^{i m (\theta-\phi)} \nonumber \\
& \qquad = \frac{1}{2 \pi} \mathcal{K}_0 \left ( \sqrt{r^2 + \rho^2 - 2 r \rho \cos(\theta-\phi)} \right ) - \frac{1}{2 \pi} \sum_{m=-\infty}^{\infty} \frac{{\mathcal{K}_m}' (R)}{{\mathcal{I}_m}' (R)} \mathcal{I}_m (\rho) \mathcal{I}_m (r) e^{i m (\theta-\phi)} \nonumber \\
& \qquad = \mbox{(main term)} - \frac{1}{2 \pi} \frac{{\mathcal{K}_0}' (R)}{{\mathcal{I}_0}' (R)} \mathcal{I}_m (\rho) \mathcal{I}_m (r) - \frac{1}{\pi} \sum_{m=1}^{\infty} \frac{{\mathcal{K}_m}' (R)}{{\mathcal{I}_m}' (R)} \mathcal{I}_m (\rho) \mathcal{I}_m (r) \cos m (\theta-\phi) \nonumber \\
& \qquad = \mbox{(main term)} + \sum_{m=0}^{\infty} h_m^{00} (R) g_m^{00} (r, \rho) \cos m (\theta-\phi) \nonumber \\
& \qquad = \mbox{(main term)} + \alpha^{01}(R,r) + \alpha^{02}(R,r) \rho^2 + \alpha^{03}(R,r) \rho \cos (\theta - \phi) +\alpha^{04}(R,r) \rho^3 \cos (\theta - \phi) \nonumber \\
& \qquad \quad + \alpha^{05}(R,r) \rho^2 \cos 2 (\theta - \phi) + \alpha^{06}(R,r) \rho^3 \cos 3 (\theta - \phi) + \mathcal{O} (\rho^4). \label{steady_state_expanded}
\end{align}
We call the term corresponding to the steady-state concentration field in an infinite system as ``main term''.

Here $k_{mn}$ and $a_{mn}$ are the wave number and the normalization constant, respectively.
The definition of them are provided in Appendix~\ref{appB}.
By changing the length scale, i.e., $\rho \to \lambda \rho$, $r \to \lambda r$, $R \to \lambda R$, $k_{m n} \to k_{m n}/\lambda$, $a_{m n} \to a_{m n}/\lambda^2$, we have
\begin{align}
&\frac{1}{2 \pi} \sum_{m=-\infty}^{\infty} \sum_{n=0}^{\infty} a_{|m|n} \frac{1}{{k_{|m|n}}^2+\lambda^2} \mathcal{J}_{|m|} (k_{|m|n} \rho) \mathcal{J}_{|m|} (k_{|m|n} r) e^{i m (\theta-\phi)} \nonumber \\
& \qquad = \mbox{(main term)} - \frac{1}{2 \pi} \frac{{\mathcal{K}_0}' (\lambda R)}{{\mathcal{I}_0}' (\lambda R)} \mathcal{I}_m (\lambda \rho) \mathcal{I}_m (\lambda r) - \sum_{m=1}^{\infty} \frac{{\mathcal{K}_m}' (\lambda R)}{{\mathcal{I}_m}' (\lambda R)} \mathcal{I}_m (\lambda \rho) \mathcal{I}_m (\lambda r) \cos m (\theta-\phi) \nonumber \\
& \qquad = \mbox{(main term)} + \sum_{m=0}^{\infty} \bar{h}_m^{00} (\lambda, R) \bar{g}_m^{00} (\lambda, r, \rho) \cos m (\theta-\phi). \label{t0r0p0}
\end{align}
Here we define 
\begin{align}
\bar{h}_m^{00} (\lambda, R) = - \frac{\sigma_{m}}{\pi} \frac{{\mathcal{K}_m}' (\lambda R)}{{\mathcal{I}_m}' (\lambda R)}, \qquad \bar{g}_m^{00} (\lambda, r, \rho) = \mathcal{I}_m (\lambda \rho) \mathcal{I}_m (\lambda r),
\end{align}
where $\sigma_m$ is equal to $1$ for $m \neq 0$ and $1/2$ for $m=0$.
In Eq.~\eqref{t0r0p0} and later, the terms transformed from the steady-state concentration field in an infinite system are also denoted as ``main term''.

%Here we do not consider the main term, since the effect by the main term corresponds to the concentration field without boundaries and is already calculated as shown in Eq.~\eqref{concentration_field_2D_infinite}.
%
By differentiating the both sides with regard to $\lambda$ and then dividing the both sides of Eq.~\eqref{t0r0p0} by $2\lambda$, we have
\begin{align}
&\frac{1}{2 \pi} \sum_{m=-\infty}^{\infty} \sum_{n=0}^{\infty} a_{|m|n} \frac{-1}{({k_{|m|n}}^2+\lambda^2)^2} \mathcal{J}_{|m|} (k_{|m|n} \rho) \mathcal{J}_{|m|} (k_{|m|n} r) e^{i m (\theta-\phi)} \nonumber \\
& \qquad = \mbox{(main term)} + \sum_{m=0}^{\infty} \left ( \bar{h}_m^{10} (\lambda, R) \bar{g}_m^{00} (\lambda, r, \rho) + \bar{h}_m^{11} (\lambda, R) \bar{g}_m^{10} (\lambda, r, \rho) \right ) \cos m (\theta-\phi). \label{l_t1r0p0}
\end{align}
By setting $\lambda$ to be 1, we have
\begin{align}
&\frac{1}{2 \pi} \sum_{m=-\infty}^{\infty} \sum_{n=0}^{\infty} a_{|m|n} \frac{-1}{({k_{|m|n}}^2+1)^2} \mathcal{J}_{|m|} (k_{|m|n} \rho) \mathcal{J}_{|m|} (k_{|m|n} r) e^{i m (\theta-\phi)} \nonumber \\
& \qquad = \mbox{(main term)} + \sum_{m=0}^{\infty} \left ( h_m^{10} (R) g_m^{00} (r, \rho) + h_m^{11} (R) g_m^{10} (r, \rho) \right ) \cos m (\theta-\phi). \label{t1r0p0}
\end{align}
By differentiating the both sides of Eq.~\eqref{t1r0p0} with regard to $\rho$, we have
\begin{align}
&\frac{1}{2 \pi} \sum_{m=-\infty}^{\infty} \sum_{n=0}^{\infty} a_{|m|n} \frac{- k_{|m|n}}{({k_{|m|n}}^2+1)^2} \mathcal{J}'_{|m|} (k_{|m|n} \rho) \mathcal{J}_{|m|} (k_{|m|n} r) e^{i m (\theta-\phi)} \nonumber \\
& \qquad = \mbox{(main term)} + \sum_{m=0}^{\infty} \left ( h_m^{10} (R) g_m^{01} (r, \rho) + h_m^{11} (R) g_m^{11} (r, \rho) \right ) \cos m (\theta-\phi) \nonumber \\
& \qquad = \mbox{(main term)} + \alpha^{11}(R,r) \rho + \alpha^{12}(R,r) \cos (\theta - \phi) + 3 \alpha^{13}(R,r) \rho^2 \cos (\theta - \phi) + \alpha^{14}(R,r) \rho \cos 2 (\theta - \phi) \nonumber \\
& \qquad \quad + \alpha^{15}(R,r) \rho^2 \cos 3 (\theta - \phi) + \mathcal{O}(\rho^3). \label{t1r1p0_expanded}
\end{align}
Similarly, by differentiating the both sides of Eq.~\eqref{t1r0p0} with regard to $\phi$, we have
\begin{align}
&\frac{1}{2 \pi} \sum_{m=-\infty}^{\infty} \sum_{n=0}^{\infty} a_{|m|n} \frac{i m}{({k_{|m|n}}^2+1)^2} \mathcal{J}_{|m|} (k_{|m|n} \rho) \mathcal{J}_{|m|} (k_{|m|n} r) e^{i m (\theta-\phi)} \nonumber \\
& \qquad = \mbox{(main term)} + \sum_{m=1}^{\infty} m \left ( h_m^{10} (R) g_m^{00} (r, \rho) + h_m^{11} (R) g_m^{10} (r, \rho) \right ) \sin m (\theta-\phi) \nonumber \\
& \qquad = \mbox{(main term)} + \alpha^{12}(R,r) \rho \sin (\theta - \phi) + \alpha^{13}(R,r) \rho^3 \sin (\theta - \phi) + \alpha^{14}(R,r) \rho^2 \sin 2 (\theta - \phi) \nonumber \\
& \qquad \quad + \alpha^{15}(R,r) \rho^3 \sin 3 (\theta - \phi) + \mathcal{O}(\rho^4). \label{t1r0p1_expanded}
\end{align}
By differentiating the both sides of Eq.~\eqref{l_t1r0p0} with regard to $\lambda$ and then dividing the both sides by $4\lambda$, we have
\begin{align}
&\frac{1}{2 \pi} \sum_{m=-\infty}^{\infty} \sum_{n=0}^{\infty} a_{|m|n} \frac{1}{({k_{|m|n}}^2+\lambda^2)^3} \mathcal{J}_{|m|} (k_{|m|n} \rho) \mathcal{J}_{|m|} (k_{|m|n} r) e^{i m (\theta-\phi)} \nonumber \\
& \qquad = \mbox{(main term)} + \sum_{m=0}^{\infty} \left ( \bar{h}_m^{20} (\lambda, R) \bar{g}_m^{00} (\lambda, r, \rho) + \bar{h}_m^{21} (\lambda, R) \bar{g}_m^{10} (\lambda, r, \rho) + \bar{h}_m^{22} (\lambda, R) \bar{g}_m^{20} (\lambda, r, \rho) \right ) \cos m (\theta-\phi). \label{l_t2r0p0}
\end{align}
By setting $\lambda$ to be 1, we have
\begin{align}
&\frac{1}{2 \pi} \sum_{m=-\infty}^{\infty} \sum_{n=0}^{\infty} a_{|m|n} \frac{1}{({k_{|m|n}}^2+1)^3} \mathcal{J}_{|m|} (k_{|m|n} \rho) \mathcal{J}_{|m|} (k_{|m|n} r) e^{i m (\theta-\phi)} \nonumber \\
& \qquad = \mbox{(main term)} + \sum_{m=0}^{\infty} \left ( h_m^{20} (R) g_m^{00} (r, \rho) + h_m^{21} (R) g_m^{10} (r, \rho) + h_m^{22} (R) g_m^{20} (r, \rho) \right ) \cos m (\theta-\phi). \label{t2r0p0}
\end{align}
By differentiating the both sides of Eq.~\eqref{t2r0p0} with regard to $\rho$, we have
\begin{align}
&\frac{1}{2 \pi} \sum_{m=-\infty}^{\infty} \sum_{n=0}^{\infty} a_{|m|n} \frac{k_{|m|n}}{({k_{|m|n}}^2+1)^3} \mathcal{J}'_{|m|} (k_{|m|n} \rho) \mathcal{J}_{|m|} (k_{|m|n} r) e^{i m (\theta-\phi)} \nonumber \\
& \qquad = \mbox{(main term)} + \sum_{m=0}^{\infty} \left ( h_m^{20} (R) g_m^{01} (r, \rho) + h_m^{21} (R) g_m^{11} (r, \rho) + h_m^{22} (R) g_m^{21} (r, \rho) \right ) \cos m (\theta-\phi) \label{t2r1p0} \\
& \qquad = \mbox{(main term)} + \alpha^{21}(R,r) \rho + \alpha^{22}(R,r) \cos (\theta - \phi) + 3 \alpha^{23}(R,r) \rho^2 \cos (\theta - \phi) + \alpha^{24}(R,r) \rho \cos 2 (\theta - \phi) \nonumber \\
& \qquad \quad + \alpha^{25}(R,r) \rho^2 \cos 3 (\theta - \phi) + \mathcal{O}(\rho^3). \label{t2r1p0_expanded}
\end{align}
By differentiating the both sides of Eq.~\eqref{t2r1p0} with regard to $\rho$, we have
\begin{align}
&\frac{1}{2 \pi} \sum_{m=-\infty}^{\infty} \sum_{n=0}^{\infty} a_{|m|n} \frac{{k_{|m|n}}^2}{({k_{|m|n}}^2+1)^3} \mathcal{J}''_{|m|} (k_{|m|n} \rho) \mathcal{J}_{|m|} (k_{|m|n} r) e^{i m (\theta-\phi)} \nonumber \\
& \qquad = \mbox{(main term)} + \sum_{m=0}^{\infty} \left ( h_m^{20} (R) g_m^{02} (r, \rho) + h_m^{21} (R) g_m^{12} (r, \rho) + h_m^{22} (R) g_m^{22} (r, \rho) \right ) \cos m (\theta-\phi) \label{t2r2p0} \\
& \qquad = \mbox{(main term)} + \alpha^{21}(R,r) + 6 \alpha^{23}(R,r) \rho \cos (\theta - \phi) + \alpha^{24}(R,r) \cos 2 (\theta - \phi) + 2 \alpha^{25}(R,r) \rho \cos 3 (\theta - \phi) + \mathcal{O}(\rho^2). \label{t2r2p0_expanded}
\end{align}
By differentiating the both sides of Eq.~\eqref{t2r1p0} with regard to $\phi$, we have
\begin{align}
&\frac{1}{2 \pi} \sum_{m=-\infty}^{\infty} \sum_{n=0}^{\infty} a_{|m|n} \frac{- i m k_{|m|n}}{({k_{|m|n}}^2+1)^3} \mathcal{J}'_{|m|} (k_{|m|n} \rho) \mathcal{J}_{|m|} (k_{|m|n} r) e^{i m (\theta-\phi)} \nonumber \\
& \qquad = \mbox{(main term)} + \sum_{m=1}^{\infty} m \left ( h_m^{20} (R) g_m^{01} (r, \rho) + h_m^{21} (R) g_m^{11} (r, \rho) + h_m^{22} (R) g_m^{21} (r, \rho) \right ) \sin m (\theta-\phi) \label{t2r1p1} \\
& \qquad = \mbox{(main term)} + \frac{1}{2} \left [ 2 \alpha^{22}(R,r) \sin (\theta - \phi) + 6 \alpha^{23}(R,r) \rho^2 \sin (\theta - \phi) + 4 \alpha^{24}(R,r) \rho \sin 2 (\theta - \phi) \right . \nonumber \\
& \qquad \quad \left . + 6 \alpha^{25}(R,r) \rho^2 \sin 3 (\theta - \phi) \right ] + \mathcal{O}(\rho^3). \label{t2r1p1_expanded}
\end{align}
By differentiating the both sides of Eq.~\eqref{t2r0p0} with regard to $\phi$, we have
\begin{align}
&\frac{1}{2 \pi} \sum_{m=-\infty}^{\infty} \sum_{n=0}^{\infty} a_{|m|n} \frac{-im}{({k_{|m|n}}^2+1)^3} \mathcal{J}_{|m|} (k_{|m|n} \rho) \mathcal{J}_{|m|} (k_{|m|n} r) e^{i m (\theta-\phi)} \nonumber \\
& \qquad = \mbox{(main term)} + \sum_{m=1}^{\infty} m \left ( h_m^{20} (R) g_m^{00} (r, \rho) + h_m^{21} (R) g_m^{10} (r, \rho) + h_m^{22} (R) g_m^{20} (r, \rho) \right ) \sin m (\theta-\phi) \label{t2r0p1} \\
& \qquad = \mbox{(main term)} + \alpha^{22}(R,r) \rho \sin (\theta - \phi) + \alpha^{23}(R,r) \rho^3 \sin (\theta - \phi) + \alpha^{24}(R,r) \rho^2 \sin 2 (\theta - \phi) \nonumber \\
& \qquad \quad + \alpha^{25}(R,r) \rho^3 \sin 3 (\theta - \phi) + \mathcal{O}(\rho^4). \label{t2r0p1_expanded}
\end{align}
By differentiating the both sides of Eq.~\eqref{t2r0p1} with regard to $\phi$, we have
\begin{align}
&\frac{1}{2 \pi} \sum_{m=-\infty}^{\infty} \sum_{n=0}^{\infty} a_{|m|n} \frac{-m^2}{({k_{|m|n}}^2+1)^3} \mathcal{J}_{|m|} (k_{|m|n} \rho) \mathcal{J}_{|m|} (k_{|m|n} r) e^{i m (\theta-\phi)} \nonumber \\
& \qquad = \mbox{(main term)} - \sum_{m=1}^{\infty} m^2 \left ( h_m^{20} (R) g_m^{00} (r, \rho) + h_m^{21} (R) g_m^{10} (r, \rho) + h_m^{22} (R) g_m^{20} (r, \rho) \right ) \cos m (\theta-\phi) \label{t2r0p2} \\
& \qquad = \mbox{(main term)} - \alpha^{22}(R,r) \rho \cos (\theta - \phi) - \alpha^{23}(R,r) \rho^3 \cos (\theta - \phi) - 2 \alpha^{24}(R,r) \rho^2 \cos 2 (\theta - \phi) \nonumber \\
& \qquad \quad - 3 \alpha^{25}(R,r) \rho^3 \cos 3 (\theta - \phi) + \mathcal{O}(\rho^4). \label{t2r0p2_expanded}
\end{align}
By differentiating the both sides of Eq.~\eqref{l_t2r0p0} with regard to $\lambda$ and then dividing the both sides by $6\lambda$, we have
\begin{align}
&\frac{1}{2 \pi} \sum_{m=-\infty}^{\infty} \sum_{n=0}^{\infty} a_{|m|n} \frac{-1}{({k_{|m|n}}^2+\lambda^2)^4} \mathcal{J}_{|m|} (k_{|m|n} \rho) \mathcal{J}_{|m|} (k_{|m|n} r) e^{i m (\theta-\phi)} \nonumber \\
& \qquad = \mbox{(main term)} + \sum_{m=0}^{\infty} \left ( \bar{h}_m^{30} (\lambda, R) \bar{g}_m^{00} (\lambda, r, \rho) + \bar{h}_m^{31} (\lambda, R) \bar{g}_m^{10} (\lambda, r, \rho) + \bar{h}_m^{32} (\lambda, R) \bar{g}_m^{20} (\lambda, r, \rho) \right . \nonumber \\
& \qquad \qquad \qquad \left . + \bar{h}_m^{33} (\lambda, R) \bar{g}_m^{30} (\lambda, r, \rho) \right ) \cos m (\theta-\phi). \label{l_t3r0p0}
\end{align}
By setting $\lambda$ to be 1, we have
\begin{align}
&\frac{1}{2 \pi} \sum_{m=-\infty}^{\infty} \sum_{n=0}^{\infty} a_{|m|n} \frac{-1}{({k_{|m|n}}^2+1)^4} \mathcal{J}_{|m|} (k_{|m|n} \rho) \mathcal{J}_{|m|} (k_{|m|n} r) e^{i m (\theta-\phi)} \nonumber \\
& \qquad = \mbox{(main term)} + \sum_{m=0}^{\infty} \left ( h_m^{30} (R) g_m^{00} (r, \rho) + h_m^{31} (R) g_m^{10} (r, \rho) + h_m^{32} (R) g_m^{20} (r, \rho) + h_m^{33} (R) g_m^{30} (r, \rho) \right ) \cos m (\theta-\phi). \label{t3r0p0}
\end{align}
By differentiating the both sides of Eq.~\eqref{t3r0p0} with regard to $\rho$, we have
\begin{align}
&\frac{1}{2 \pi} \sum_{m=-\infty}^{\infty} \sum_{n=0}^{\infty} a_{|m|n} \frac{-k_{|m|n}}{({k_{|m|n}}^2+1)^4} \mathcal{J}'_{|m|} (k_{|m|n} \rho) \mathcal{J}_{|m|} (k_{|m|n} r) e^{i m (\theta-\phi)} \nonumber \\
& \qquad = \mbox{(main term)} + \sum_{m=0}^{\infty} \left ( h_m^{30} (R) g_m^{01} (r, \rho) + h_m^{31} (R) g_m^{11} (r, \rho) + h_m^{32} (R) g_m^{21} (r, \rho) + h_m^{33} (R) g_m^{31} (r, \rho) \right ) \cos m (\theta-\phi) \label{t3r1p0} \\
& \qquad = \mbox{(main term)} + \alpha^{31}(R,r) \rho + \alpha^{32}(R,r) \rho \cos (\theta - \phi) + 3 \alpha^{33}(R,r) \rho^2 \cos (\theta - \phi) + \alpha^{34}(R,r) \rho \cos 2 (\theta - \phi) \nonumber \\
& \qquad \quad + \alpha^{35}(R,r) \rho^2 \cos 3 (\theta - \phi) + \mathcal{O}(\rho^3). \label{t3r1p0_expanded}
\end{align}
By differentiating the both sides of Eq.~\eqref{t3r1p0} with regard to $\rho$, we have
\begin{align}
&\frac{1}{2 \pi} \sum_{m=-\infty}^{\infty} \sum_{n=0}^{\infty} a_{|m|n} \frac{-{k_{|m|n}}^2}{({k_{|m|n}}^2+1)^4} \mathcal{J}''_{|m|} (k_{|m|n} \rho) \mathcal{J}_{|m|} (k_{|m|n} r) e^{i m (\theta-\phi)} \nonumber \\
& \qquad = \mbox{(main term)} + \sum_{m=0}^{\infty} \left ( h_m^{30} (R) g_m^{02} (r, \rho) + h_m^{31} (R) g_m^{12} (r, \rho) + h_m^{32} (R) g_m^{22} (r, \rho) + h_m^{33} (R) g_m^{32} (r, \rho) \right ) \cos m (\theta-\phi) \label{t3r2p0} \\
& \qquad = \mbox{(main term)} + \alpha^{31}(R,r) + 6 \alpha^{33}(R,r) \rho \cos (\theta - \phi) + \alpha^{34}(R,r) \cos 2 (\theta - \phi) + 2 \alpha^{35}(R,r) \rho \cos 3 (\theta - \phi) + \mathcal{O}(\rho^2). \label{t3r2p0_expanded}
\end{align}
By differentiating the both sides of Eq.~\eqref{t3r2p0} with regard to $\rho$, we have
\begin{align}
&\frac{1}{2 \pi} \sum_{m=-\infty}^{\infty} \sum_{n=0}^{\infty} a_{|m|n} \frac{-{k_{|m|n}}^3}{({k_{|m|n}}^2+1)^4} \mathcal{J}'''_{|m|} (k_{|m|n} \rho) \mathcal{J}_{|m|} (k_{|m|n} r) e^{i m (\theta-\phi)} \nonumber \\
& \qquad = \mbox{(main term)} + \sum_{m=0}^{\infty} \left ( h_m^{30} (R) g_m^{03} (r, \rho) + h_m^{31} (R) g_m^{13} (r, \rho) + h_m^{32} (R) g_m^{23} (r, \rho) + h_m^{33} (R) g_m^{33} (r, \rho) \right ) \cos m (\theta-\phi) \label{t3r3p0} \\
& \qquad = \mbox{(main term)} + 6 \alpha^{33}(R,r) \cos (\theta - \phi) + 2 \alpha^{35}(R,r) \cos 3 (\theta - \phi) + \mathcal{O}(\rho). \label{t3r3p0_expanded}
\end{align}
By differentiating the both sides of Eq.~\eqref{t3r1p0} with regard to $\phi$, we have
\begin{align}
&\frac{1}{2 \pi} \sum_{m=-\infty}^{\infty} \sum_{n=0}^{\infty} a_{|m|n} \frac{-i m k_{|m|n}}{({k_{|m|n}}^2+1)^4} \mathcal{J}'_{|m|} (k_{|m|n} \rho) \mathcal{J}_{|m|} (k_{|m|n} r) e^{i m (\theta-\phi)} \nonumber \\
& \qquad = \mbox{(main term)} + \sum_{m=1}^{\infty} m \left ( h_m^{30} (R) g_m^{01} (r, \rho) + h_m^{31} (R) g_m^{11} (r, \rho) + h_m^{32} (R) g_m^{21} (r, \rho) + h_m^{33} (R) g_m^{31} (r, \rho) \right ) \sin m (\theta-\phi) \label{t3r1p1} \\
& \qquad = \mbox{(main term)} + \alpha^{32}(R,r) \sin (\theta - \phi) + 3a^{33}(R,r) \rho^2 \sin (\theta - \phi) + 2 \alpha^{34}(R,r) \rho \sin 2 (\theta - \phi) \nonumber \\
& \qquad \quad + 3 \alpha^{35}(R,r) \rho^2 \sin 3 (\theta - \phi) + \mathcal{O}(\rho^3). \label{t3r1p1_expanded}
\end{align}
By differentiating the both sides of Eq.~\eqref{t3r1p1} with regard to $\phi$, we have
\begin{align}
&\frac{1}{2 \pi} \sum_{m=-\infty}^{\infty} \sum_{n=0}^{\infty} a_{|m|n} \frac{- m^2 k_{|m|n}}{({k_{|m|n}}^2+1)^4} \mathcal{J}'_{|m|} (k_{|m|n} \rho) \mathcal{J}_{|m|} (k_{|m|n} r) e^{i m (\theta-\phi)} \nonumber \\
& \qquad = \mbox{(main term)} - \sum_{m=1}^{\infty} m^2 \left ( h_m^{30} (R) g_m^{01} (r, \rho) + h_m^{31} (R) g_m^{11} (r, \rho) + h_m^{32} (R) g_m^{21} (r, \rho) + h_m^{33} (R) g_m^{31} (r, \rho) \right ) \cos m (\theta-\phi) \label{t3r1p2} \\
& \qquad = \mbox{(main term)} - \alpha^{32}(R,r) \cos (\theta - \phi) - 3 \alpha^{33}(R,r) \rho^2 \cos (\theta - \phi) - 4 \alpha^{34}(R,r) \rho \cos 2 (\theta - \phi) \nonumber \\
& \qquad \quad - 9 \alpha^{35}(R,r) \rho^2 \cos 3 (\theta - \phi) + \mathcal{O}(\rho^3). \label{t3r1p2_expanded}
\end{align}
By differentiating the both sides of Eq.~\eqref{t3r2p0} with regard to $\phi$, we have
\begin{align}
&\frac{1}{2 \pi} \sum_{m=-\infty}^{\infty} \sum_{n=0}^{\infty} a_{|m|n} \frac{-i m {k_{|m|n}}^2}{({k_{|m|n}}^2+1)^4} \mathcal{J}''_{|m|} (k_{|m|n} \rho) \mathcal{J}_{|m|} (k_{|m|n} r) e^{i m (\theta-\phi)} \nonumber \\
& \qquad = \mbox{(main term)} + \sum_{m=1}^{\infty} m \left ( h_m^{30} (R) g_m^{02} (r, \rho) + h_m^{31} (R) g_m^{12} (r, \rho) + h_m^{32} (R) g_m^{22} (r, \rho) + h_m^{33} (R) g_m^{32} (r, \rho) \right ) \sin m (\theta-\phi) \label{t3r2p1} \\
& \qquad = \mbox{(main term)} + 6 \alpha^{33}(R,r) \rho \sin (\theta - \phi) + 2 \alpha^{34}(R,r) \sin 2 (\theta - \phi) + 6 \alpha^{35}(R,r) \rho \sin 3 (\theta - \phi) + \mathcal{O}(\rho^2). \label{t3r2p1_expanded}
\end{align}
By differentiating the both sides of Eq.~\eqref{t3r0p0} with regard to $\phi$, we have
\begin{align}
&\frac{1}{2 \pi} \sum_{m=-\infty}^{\infty} \sum_{n=0}^{\infty} a_{|m|n} \frac{-im}{({k_{|m|n}}^2+1)^4} \mathcal{J}_{|m|} (k_{|m|n} \rho) \mathcal{J}_{|m|} (k_{|m|n} r) e^{i m (\theta-\phi)} \nonumber \\
& \qquad = \mbox{(main term)} + \sum_{m=1}^{\infty} m \left ( h_m^{30} (R) g_m^{00} (r, \rho) + h_m^{31} (R) g_m^{10} (r, \rho) + h_m^{32} (R) g_m^{20} (r, \rho) + h_m^{33} (R) g_m^{30} (r, \rho) \right ) \sin m (\theta-\phi) \label{t3r0p1} \\
& \qquad = \mbox{(main term)} + \alpha^{32}(R,r) \rho \sin (\theta - \phi) + \alpha^{33}(R,r) \rho^3 \sin (\theta - \phi) + \alpha^{34}(R,r) \rho^2 \sin 2 (\theta - \phi) \nonumber \\
& \qquad \quad + \alpha^{35}(R,r) \rho^3 \sin 3 (\theta - \phi) + \mathcal{O}(\rho^4). \label{t3r0p1_expanded}
\end{align}
By differentiating the both sides of Eq.~\eqref{t3r0p1} with regard to $\phi$, we have
\begin{align}
&\frac{1}{2 \pi} \sum_{m=-\infty}^{\infty} \sum_{n=0}^{\infty} a_{|m|n} \frac{-m^2}{({k_{|m|n}}^2+1)^4} \mathcal{J}_{|m|} (k_{|m|n} \rho) \mathcal{J}_{|m|} (k_{|m|n} r) e^{i m (\theta-\phi)} \nonumber \\
& \qquad = \mbox{(main term)} - \sum_{m=1}^{\infty} m^2 \left ( h_m^{30} (R) g_m^{00} (r, \rho) + h_m^{31} (R) g_m^{10} (r, \rho) + h_m^{32} (R) g_m^{20} (r, \rho) + h_m^{33} (R) g_m^{30} (r, \rho) \right ) \cos m (\theta-\phi) \label{t3r0p2} \\
& \qquad = \mbox{(main term)} - \alpha^{32}(R,r) \rho \cos (\theta - \phi) - \alpha^{33}(R,r) \rho^3 \cos (\theta - \phi) - 2 \alpha^{34}(R,r) \rho^2 \cos 2 (\theta - \phi) \nonumber \\
& \qquad \quad - 3 \alpha^{35}(R,r) \rho^3 \cos 3 (\theta - \phi) + \mathcal{O}(\rho^4). \label{t3r0p2_expanded}
\end{align}
By differentiating the both sides of Eq.~\eqref{t3r0p2} with regard to $\phi$, we have
\begin{align}
&\frac{1}{2 \pi} \sum_{m=-\infty}^{\infty} \sum_{n=0}^{\infty} a_{|m|n} \frac{i m^3}{({k_{|m|n}}^2+1)^4} \mathcal{J}_{|m|} (k_{|m|n} \rho) \mathcal{J}_{|m|} (k_{|m|n} r) e^{i m (\theta-\phi)} \nonumber \\
& \qquad = \mbox{(main term)} - \sum_{m=1}^{\infty} m^3 \left ( h_m^{30} (R) g_m^{00} (r, \rho) + h_m^{31} (R) g_m^{10} (r, \rho) + h_m^{32} (R) g_m^{20} (r, \rho) + h_m^{33} (R) g_m^{30} (r, \rho) \right ) \sin m (\theta-\phi) \label{t3r0p3} \\
& \qquad = \mbox{(main term)} - \alpha^{32}(R,r) \rho \sin (\theta - \phi) - \alpha^{33}(R,r) \rho^2 \sin (\theta - \phi) - 4 \alpha^{34}(R,r) \rho \sin 2 (\theta - \phi) \nonumber \\
& \qquad \quad - 9 \alpha^{35}(R,r) \rho^3 \sin 3 (\theta - \phi) + \mathcal{O}(\rho^4). \label{t3r0p3_expanded}
\end{align}
\allowdisplaybreaks[1]
Here, we define $h_{m}^{ij} (R) = \bar{h}_{m}^{ij} (1, R)$ and the explicit forms of $\bar{h}_{m}^{ij} (\lambda, R)$ are given as follows:
\begin{align}
\bar{h}_m^{00} (\lambda, R) =
- \frac{\sigma_m}{\pi} \frac{\mathcal{K}'_m (\lambda R)}{\mathcal{I}'_m (\lambda R)},
\end{align}
\begin{align}
\bar{h}_m^{10} (\lambda, R) =
\frac{1}{2 \lambda} \frac{d}{d\lambda} \bar{h}_m^{00}(\lambda, R)
= - \frac{\sigma_m}{2 \pi} \left ( \frac{1}{\lambda^2} + \frac{m^2}{\lambda^4 R^2} \right ) \frac{1}{(\mathcal{I}'_m (\lambda R))^2},
\end{align}
\begin{align}
\bar{h}_m^{11} (\lambda, R)
= \frac{1}{2 \lambda} \bar{h}_m^{00}(\lambda, R)
= - \frac{\sigma_m}{2 \pi \lambda} \frac{\mathcal{K}'_m (\lambda R)}{\mathcal{I}'_m (\lambda R)},
\end{align}
\begin{align}
\bar{h}_m^{20} (\lambda, R)
= \frac{1}{4 \lambda} \frac{d}{d\lambda} \left ( \frac{1}{2 \lambda} \frac{d}{d\lambda} \bar{h}_m^{00}(\lambda, R) \right ) 
= \frac{\sigma_m}{4 \pi} \left ( \left ( \frac{1}{\lambda^4} + \frac{2 m^2}{\lambda^6 R^2} \right ) \frac{1}{(\mathcal{I}'_m (\lambda R))^2} + \left ( \frac{1}{\lambda^3} + \frac{m^2}{\lambda^5 R^2} \right ) \frac{R \mathcal{I}''_m (\lambda R)}{(\mathcal{I}'_m (\lambda R))^3} \right ),
\end{align}
\begin{align}
\bar{h}_m^{21} (\lambda, R)
= \frac{1}{4 \lambda} \left ( \frac{1}{2 \lambda} \frac{d}{d\lambda} \bar{h}_0^{00} (\lambda, R) \right ) + \frac{1}{4 \lambda} \frac{d}{d\lambda} \left ( \frac{1}{2 \lambda} \bar{h}_0^{00}(\lambda, R) \right ) 
= \frac{\sigma_m}{8 \pi} \left ( - 2 \left ( \frac{1}{\lambda^3} + \frac{m^2}{\lambda^5 R^2} \right ) \frac{1}{(\mathcal{I}'_m (\lambda R))^2} + \frac{\mathcal{K}'_m (\lambda R)}{\lambda^3 \mathcal{I}'_m (\lambda R)} \right ),
\end{align}
\begin{align}
\bar{h}_m^{22} (\lambda, R)
= \frac{1}{4 \lambda} h_m^{11}(\lambda, R)
= - \frac{\sigma_m}{8 \pi \lambda^2} \frac{\mathcal{K}'_m (\lambda R)}{\mathcal{I}'_m (\lambda R)},
\end{align}
\begin{align}
\bar{h}_m^{30} (\lambda, R)
=& \frac{1}{6 \lambda} \frac{d}{d\lambda} \left ( \frac{1}{4 \lambda} \frac{d}{d\lambda} \left ( \frac{1}{2 \lambda} \frac{d}{d\lambda} \bar{h}_m^{00}(\lambda, R) \right ) \right ) \nonumber \\
=& -\frac{\sigma_m}{24 \pi} \left ( 4 \left ( \frac{1}{\lambda^6} + \frac{3 m^2}{\lambda^8 R^2} \right ) \frac{1}{(\mathcal{I}'_m (\lambda R))^2} + \left ( \frac{5}{\lambda^5} + \frac{9 m^2}{\lambda^7 R^2} \right ) \frac{R \mathcal{I}''_m (\lambda R)}{(\mathcal{I}'_m (\lambda R))^3} \right . \nonumber \\
& \qquad \qquad \left . - \left ( \frac{1}{\lambda^4} + \frac{m^2}{\lambda^6 R^2} \right ) \frac{R^2 \mathcal{I}'''_m (\lambda R)}{(\mathcal{I}'_m (\lambda R))^3} + 3 \left ( \frac{1}{\lambda^4} + \frac{m^2}{\lambda^6 R^2} \right ) \frac{R^2 (\mathcal{I}''_m (\lambda R))^2}{(\mathcal{I}'_m (\lambda R))^4}\right ),
\end{align}
\begin{align}
\bar{h}_m^{31} (\lambda, R)
=& \frac{1}{6 \lambda} \left ( \frac{1}{4 \lambda} \frac{d}{d\lambda} \left ( \frac{1}{2 \lambda} \frac{d}{d\lambda} \bar{h}_m^{00}(\lambda, R) \right ) \right ) + \frac{1}{6 \lambda} \frac{d}{d\lambda} \left ( \frac{1}{4 \lambda} \left ( \frac{1}{2 \lambda} \frac{d}{d\lambda} \bar{h}_m^{00}(\lambda, R) \right ) \right ) + \frac{1}{6 \lambda} \frac{d}{d\lambda} \left ( \frac{1}{4 \lambda} \frac{d}{d\lambda} \left ( \frac{1}{2 \lambda} \bar{h}_m^{00}(\lambda, R) \right ) \right ) \nonumber \\
=& \frac{\sigma_m}{16 \pi} \left ( \left ( \frac{3}{\lambda^5} + \frac{5m^2}{\lambda^7 R^2} \right ) \frac{1}{(\mathcal{I}'_m (\lambda R))^2} + 2 \left ( \frac{1}{\lambda^4} + \frac{m^2}{\lambda^6 R^2} \right ) \frac{R \mathcal{I}''_m (\lambda R)}{(\mathcal{I}'_m (\lambda R))^3} - \frac{\mathcal{K}'_m (\lambda R)}{\lambda^5 \mathcal{I}'_m (\lambda R)} \right ),
\end{align}
\begin{align}
\bar{h}_m^{32} (\lambda, R) 
& = \frac{1}{6 \lambda} \left ( \frac{1}{4 \lambda} \left ( \frac{1}{2 \lambda} \frac{d}{d\lambda} \bar{h}_m^{00}(\lambda, R) \right ) \right ) + \frac{1}{6 \lambda} \left ( \frac{1}{4 \lambda} \frac{d}{d\lambda} \left ( \frac{1}{2 \lambda} \bar{h}_m^{00}(\lambda, R) \right ) \right ) + \frac{1}{6 \lambda} \frac{d}{d\lambda} \left ( \frac{1}{4 \lambda} \left ( \frac{1}{2 \lambda} \bar{h}_m^{00}(\lambda, R) \right ) \right ) \nonumber \\
& = \frac{\sigma_m}{16 \pi} \left ( \frac{\mathcal{K}'_m (\lambda R)}{\lambda^4 \mathcal{I}'_m (\lambda R)} - \left ( \frac{1}{\lambda^4} + \frac{m^2}{\lambda^6 R^2} \right ) \frac{1}{(\mathcal{I}'_m (\lambda R))^2} \right ),
\end{align}
\begin{align}
\bar{h}_m^{33} (\lambda, R)
= \frac{1}{6 \lambda} \bar{h}_m^{22}(\lambda, R)
= - \frac{\sigma_m}{48 \pi \lambda^3} \frac{\mathcal{K}'_m (\lambda R)}{\mathcal{I}'_m (\lambda R)},
\end{align}
We also define $g_{m}^{i0} (r, \rho) = \bar{g}_{m}^{i0} (1, r, \rho)$ and the explicit forms of $\bar{g}_{m}^{i0} (\lambda, r, \rho)$ are as follows:
\begin{align}
\bar{g}_m^{00}(\lambda, r, \rho) =& \mathcal{I}_m (\lambda r) \mathcal{I}_m (\lambda \rho), \\
\bar{g}_m^{10}(\lambda, r, \rho) =& \frac{d}{d\lambda} \bar{g}_m^{00}(\lambda, r, \rho)
= r \mathcal{I}'_m (\lambda r) \mathcal{I}_m (\lambda \rho) + \rho \mathcal{I}_m (\lambda r) \mathcal{I}'_m (\lambda \rho), \\
\bar{g}_m^{20}(\lambda, r, \rho) =& \frac{d^2}{d\lambda^2} \bar{g}_m^{00}(\lambda, r, \rho)
= r^2 \mathcal{I}''_m (\lambda r) \mathcal{I}_m (\lambda \rho) + 2 r \rho \mathcal{I}'_m (\lambda r) \mathcal{I}'_m (\lambda \rho) + \rho^2 \mathcal{I}_m (\lambda r) \mathcal{I}''_m (\lambda \rho), \\
\bar{g}_m^{30}(\lambda, r, \rho) =& \frac{d^3}{d\lambda^3} \bar{g}_m^{00}(\lambda, r, \rho) = r^3 \mathcal{I}'''_m (\lambda r) \mathcal{I}_m (\lambda \rho) + 3 r^2 \rho \mathcal{I}''_m (\lambda r) \mathcal{I}'_m (\lambda \rho) + 3 r \rho^2 \mathcal{I}'_m (\lambda r) \mathcal{I}''_m (\lambda \rho) + \rho^3 \mathcal{I}_m (\lambda r) \mathcal{I}'''_m (\lambda \rho).
\end{align}
The functions $g_{m}^{ij} (r, \rho)$ ($j \neq 0$) are defined by the derivatives of $g_{m}^{i0} (r, \rho)$ with regard to $r$ and/or $\rho$ as follows:
\begin{align}
g_m^{01}(r, \rho) =& \frac{d}{d\rho} \left ( g_m^{00}(r, \rho) \right )
= \mathcal{I}_m (r) \mathcal{I}'_m (\rho), \\
g_m^{02}(r, \rho) =& \frac{d^2}{d\rho^2} \left ( g_m^{00}(r, \rho) \right )
= \mathcal{I}_m (r) \mathcal{I}''_m (\rho), \\
g_m^{03}(r, \rho) =& \frac{d^3}{d\rho^3} \left ( g_m^{00}(r, \rho) \right )
= \mathcal{I}_m (r) \mathcal{I}'''_m (\rho),
\end{align}
\begin{align}
g_m^{11}(r, \rho) =& \frac{d}{d\rho} g_m^{10}(r, \rho)
= r \mathcal{I}'_m (r) \mathcal{I}'_m (\rho) + \mathcal{I}_m (r) \mathcal{I}'_m (\rho) + \rho \mathcal{I}_m (r) \mathcal{I}''_m (\rho), \\
g_m^{12}(r, \rho) =& \frac{d^2}{d\rho^2} g_m^{10}(r, \rho)
= r \mathcal{I}'_m (r) \mathcal{I}''_m (\rho) + 2 \mathcal{I}_m (r) \mathcal{I}''_m (\rho) + \rho \mathcal{I}_m (r) \mathcal{I}'''_m (\rho), \\
g_m^{13}(r, \rho) =& \frac{d^3}{d\rho^3} g_m^{10}(r, \rho)
= r \mathcal{I}'_m (r) \mathcal{I}'''_m (\rho) + 3 \mathcal{I}_m (r) \mathcal{I}'''_m (\rho) + \rho \mathcal{I}_m (r) \mathcal{I}^{(4)}_m (\rho),
\end{align}
\begin{align}
g_m^{21}(r, \rho) =& \frac{d}{d\rho} g_m^{20}(r, \rho) \nonumber \\
=& r^2 \mathcal{I}''_m (r) \mathcal{I}'_m (\rho) + 2 r \mathcal{I}'_m (r) \mathcal{I}'_m (\rho) + 2 r \rho \mathcal{I}'_m (r) \mathcal{I}''_m (\rho) + 2 \rho \mathcal{I}_m (r) \mathcal{I}''_m (\rho) + \rho^2 \mathcal{I}_m (r) \mathcal{I}'''_m (\rho), \\
g_m^{22}(r, \rho) =& \frac{d^2}{d\rho^2} g_m^{20}(r, \rho) \nonumber \\
=& r^2 \mathcal{I}''_m (r) \mathcal{I}''_m (\rho) + 4 r \mathcal{I}'_m (r) \mathcal{I}''_m (\rho) + 2 r \rho \mathcal{I}'_m (r) \mathcal{I}'''_m (\rho) + 2 \mathcal{I}_m (r) \mathcal{I}''_m (\rho) + 4 \rho \mathcal{I}_m (r) \mathcal{I}'''_m (\rho) + \rho^2 \mathcal{I}_m (r) \mathcal{I}^{(4)}_m (\rho), \\
g_m^{23}(r, \rho) =& \frac{d^3}{d\rho^3} g_m^{20}(r, \rho) \nonumber \\
=& r^2 \mathcal{I}''_m (r) \mathcal{I}'''_m (\rho) + 6 r \mathcal{I}'_m (r) \mathcal{I}'''_m (\rho) + 2 r \rho \mathcal{I}'_m (r) \mathcal{I}^{(4)}_m (\rho) + 6 \mathcal{I}_m (r) \mathcal{I}'''_m (\rho) + 6 \rho \mathcal{I}_m (r) \mathcal{I}^{(4)}_m (\rho) + \rho^2 \mathcal{I}_m (r) \mathcal{I}^{(5)}_m (\rho),
\end{align}
\begin{align}
g_m^{31}(r, \rho) =& \frac{d}{d\rho} g_m^{30}(r, \rho) \nonumber \\
=& r^3 \mathcal{I}'''_m (r) \mathcal{I}'_m (\rho) + 3 r^2 \mathcal{I}''_m (r) \mathcal{I}'_m (\rho) + 3 r^2 \rho \mathcal{I}''_m (r) \mathcal{I}''_m (\rho) + 6 r \rho \mathcal{I}'_m (r) \mathcal{I}''_m (\rho) + 3 r \rho^2 \mathcal{I}'_m (r) \mathcal{I}'''_m (\rho) \nonumber \\
& + 3 \rho^2 \mathcal{I}_m (r) \mathcal{I}'''_m (\rho) + \rho^3 \mathcal{I}_m (r) \mathcal{I}^{(4)}_m (\rho), \\
g_m^{32}(r, \rho) =& \frac{d^2}{d\rho^2} g_m^{30}(r, \rho) \nonumber \\
=& r^3 \mathcal{I}'''_m (r) \mathcal{I}''_m (\rho) + 6 r^2 \mathcal{I}''_m (r) \mathcal{I}''_m (\rho) + 3 r^2 \rho \mathcal{I}''_m (r) \mathcal{I}'''_m (\rho) + 6 r \mathcal{I}'_m (r) \mathcal{I}''_m (\rho) + 12 r \rho \mathcal{I}'_m (r) \mathcal{I}'''_m (\rho) \nonumber \\
& + 3 r \rho^2 \mathcal{I}'_m (r) \mathcal{I}^{(4)}_m (\rho) + 6 \rho \mathcal{I}_m (r) \mathcal{I}'''_m (\rho) + 6 \rho^2 \mathcal{I}_m (r) \mathcal{I}^{(4)}_m (\rho) + \rho^3 \mathcal{I}_m (r) \mathcal{I}^{(5)}_m (\rho), \\
g_m^{33}(r, \rho) =& \frac{d^3}{d\rho^3} g_m^{30}(r, \rho) \nonumber \\
=& r^3 \mathcal{I}'''_m (r) \mathcal{I}'''_m (\rho) + 9 r^2 \mathcal{I}''_m (r) \mathcal{I}'''_m (\rho) + 3 r^2 \rho \mathcal{I}''_m (r) \mathcal{I}^{(4)}_m (\rho) + 18 r \mathcal{I}'_m (r) \mathcal{I}'''_m (\rho) + 18 r \rho \mathcal{I}'_m (r) \mathcal{I}^{(4)}_m (\rho) \nonumber \\
& + 3 r \rho^2 \mathcal{I}'_m (r) \mathcal{I}^{(5)}_m (\rho) + 6 \mathcal{I}_m (r) \mathcal{I}'''_m (\rho) + 18 \rho \mathcal{I}_m (r) \mathcal{I}^{(4)}_m (\rho) + 9 \rho^2 \mathcal{I}_m (r) \mathcal{I}^{(5)}_m (\rho) + \rho^3 \mathcal{I}_m (r) \mathcal{I}^{(6)}_m (\rho),
\end{align}
By expanding the explicit forms of $g_{m}^{ij}$ with respect to $\rho$, the functions $a^{kl}(R,r)$ are obtained as follows:
\begin{align}
&\alpha^{01}(R,r) = h_{0}^{00} (R) \mathcal{I}_0 (r), && \alpha^{02}(R,r) = \frac{1}{4} h_{0}^{00} (R) \mathcal{I}_0 (r), \nonumber \\
&\alpha^{03}(R,r) = \frac{1}{2} h_{1}^{00} (R) \mathcal{I}_1 (r), && \alpha^{04}(R,r) = \frac{1}{16} h_{1}^{00} (R) \mathcal{I}_1 (r), \nonumber \\
&\alpha^{05}(R,r) = \frac{1}{8} h_{2}^{00} (R) \mathcal{I}_2 (r), && \alpha^{06}(R,r) = \frac{1}{48} h_{3}^{00} (R) \mathcal{I}_3 (r),
\end{align}
\begin{align}
\alpha^{11}(R,r) =& \frac{1}{2} \left [ (h_{0}^{10} (R) + 2 h_{0}^{11} (R) \mathcal{I}_0 (r) + h_{0}^{11} (R) r \mathcal{I}'_0 (r) \right ], \\
\alpha^{12}(R,r) =& \frac{1}{2} \left [ (h_{1}^{10} (R) + h_{1}^{11} (R) \mathcal{I}_1 (r) + h_{1}^{11} (R) r \mathcal{I}'_1 (r) \right ], \\
\alpha^{13}(R,r) =& \frac{3}{16} \left [ (h_{1}^{10} (R) + 3 h_{1}^{11} (R) \mathcal{I}_1 (r) + h_{1}^{11} (R) r \mathcal{I}'_1 (r) \right ], \\
\alpha^{14}(R,r) =& \frac{1}{4} \left [ (h_{2}^{10} (R) + 2 h_{2}^{11} (R) \mathcal{I}_2 (r) + h_{2}^{11} (R) r \mathcal{I}'_2 (r) \right ], \\
\alpha^{15}(R,r) =& \frac{1}{16} \left [ (h_{3}^{10} (R) + 3 h_{3}^{11} (R) \mathcal{I}_3 (r) + h_{3}^{11} (R) r \mathcal{I}'_3 (r) \right ],
\end{align}
\begin{align}
\alpha^{21}(R,r) =& \frac{1}{2} \left [ (h_{0}^{20} (R) + 2 h_{0}^{21} (R) + 2 h_{0}^{22} (R)) \mathcal{I}_0 (r) + (h_{0}^{21} (R) + 4 h_{0}^{22} (R)) r \mathcal{I}'_0 (r) + h_{0}^{22} (R) r^2 \mathcal{I}''_0 (r) \right ], \\
\alpha^{22}(R,r) =& \frac{1}{2} \left [ (h_{1}^{20} (R) + h_{1}^{21} (R)) \mathcal{I}_1 (r) + (h_{1}^{21} (R) + 2 h_{1}^{22} (R)) r \mathcal{I}'_1 (r) + h_{1}^{22} (R) r^2 \mathcal{I}''_1 (r) \right ], \\
\alpha^{23}(R,r) =& \frac{1}{16} \left [ (h_{1}^{20} (R) + 3 h_{1}^{21} (R) + 6 h_{1}^{22} (R)) \mathcal{I}_1 (r) + (h_{1}^{21} (R) + 6 h_{1}^{22} (R)) r \mathcal{I}'_1 (r) + h_{1}^{22} (R) r^2 \mathcal{I}''_1 (r) \right ], \\
\alpha^{24}(R,r) =& \frac{1}{4} \left [ (h_{2}^{20} (R) + 2 h_{2}^{21} (R) + 2 h_{2}^{22} (R)) \mathcal{I}_2 (r) + (h_{2}^{21} (R) + 4 h_{2}^{22} (R)) r \mathcal{I}'_2 (r) + h_{2}^{22} (R) r^2 \mathcal{I}''_2 (r) \right ], \\
\alpha^{25}(R,r) =& \frac{1}{16} \left [ (h_{3}^{20} (R) + 3 h_{3}^{21} (R) + 6 h_{3}^{22} (R)) \mathcal{I}_1 (r) + (h_{3}^{21} (R) + 6 h_{3}^{22} (R)) r \mathcal{I}'_1 (r) + h_{3}^{22} (R) r^2 \mathcal{I}''_3 (r) \right ],
\end{align}
\begin{align}
\alpha^{31}(R,r) =& \frac{1}{2} \left [ (h_{0}^{30} (R) + 2 h_{0}^{31} (R) + 2 h_{0}^{32} (R)) \mathcal{I}_0 (r) + (h_{0}^{31} (R) + 4 h_{0}^{32} (R) + 6 h_{0}^{36} (R)) r \mathcal{I}'_0 (r) \right . \nonumber \\
& \qquad \left . + (h_{0}^{32} (R) + 6 h_{0}^{32} (R)) r^2 \mathcal{I}''_0 (r) + h_{0}^{31} (R)) r^3 \mathcal{I}'''_0 (r) \right ], \\
\alpha^{32}(R,r) =& \frac{1}{2} \left [ (h_{1}^{30} (R) + h_{1}^{31} (R)) \mathcal{I}_1 (r) + (h_{1}^{31} (R) + 2 h_{0}^{32} (R)) r \mathcal{I}'_1 (r) + (h_{0}^{32} (R) + 3 h_{0}^{32} (R)) r^2 \mathcal{I}''_1 (r) + h_{0}^{33} (R)) r^3 \mathcal{I}'''_1 (r) \right ], \\
\alpha^{33}(R,r) =& \frac{1}{16} \left [ (h_{1}^{30} (R) + 3 h_{1}^{31} (R) + 6 h_{1}^{32} (R) + 6 h_{1}^{33} (R)) \mathcal{I}_1 (r) + (h_{1}^{31} (R) + 6 h_{1}^{32} (R) + 18 h_{1}^{36} (R)) r \mathcal{I}'_1 (r) \right . \nonumber \\
& \qquad \left . + (h_{1}^{32} (R) + 9 h_{1}^{33} (R)) r^2 \mathcal{I}''_1 (r) + h_{1}^{33} (R)) r^3 \mathcal{I}'''_1 (r) \right ], \\
\alpha^{34}(R,r) =& \frac{1}{4} \left [ (h_{2}^{30} (R) + 2 h_{2}^{31} (R) + 2 h_{2}^{32} (R)) \mathcal{I}_2 (r) + (h_{2}^{31} (R) + 4 h_{2}^{32} (R) + 6 h_{2}^{36} (R)) r \mathcal{I}'_2 (r) \right . \nonumber \\
& \qquad \left . + (h_{2}^{32} (R) + 6 h_{2}^{32} (R)) r^2 \mathcal{I}''_2 (r) + h_{2}^{31} (R)) r^3 \mathcal{I}'''_2 (r) \right ], \\
\alpha^{35}(R,r) =& \frac{1}{16} \left [ (h_{3}^{30} (R) + 3 h_{3}^{31} (R) + 6 h_{3}^{32} (R) + 6 h_{3}^{33} (R)) \mathcal{I}_3 (r) + (h_{3}^{31} (R) + 6 h_{3}^{32} (R) + 18 h_{3}^{36} (R)) r \mathcal{I}'_3 (r) \right . \nonumber \\
& \qquad \left . + (h_{3}^{32} (R) + 9 h_{3}^{33} (R)) r^2 \mathcal{I}''_3 (r) + h_{3}^{33} (R)) r^3 \mathcal{I}'''_3 (r) \right ].
\end{align}
\allowdisplaybreaks[0]

From the symmetric property of the system, the concentration field expanded with regard to $\bm{\rho}$ should have the following form:
\begin{align}
& c_{\mathrm{correction}} (\bm{r}; \bm{\rho}) \nonumber \\
& = c_{0}^{00} (R,r) + c_{0}^{10} (R,r) (\bm{r} \cdot \bm{\rho}) + c_{0}^{20} (R,r) (\bm{r} \cdot \bm{\rho})^2 + c_{1}^{20} (R,r) |\bm{\rho}|^2 + c_{0}^{11} (R,r) \left (\bm{r} \cdot \dot{\bm{\rho}} \right ) + c_{0}^{30} (R,r) (\bm{r} \cdot \bm{\rho})^3 \nonumber \\
& \quad + c_{1}^{30} (R,r) |\bm{\rho}|^2 (\bm{r} \cdot \bm{\rho}) + c_{0}^{21} (R,r) \left ( \bm{\rho} \cdot \dot{\bm{\rho}} \right ) + c_{1}^{21} (R,r) \left ( \bm{r} \cdot \bm{\rho} \right ) \left ( \bm{r} \cdot \dot{\bm{\rho}} \right ) + c_{0}^{12} (R,r) \left ( \bm{r} \cdot \ddot{\bm{\rho}} \right ) + c_{0}^{31} (R,r) |\bm{\rho}|^2 \left (\bm{r} \cdot \dot{\bm{\rho}} \right ) \nonumber \\
& \quad + c_{1}^{31} (R,r) \left (\bm{r} \cdot \bm{\rho} \right ) \left (\bm{\rho} \cdot \dot{\bm{\rho}} \right ) + c_{2}^{31} (R,r) \left (\bm{r} \cdot \bm{\rho} \right )^2 \left (\bm{r} \cdot \dot{\bm{\rho}} \right ) + c_{0}^{22} (R,r) \left (\bm{\rho} \cdot \ddot{\bm{\rho}} \right ) + c_{1}^{22} (R,r) \left | \dot{\bm{\rho}} \right |^2 + c_{2}^{22} (R,r) \left (\bm{r} \cdot \bm{\rho} \right ) \left (\bm{r} \cdot \ddot{\bm{\rho}} \right ) \nonumber \\
& \quad + c_{3}^{22} (R,r) \left (\bm{r} \cdot \dot{\bm{\rho}} \right )^2 + c_{0}^{13} (R,r) \left (\bm{r} \cdot \dddot{\bm{\rho}} \right ) + c_{0}^{32} (R,r) \left | \dot{\bm{\rho}} \right |^2 \left (\bm{r} \cdot \bm{\rho} \right ) + c_{1}^{32} (R,r) \left (\bm{r} \cdot \dot{\bm{\rho}} \right ) \left (\bm{\rho} \cdot \dot{\bm{\rho}} \right ) + c_{2}^{32} (R,r) \left (\bm{r} \cdot \bm{\rho} \right ) \left (\bm{r} \cdot \dot{\bm{\rho}} \right )^2 \nonumber \\
& \quad + c_{3}^{32} (R,r) \left (\bm{r} \cdot \bm{\rho} \right ) \left (\bm{\rho} \cdot \ddot{\bm{\rho}} \right ) + c_{4}^{32} (R,r) \left | \bm{\rho} \right |^2 \left (\bm{r} \cdot \ddot{\bm{\rho}} \right ) + c_{5}^{32} (R,r) \left (\bm{r} \cdot \bm{\rho} \right )^2 \left (\bm{r} \cdot \ddot{\bm{\rho}} \right ) + c_{0}^{23} (R,r) \left (\bm{\rho} \cdot \dddot{\bm{\rho}} \right ) + c_{1}^{23} (R,r) \left (\dot{\bm{\rho}} \cdot \ddot{\bm{\rho}} \right ) \nonumber \\
& \quad + c_{2}^{23} (R,r) \left (\bm{r} \cdot \dot{\bm{\rho}} \right ) \left (\bm{r} \cdot \ddot{\bm{\rho}} \right ) + c_{3}^{23} (R,r) \left (\bm{r} \cdot \bm{\rho} \right ) \left (\bm{r} \cdot \dddot{\bm{\rho}} \right ) + c_{0}^{33} (R,r) \left | \bm{\rho} \right |^2 \left (\bm{r} \cdot \dddot{\bm{\rho}} \right ) + c_{1}^{33} (R,r) \left | \dot{\bm{\rho}} \right |^2 \left (\bm{r} \cdot \dot{\bm{\rho}} \right ) \nonumber \\
& \quad + c_{2}^{33} (R,r) \left (\bm{r} \cdot \bm{\rho} \right ) \left (\bm{\rho} \cdot \dddot{\bm{\rho}} \right ) + c_{3}^{33} (R,r) \left (\bm{r} \cdot \dot{\bm{\rho}} \right )^3 + c_{4}^{33} (R,r) \left (\bm{r} \cdot \bm{\rho} \right ) \left (\dot{\bm{\rho}} \cdot \ddot{\bm{\rho}} \right ) + c_{5}^{33} (R,r) \left (\bm{r} \cdot \bm{\rho} \right )^2 \left (\bm{r} \cdot \dddot{\bm{\rho}} \right ) \nonumber \\
& \quad + c_{6}^{33} (R,r) \left (\bm{r} \cdot \dot{\bm{\rho}} \right ) \left (\bm{\rho} \cdot \ddot{\bm{\rho}} \right ) + c_{7}^{33} (R,r) \left (\bm{r} \cdot \ddot{\bm{\rho}} \right ) \left (\bm{\rho} \cdot \dot{\bm{\rho}} \right ) + c_{8}^{33} (R,r) \left (\bm{r} \cdot \bm{\rho} \right ) \left (\bm{r} \cdot \dot{\bm{\rho}} \right ) \left (\bm{r} \cdot \ddot{\bm{\rho}} \right ). \label{s_result_conc_app}
\end{align}
Equation \eqref{s_result_conc_app} is the same as Eq.~\eqref{eqc4}.
The coefficients in Eq.~\eqref{s_result_conc_app} are given as follows:
Comparing Eq.~\eqref{s_result_conc_app} with Eq.~\eqref{steady_state_expanded}, we have
\begin{align}
&c_{0}^{00}(R,r) = \alpha^{01}(R,r), && c_{0}^{10}(R,r) = \frac{1}{r} \alpha^{03}(R,r), \nonumber \\
&c_{0}^{20}(R,r) = \frac{2}{r^2} \alpha^{05}(R,r), &&c_{1}^{20}(R,r) = \alpha^{02}(R,r) - \alpha^{05}(R,r), \nonumber \\
&c_{0}^{30}(R,r) = \frac{4}{r^3} \alpha^{06}(R,r), && c_{1}^{30}(R,r) = \frac{1}{r} \left [ \alpha^{04}(R,r) - 3 \alpha^{06}(R,r) \right ]. \label{eq_concentration_2dcircle_ss}
\end{align}
Comparing Eq.~\eqref{s_result_conc_app} with Eqs.~\eqref{t1r1p0_expanded} and \eqref{t1r0p1_expanded}, we have
\begin{align}
&c_{0}^{11}(R,r) = \frac{1}{r} \alpha^{12}(R,r), && c_{0}^{21}(R,r) = \alpha^{11}(R,r) - \alpha^{14}(R,r), \nonumber \\
&c_{1}^{21}(R,r) = \frac{2}{r^2} \alpha^{14}(R,r), && c_{0}^{31}(R,r) = \frac{1}{r} \left [ \alpha^{13}(R,r) - \alpha^{15}(R,r) \right ], \nonumber \\
&c_{1}^{31}(R,r) = \frac{2}{r} \left [ \alpha^{13}(R,r) - \alpha^{15}(R,r) \right ], && c_{2}^{31}(R,r) = \frac{4}{r^3} \alpha^{15}(R,r). \label{eq_concentration_2dcircle_v1}
\end{align}
Comparing Eq.~\eqref{s_result_conc_app} with Eqs.~\eqref{t2r1p0_expanded}, \eqref{t2r2p0_expanded}, \eqref{t2r1p1_expanded}, \eqref{t2r0p1_expanded}, and \eqref{t2r0p2_expanded}, we have
\begin{align}
&c_{0}^{12}(R,r) = \frac{1}{r} \alpha^{22}(R,r), && c_{0}^{22}(R,r) = \alpha^{21}(R,r) - \alpha^{24}(R,r), \nonumber \\
&c_{1}^{22}(R,r) = \alpha^{21}(R,r) - \alpha^{24}(R,r), && c_{2}^{22}(R,r) = \frac{2}{r^2} \alpha^{24}(R,r), \nonumber \\
&c_{3}^{22}(R,r) = \frac{2}{r^2} \alpha^{24}(R,r), && c_{0}^{32}(R,r) = \frac{2}{r} \left [ \alpha^{23}(R,r) - \alpha^{25}(R,r) \right ], \nonumber \\
&c_{1}^{32}(R,r) = \frac{4}{r} \left [ \alpha^{23}(R,r) - \alpha^{25}(R,r) \right ], && c_{2}^{32}(R,r) = \frac{8}{r^3} \alpha^{25}(R,r), \nonumber \\
&c_{3}^{32}(R,r) = \frac{2}{r} \left [ \alpha^{23}(R,r) - \alpha^{25}(R,r) \right ], && c_{4}^{32}(R,r) = \frac{1}{r} \left [ \alpha^{23}(R,r) - \alpha^{25}(R,r) \right ], \nonumber \\
&c_{5}^{32}(R,r) = \frac{4}{r^3} \alpha^{25}(R,r). && \label{eq_concentration_2dcircle_v2_a1}
\end{align}
Comparing Eq.~\eqref{s_result_conc_app} with Eqs.~\eqref{t3r1p0_expanded}, \eqref{t3r2p0_expanded}, \eqref{t3r3p0_expanded}, \eqref{t3r1p1_expanded}, \eqref{t3r1p2_expanded}, \eqref{t3r2p1_expanded}, \eqref{t3r0p1_expanded}, \eqref{t3r0p2_expanded}, and \eqref{t3r0p3_expanded}, we have
\begin{align}
&c_{0}^{13}(R,r) = \frac{1}{r} \alpha^{32}(R,r), && c_{0}^{23}(R,r) = \alpha^{31}(R,r) - \alpha^{34}(R,r), \nonumber \\
&c_{1}^{23}(R,r) = 3 \left [ \alpha^{31}(R,r) - \alpha^{34}(R,r) \right ], && c_{2}^{23}(R,r) = \frac{6}{r^2} \alpha^{34}(R,r),\nonumber \\
&c_{3}^{23}(R,r) = \frac{2}{r^2} \alpha^{34}(R,r), && c_{0}^{33}(R,r) = \frac{1}{r} \left [ \alpha^{33}(R,r) - \alpha^{35}(R,r) \right ], \nonumber \\
&c_{1}^{33}(R,r) = \frac{6}{r} \left [ \alpha^{33}(R,r) - \alpha^{35}(R,r) \right ], && c_{2}^{33}(R,r) = \frac{2}{r} \left [ \alpha^{33}(R,r) - \alpha^{35}(R,r) \right ], \nonumber \\
&c_{3}^{33}(R,r) = \frac{8}{r^3} \alpha^{35}(R,r), && c_{4}^{33}(R,r) = \frac{6}{r} \left [ \alpha^{33}(R,r) - \alpha^{35}(R,r) \right ], \nonumber \\
&c_{5}^{33}(R,r) = \frac{4}{r^3} \alpha^{35}(R,r), && c_{6}^{33}(R,r) = \frac{6}{r} \left [ \alpha^{33}(R,r) - \alpha^{35}(R,r) \right ], \nonumber \\
&c_{7}^{33}(R,r) = \frac{6}{r} \left [ \alpha^{33}(R,r) - \alpha^{35}(R,r) \right ], && c_{8}^{33}(R,r) = \frac{24}{r^3} \alpha^{35}(R,r). \label{eq_concentration_2dcircle_v3_v1a1_j1}
\end{align}
Each term in Eq.~\eqref{s_result_conc_app} for the camphor particle located at $\bm{\rho} = (\rho, \phi) = (0.1, 0)$ in the water chamber with a radius of $R=1$ is plotted in Figs.~\ref{concentration_2dcircle_ss} to \ref{concentration_2dcircle_v3}.
\begin{figure}[h]
	\begin{center}
		\includegraphics[bb=0 0 442 252]{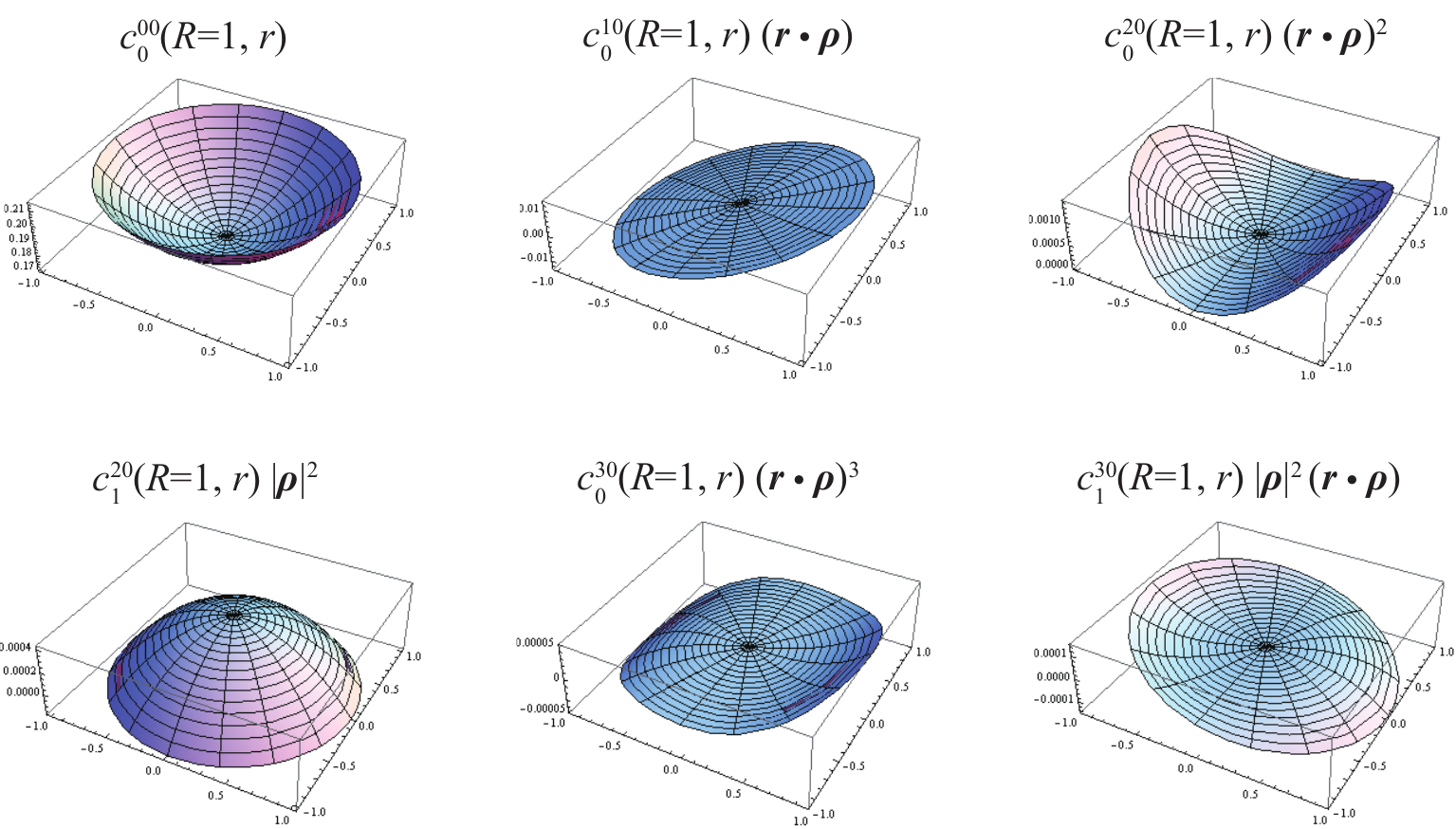}
		\caption{Concentration fields in the steady state for the resting camphor particle at $\bm{r} = \bm{\rho}$.
		The explicit expressions for the components of the concentration field are in Eq.~\eqref{eq_concentration_2dcircle_ss}.
		The radius of the water chamber $R$ is $R = 1$.
		Here we set $\bm{\rho} = (\rho, \phi) = (0.1, 0)$.
		}
		\label{concentration_2dcircle_ss}
	\end{center}
\end{figure}
\begin{figure}[h]
	\begin{center}
		\includegraphics[bb=0 0 451 260]{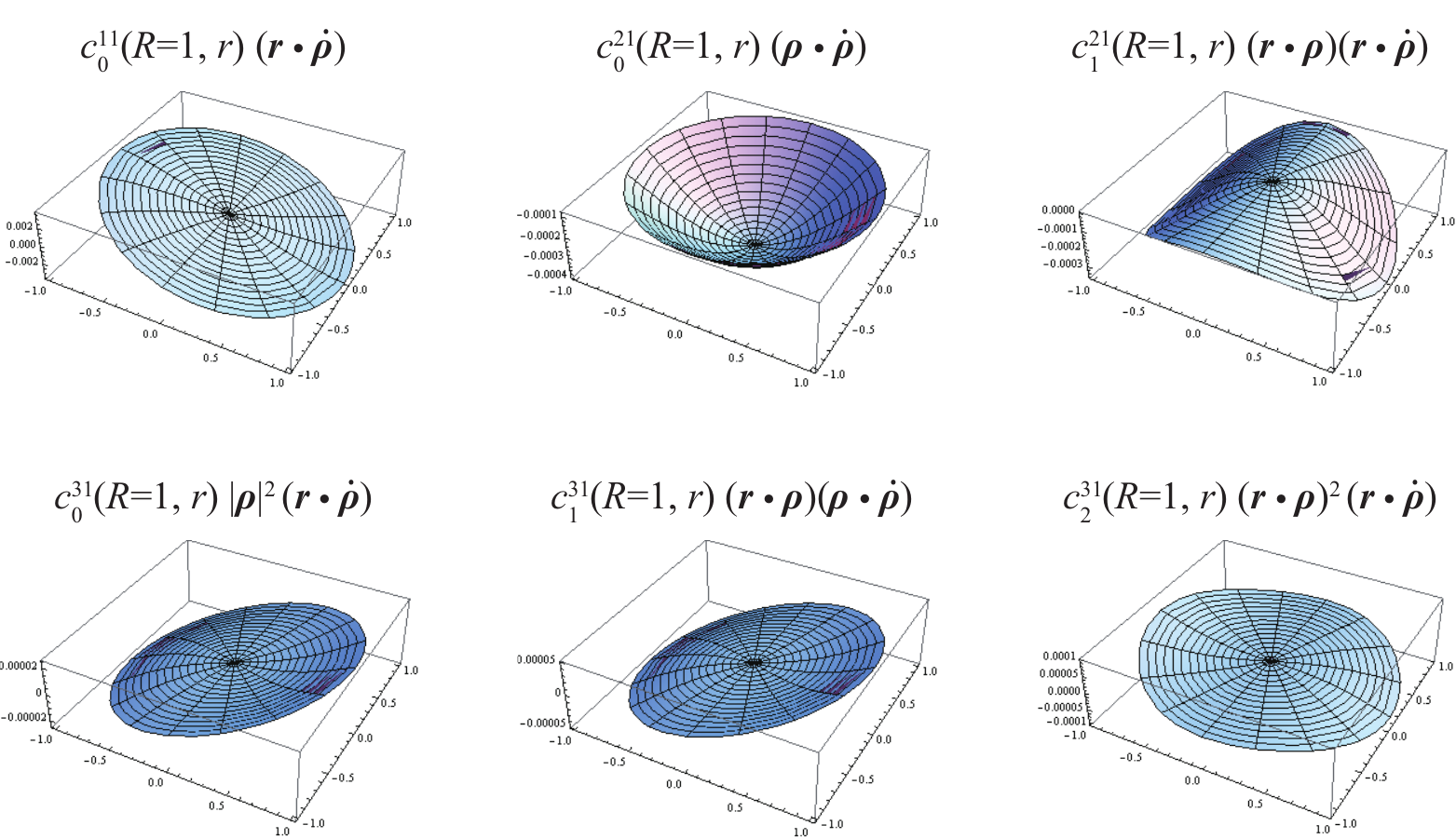}
		\caption{Concentration fields related to the first order of the velocity.
		The explicit expressions for the components of the concentration field are in Eq.~\eqref{eq_concentration_2dcircle_v1}.
		The radius of the water chamber $R$ is $R = 1$.
		Here we set $\bm{\rho} = (\rho, \phi) = (0.1, 0)$ and $\dot{\bm{\rho}} = (\dot{\rho}, \dot{\phi}) = (0.1, 0)$.}
		\label{concentration_2dcircle_v1}
	\end{center}
\end{figure}
\begin{figure}[h]
	\begin{center}
		\includegraphics[bb=0 0 437 258]{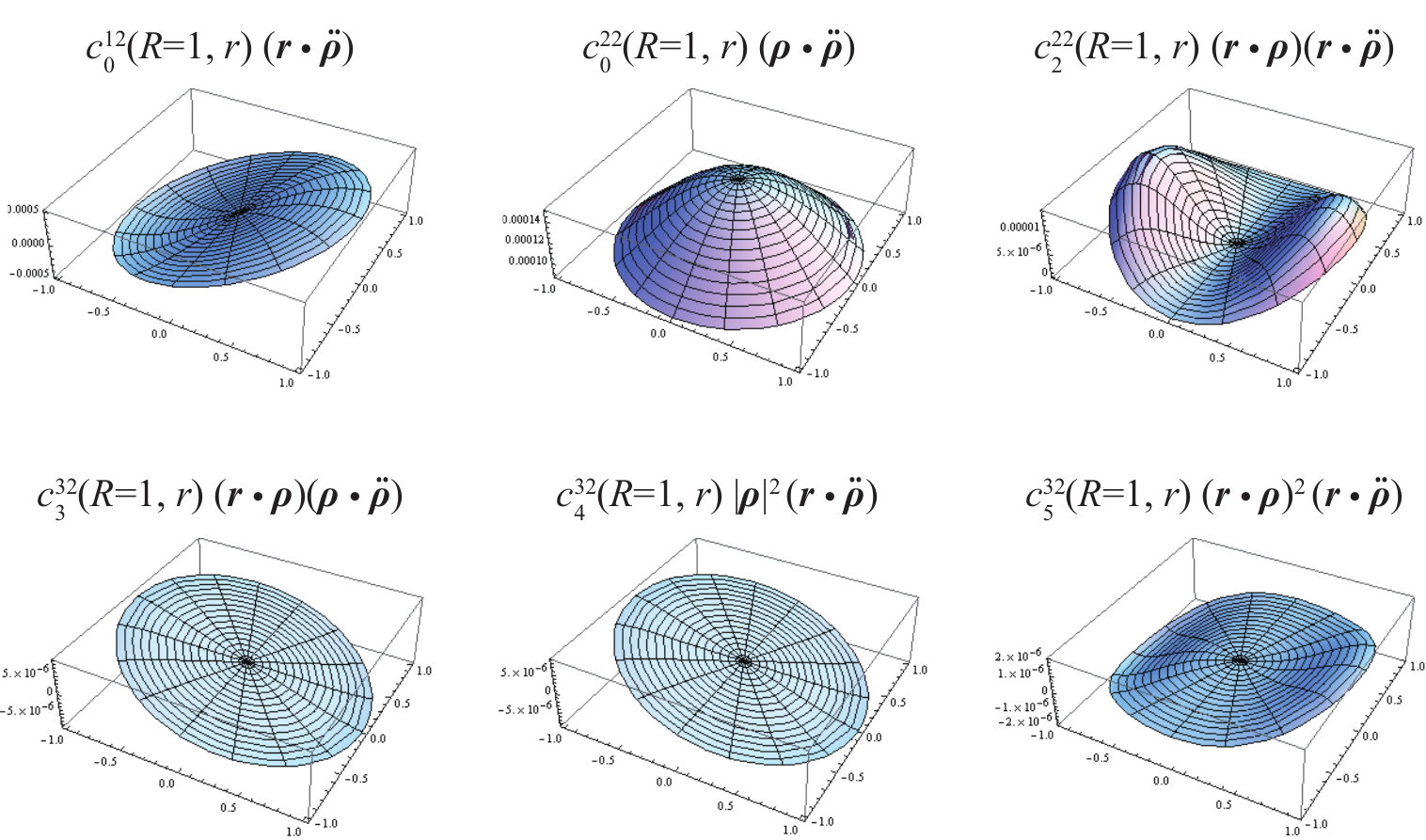}
		\caption{Concentration fields related to the first order of the acceleration.
		The explicit expressions for the components of the concentration field are in Eq.~\eqref{eq_concentration_2dcircle_v2_a1}.
		The radius of the water chamber $R$ is $R = 1$.
		Here we set $\bm{\rho} = (\rho, \phi) = (0.1, 0)$ and $\ddot{\bm{\rho}} = (\ddot{\rho}, \ddot{\phi}) = (0.1, 0)$.}
		\label{concentration_2dcircle_a1}
	\end{center}
\end{figure}
\begin{figure}[h]
	\begin{center}
		\includegraphics[bb=0 0 439 259]{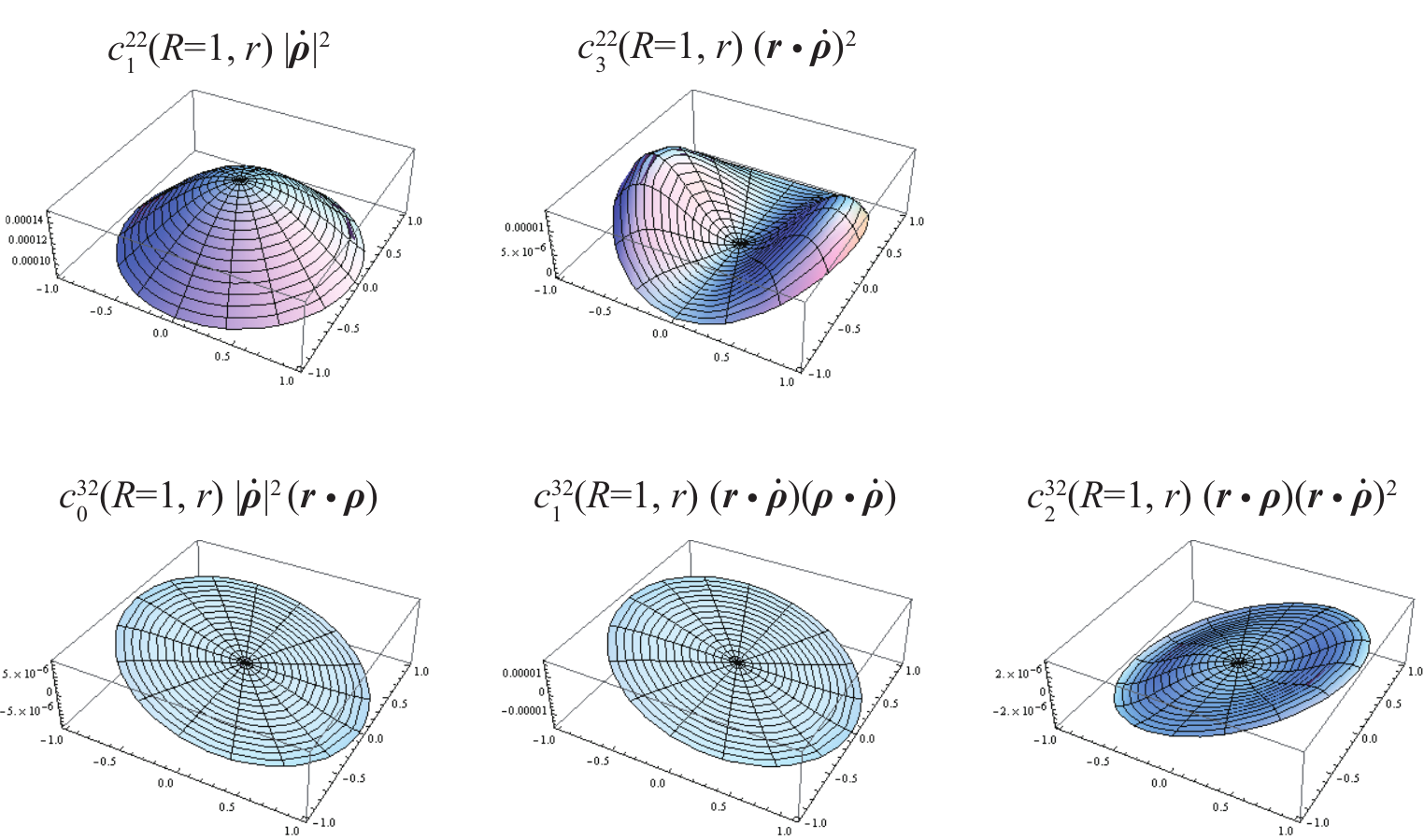}
		\caption{Concentration fields related to the second order of the velocity.
		The explicit expressions for the components of the concentration field are in Eq.~\eqref{eq_concentration_2dcircle_v2_a1}.
		The radius of the water chamber $R$ is $R = 1$.
		Here we set $\bm{\rho} = (\rho, \phi) = (0.1, 0)$ and $\dot{\bm{\rho}} = (\dot{\rho}, \dot{\phi}) = (0.1, 0)$.}
		\label{concentration_2dcircle_v2}
	\end{center}
\end{figure}
\begin{figure}[h]
	\begin{center}
		\includegraphics[bb=0 0 442 261]{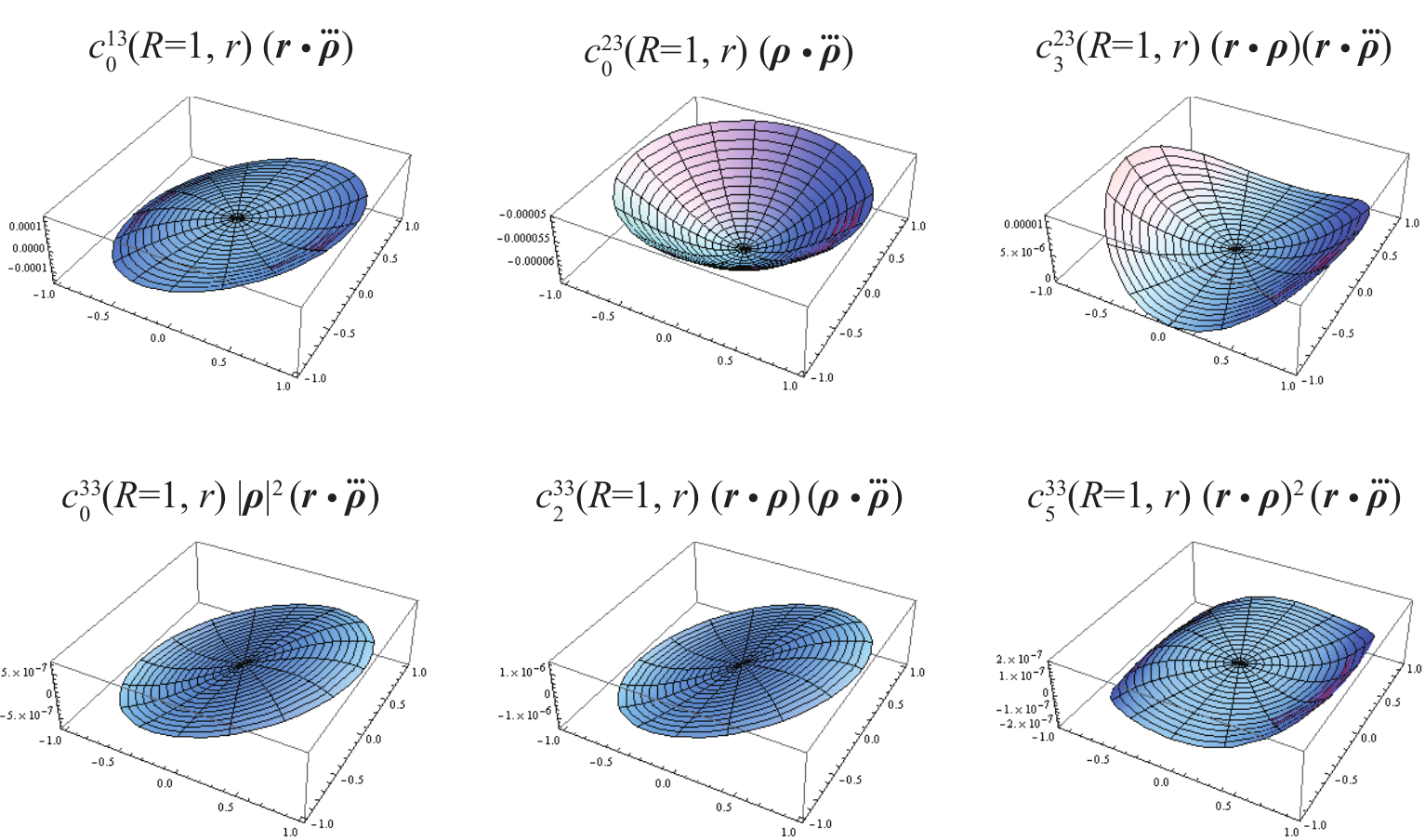}
		\caption{Concentration fields related to the first order of the jerk (time derivative of acceleration).
		The explicit expressions for the components of the concentration field are in Eq.~\eqref{eq_concentration_2dcircle_v3_v1a1_j1}.
		The radius of the water chamber $R$ is $R = 1$.
		Here we set $\bm{\rho} = (\rho, \phi) = (0.1, 0)$ and $\dddot{\bm{\rho}} = (\dddot{\rho}, \dddot{\phi}) = (0.1, 0)$.}
		\label{concentration_2dcircle_j1}
	\end{center}
\end{figure}
\begin{figure}[h]
	\begin{center}
		\includegraphics[bb=0 0 451 259]{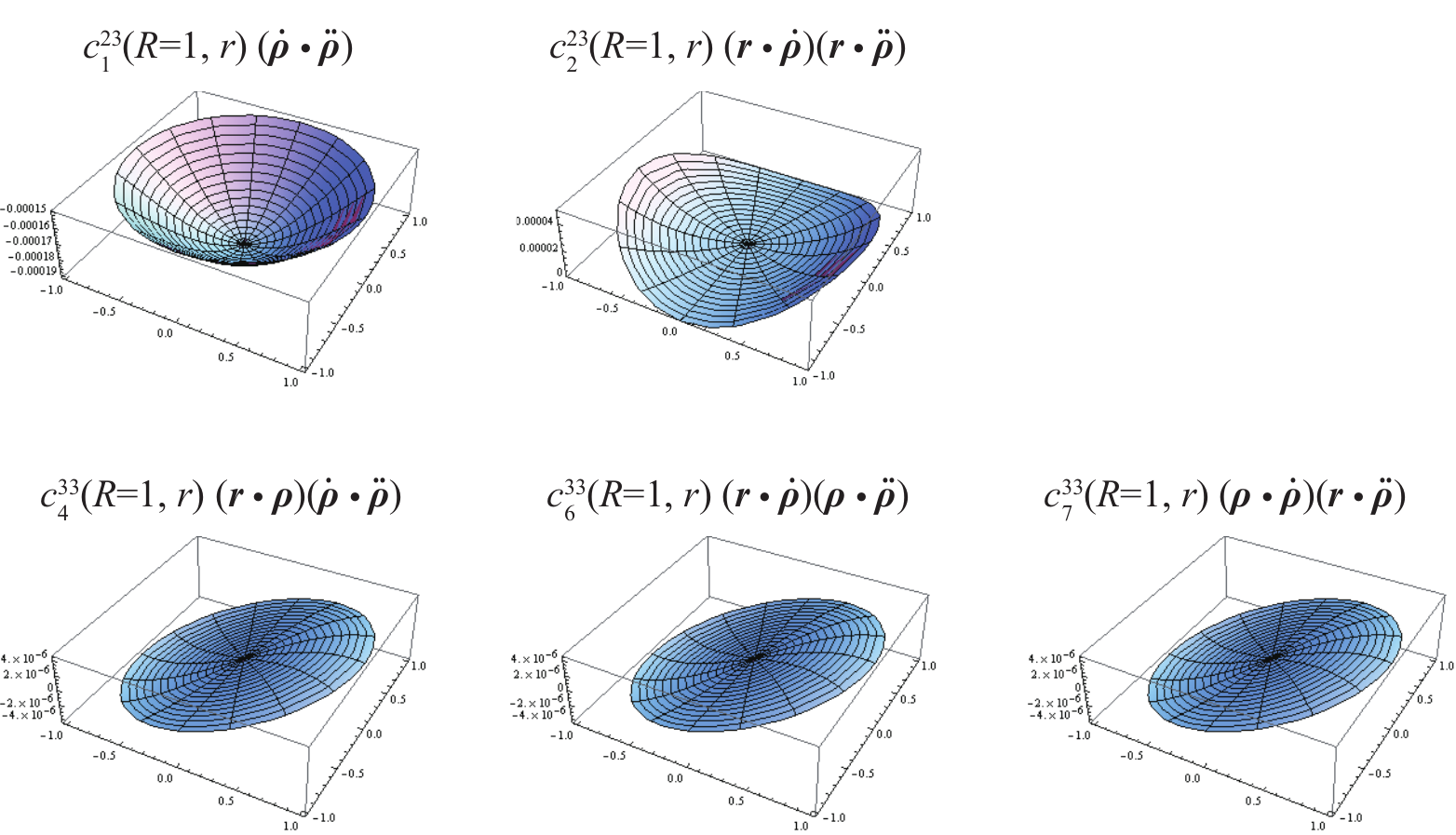}
		\caption{Concentration fields related to the cross term of the first order of the product of the velocity and acceleration.
		The explicit expressions for the components of the concentration field are in Eq.~\eqref{eq_concentration_2dcircle_v3_v1a1_j1}.
		The radius of the water chamber $R$ is $R = 1$.
		Here we set $\bm{\rho} = (\rho, \phi) = (0.1, 0)$ and $\dot{\bm{\rho}} = (\dot{\rho}, \dot{\phi}) = (0.1, 0)$.}
		\label{concentration_2dcircle_v1a1}
	\end{center}
\end{figure}
\begin{figure}[h]
	\begin{center}
		\includegraphics[bb=0 0 290 121]{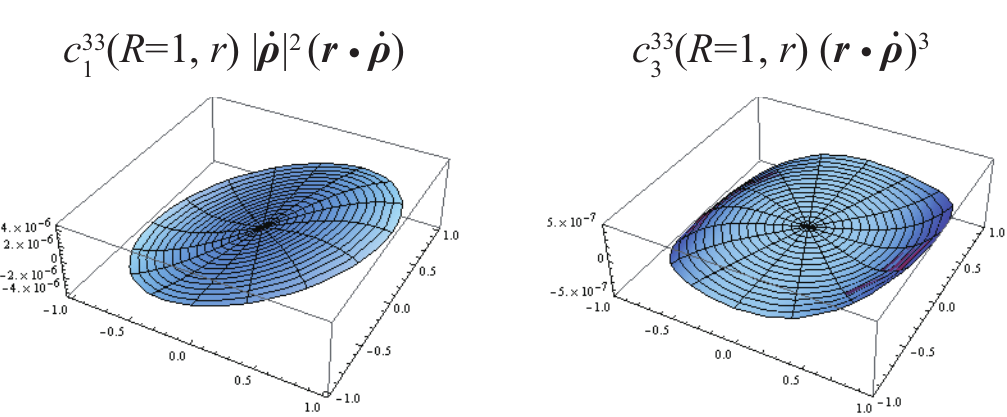}
		\caption{Concentration fields related to the third order of the velocity.
		The explicit expressions for the components of the concentration field are in Eq.~\eqref{eq_concentration_2dcircle_v3_v1a1_j1}.
		The radius of the water chamber $R$ is $R = 1$.
		Here we set $\bm{\rho} = (\rho, \phi) = (0.1, 0)$ and $\dot{\bm{\rho}} = (\dot{\rho}, \dot{\phi}) = (0.1, 0)$.}
		\label{concentration_2dcircle_v3}
	\end{center}
\end{figure}

\clearpage

\section{$R$-Dependence of the coefficients in Eq.~\eqref{DRIVING_FORCE}}

The coefficients of the terms in the driving force in Eq.~\eqref{DRIVING_FORCE} depends on $R$.
Here we show the dependence of $a(R)$, $b(R)$, $c(R)$, $g(R)$, $h(R)$, $j(R)$, $k(R)$, $n(R)$, and $p(R)$ on the radius of circular region $R$ in Fig.~\ref{coefficients_2D}.
\begin{figure}[h]
	\begin{center}
		\includegraphics[bb=0 0 503 409]{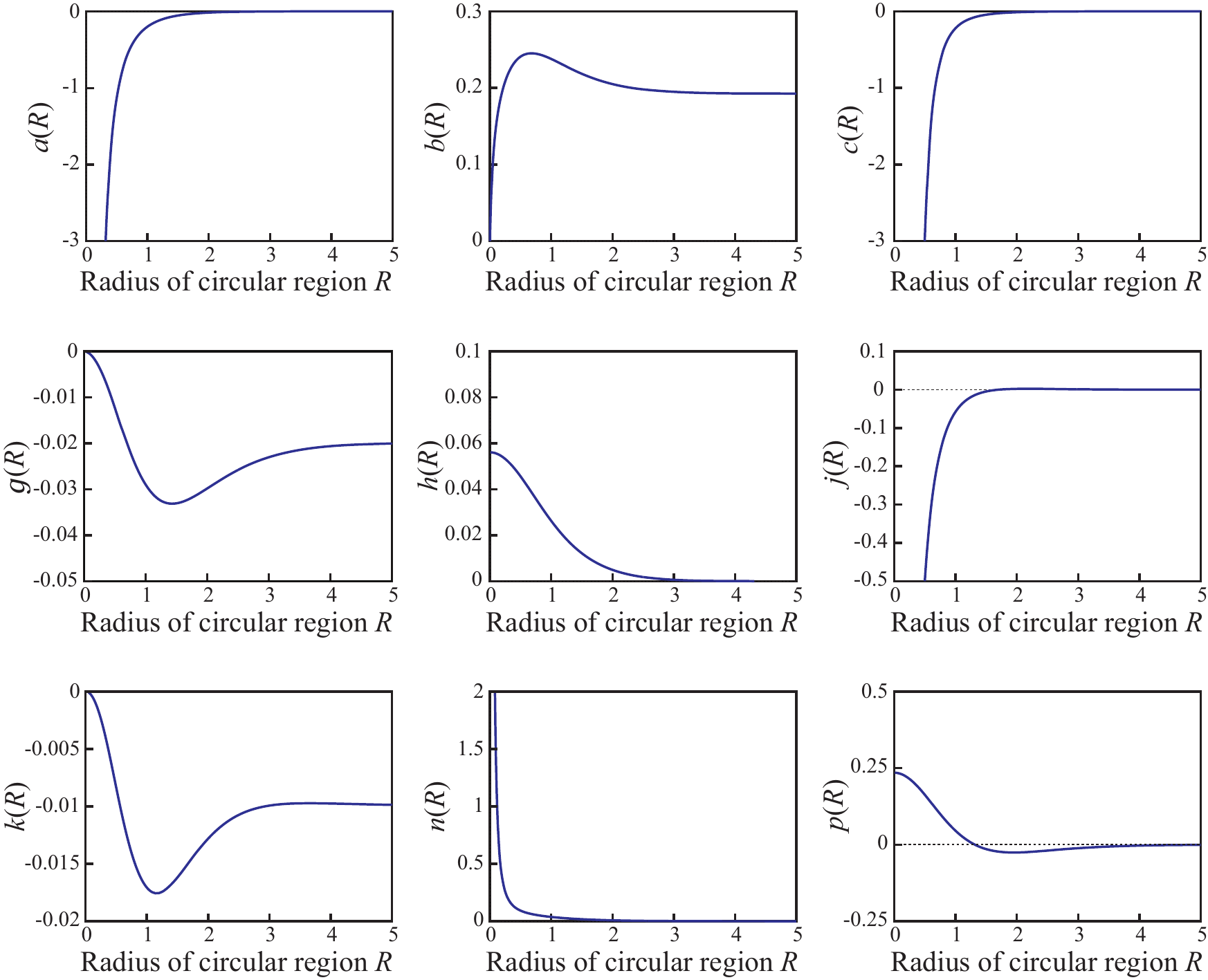}
		\caption{Plots of the coefficients $a(R)$, $b(R)$, $c(R)$, $g(R)$, $h(R)$, $j(R)$, $k(R)$, $n(R)$, and $p(R)$ against the radius of circular region $R$, which are explicitly shown in Eqs.~\eqref{AR} to \eqref{PR}.}
		\label{coefficients_2D}
	\end{center}
\end{figure}

\end{widetext}
\end{document}